\documentclass[english]{emulateapj}
\usepackage[T1]{fontenc}
\usepackage[latin1]{inputenc}
\usepackage{graphicx}
\usepackage{amssymb}
\usepackage{enumerate}

\makeatletter
\usepackage{times}
\usepackage{amsmath}
\usepackage{subfigure}
 \usepackage{multirow}

\newcommand{\Mbh}{M_{\bullet}}

\newcommand{\Mo}{M_{\odot}}

\mathchardef\mhyphen="2D

\def\arcdeg{\hbox{$^\circ$}}

\def\arcsec{\hbox{$^{\prime\prime}$}}

\newcommand{\peryr}{\,{\rm yr^{-1}}}

\newcommand{\yr}{{\,\rm yr}}

\newcommand{\Gyr}{{\,\rm Gyr}}
\newcommand{\Myr}{{\,\rm Myr}}

\newcommand{\pc}{\,\mathrm{pc}}

\shorttitle{B-STARS IN THE GALACTIC CENTER}
\shortauthors{MADIGAN ET AL.}

\usepackage{babel}
\makeatother
\begin{document}
\bibliographystyle{apj} 

\title{On the origin of the B-stars in the Galactic center}

\author{Ann-Marie Madigan$^{1,2,3}$, Oliver Pfuhl$^{4}$, Yuri Levin$^{3,5}$, Stefan Gillessen$^{4}$, Reinhard Genzel$^{4,6}$ and Hagai B. Perets$^{7}$}

\affiliation{$^1$Astronomy Department and Theoretical Astrophysics Center, University of California, Berkeley, CA 94720, USA \\$^2$Einstein Postdoctoral Fellow; ann-marie@astro.berkeley.edu \\$^3$Leiden Observatory, Leiden University, P.O. Box 9513, NL-2300 RA Leiden, The Netherlands \\$^4$Max-Planck Institut f\"ur Extraterrestrische Physik, 85748 Garching, Germany \\$^5$School of Physics, Monash University, Clayton, Victoria 3800, Australia \\$^6$Department of Physics, University of California, Berkeley, CA 94720, USA \\$^7$ Deloro Fellow; Physics Department, Technion - Israel Institute of Technology, Haifa, Israel 32000}

\begin{abstract}

We present a new directly-observable statistic which uses sky position ($x, y$) and proper motion ($v_x, v_y$) of stars near the Galactic center massive black hole to identify populations with high orbital eccentricities. It is most useful for stars with large orbital periods for which dynamical accelerations are difficult to determine. 

We apply this statistic to a data set of B-stars with projected radii $0.1 \arcsec\! < p \!< 25 \arcsec$ ($\sim0.004 - 1 \pc$) from the massive black hole in the Galactic center. We compare the results with those from $N$-body simulations to distinguish between scenarios for their formation. We find that the scenarios favored by the data correlate strongly with particular $K$-magnitude intervals, corresponding to different zero-age main-sequence (MS) masses and lifetimes. Stars with $14 \lesssim m_K \lesssim 15$ ($15 - 20 \Mo$, $t_{\rm MS} = 8-13 \Myr$) match well to a disk formation origin, while those with $m_K \geq 15$ ($<15 \Mo$, $t_{\rm MS} >13 \Myr$), if isotropically distributed, form a population that is more eccentric than thermal, which suggests a Hills binary-disruption origin. 

\end{abstract}


\section{Introduction}

The nuclear star cluster (NSC) within the central few parsecs of our galaxy contains a massive black hole (MBH), SgrA*, and $\sim 10^7 \Mo$ in stellar populations of various ages. The bulk of the stars are old, $\sim 80\%$ forming more than 5 Gyr ago, possibly at the same time as the galactic bulge. After a period of reduced star formation, the star formation rate increased during the last $200 - 300 \Myr$ \citep{Blu03,Pfu11}. While the late type stars may be too old to retain memory of their initial orbital configuration, and hence formation mechanism, the kinematics of the early type stars should reflect their original distribution. One can resolve these stars individually due to our proximity to the Galactic center (GC)\footnote{We adopt a distance to SgrA* of 8.3 kpc \citep{Eis03,Ghe08a,Gil09a,Gil09b} for which $1\arcsec \sim 0.04 \pc$, an average extinction of $A_{K_s} = 2.7$ \citep{Fri11}, and refer the reader to \citet{Gen11} for a complete review of the GC.}, and use their phase space parameters to constrain formation scenarios. This is a useful way to understand stellar mass accumulation in NSCs in general, which is very likely connected to the formation and growth of MBHs at their centers \citep{Hop10a, Hop10b}. 

There are two fundamental scenarios for the accumulation of stars in NSCs. The first scenario is the merger of multiple star clusters following migration towards the center of a galaxy via dynamical friction with background stars \citep{Tre75,Cap93,Aga11,Ant12a,Ant13a}. The second scenario is in-situ formation in nuclear stellar disks as a result of gas migration into the center of galaxies \citep{Mil04b}. Observational results and theoretical arguments suggest that both mechanisms are necessary to explain the morphology, kinematics and complex star formation history of NSC stellar populations \citep{Har11, Lei12, Ant12a}. 

Massive young stars in the central parsec of the GC provide evidence of the second mechanism, that is formation in a nuclear stellar disk. Roughly half of the brightest young stars \--- a population of O- and Wolf-Rayet (WR) stars, $\sim 6 \pm 2 \Myr$ old\footnote{\citet{Lu13} analyze the entire population of young stars as a single starburst cluster and find an age between $2.5-5.8 \Myr$ with $95\%$ confidence. Though a younger age than commonly adopted for the O/WR stars, it is consistent within the uncertainty range reported by \citet{Pau06}.}, with masses $\gtrsim 20 \Mo$ \citep{Pau06} \--- form a thin, clockwise(CW)-rotating disk with projected radii $0.8 \arcsec \-- 12\arcsec$ \citep{Lev03, Gen03a,Lu09, Bar09}, though more young star candidates have been detected at larger distances \citep{Bar10,Nis12}. This disk is thought to have formed in-situ from the fragmentation of in-falling or colliding gas clumps at the GC \citep{Mor93, San98, Lev03, Nay05b,Lev07,War08,War12}. We refer to this structure as the ``young CW disk''.  
 
One also observes a population of fainter B-stars, $m_{\rm k} \gtrsim 14$ ($m_K=14$ corresponds to a B0V star), which are not obviously associated with this disk. They appear more isotropically distributed than the brighter stars, though a number may be members of the young CW disk \citep{Bar10}. They are not truncated in projected radius at the disk inner edge, but continue inwards to the MBH. Those that lie within the central $0.8 \arcsec$ are collectively referred to as the ``S-stars''; their kinematics reveal randomly-inclined and near-thermal eccentricity orbits \citep{Ghe05,Eis05,Gil09a}.

The orbits of the B-stars further out have not yet been determined and it is unclear whether or not the S-stars and the outermost B-stars form distinct populations. Their ages range from a spectroscopically confirmed $<\!  10 \Myr$ for the S-star S2/S0-2 \citep{Ghe03a,Eis05,Mar08}, to an upper-limit of $\sim \! 100 - 200 \Myr$ on the main sequence lifetime of the lower-mass B-stars. This upper limit does not preclude the B-stars forming contemporaneously with the young CW stars, but they may well derive from an older starburst or even form a continuous distribution in age. 

The proximity of the S-stars to the MBH, which prohibits in-situ star formation due to its immense tidal force \citep{Mor93}, combined with their young ages imply a ``paradox of youth'' \citep{Ghe03a}. Arguably, the most plausible theory for their origin is the tidal capture by SgrA* of in-falling B-star binaries by Hills mechanism \citep{Hil88,Hil91,Gou03}, following dynamical relaxation by massive perturbers such as giant molecular clouds \citep{Per07} within the central $10 \-- \!100 \pc$. In this theory, the S-stars are captured on orbits of very high orbital eccentricity whilst their binary companions may be ejected as hypervelocity stars \citep[see e.g., ][and references therein]{Bro12b}. \citet{Ant13b} show that post-capture dynamical evolution via resonant relaxation \citep{Rau96} can bring the highly-eccentric population of S-stars close to their observed near-thermal eccentricity distribution \citep{Gil09a} within $50 \Myr$ for models of the GC with relaxed NSC, or $\sim 10 \Myr$ for models with a dense cluster of $10 \Mo$ black holes \citep[see also][]{Per09b}. It is possible that the B-stars outside the central arcsecond also formed via Hills mechanism, as proposed by \citet[][hereafter PG10]{Per10}. However, as the latter authors point out, the initial high-eccentricity distribution must persist, since neither two-body nor resonant relaxation will be able to significantly change the orbital eccentricities of the B-stars at large radii within their lifetimes. This sets up a prediction which observational data from this population can verify or refute.

Outside of the central arcsecond, the accelerations of the B-stars are too small to be reliably detected within $\sim 10 \yr$ of observations. Thus we do not get a full orbital solution for each star. In this paper we devise a statistic which uses only the star's sky position and proper motion velocity and is particularly sensitive in identifying distributions with high orbital eccentricities. We present this high-eccentricity statistic in Section \ref{S:h-stat}. In Section \ref{S:simulations} we explore another mechanism for dynamically relaxing the orbital eccentricity distribution of B-stars \--- the formation and gravitational influence of the young CW stellar disk \--- and investigate two scenarios with $N$-body simulations: the massive perturber plus binary disruption scenario \citep{Per07}, and one based on the proposed model by \citet{Set06} of episodic in-situ star formation, wherein the B-stars formed in a nuclear stellar disk $\sim \! 100 \Myr$ ago. We examine the resulting orbital eccentricities of the B-stars after $6 \Myr$ of interaction with the young CW disk. In Section \ref{S:obs} we introduce our observations and use direct observables and the high-eccentricity statistic to compare them with simulations in Section \ref{S:comp}. We discuss our findings in Section \ref{S:discussion}. In a follow-up paper we will expand our current analysis on the orbital parameters of the B-stars including radial velocity information. 


\section{The high-eccentricity statistic}\label{S:h-stat}

\begin{figure*}[th!]
\begin{minipage}[b]{0.5\linewidth}
\centering   
   \includegraphics[trim=0cm  0.0cm 0cm 0cm, clip=true, angle = -90, scale=0.35]{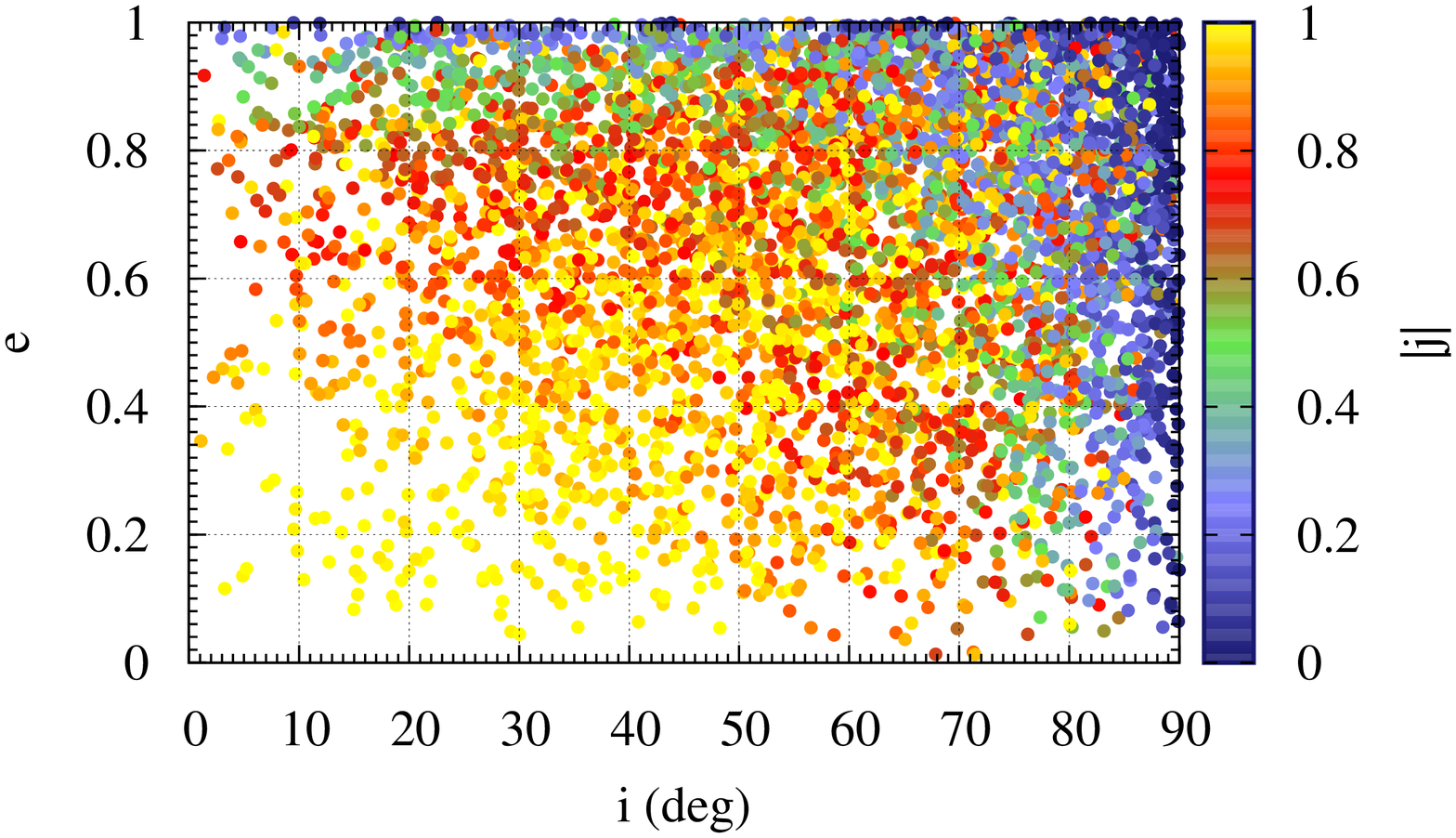}
    		\vspace{-0 pt}
				\end{minipage}
\begin{minipage}[b]{0.5\linewidth}
\centering
  \includegraphics[trim=0cm 0cm 0.cm 0cm, clip=true, angle = -90, scale=0.35]{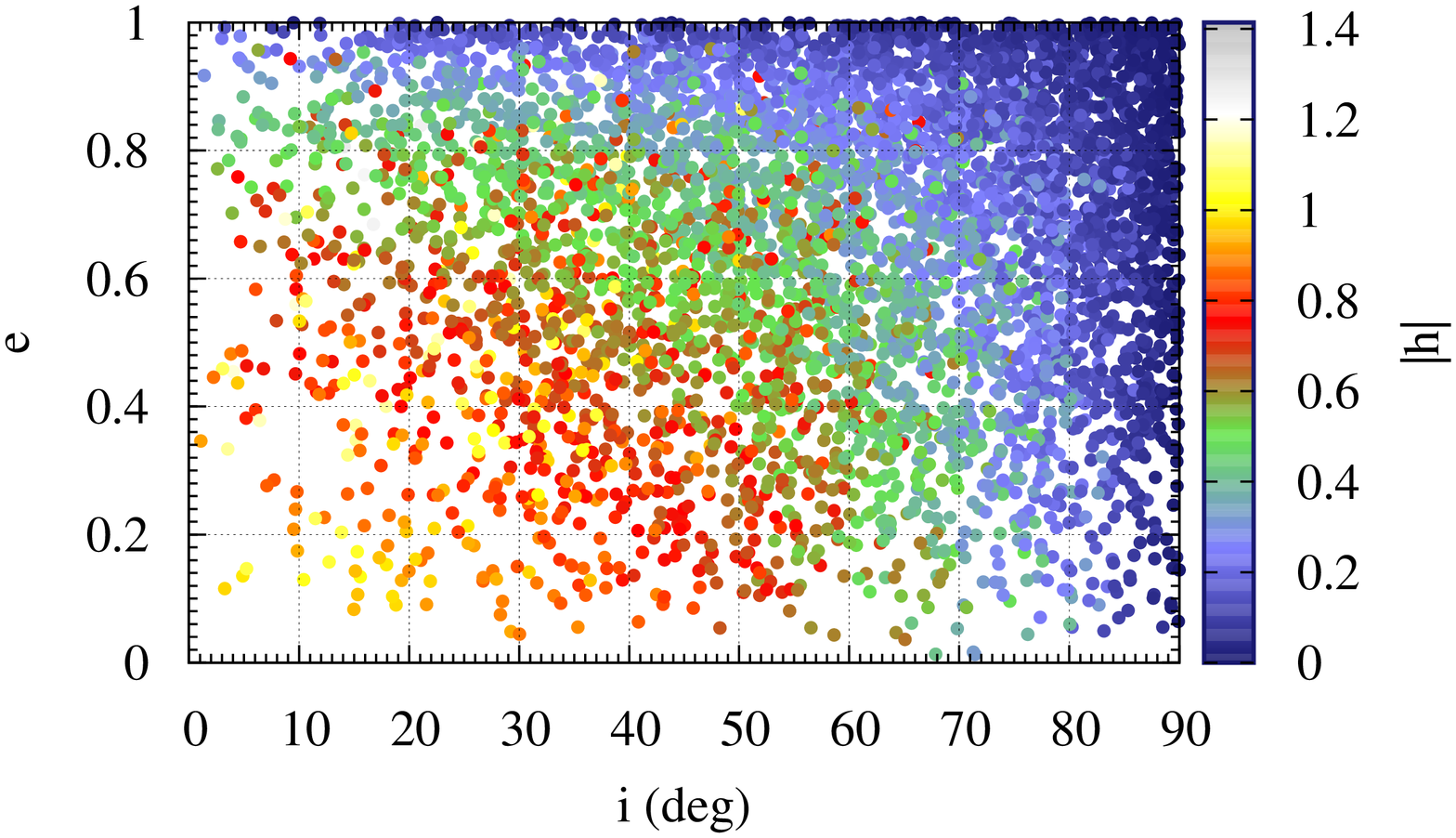}
    		\vspace{-0 pt}
\end{minipage}
\caption{Plot of $|j|$-values (left) and $|h|$-values (right). Each point represents a star drawn from an isotropic stellar distribution, with $i = 0 \arcdeg$ ($90 \arcdeg$) corresponding to a face-on (edge-on) orbit, and with orbital eccentricity $e$. From this figure it is clear that the colors are much better stratified in the right panel than in the left panel, because the $h$-statistic differentiates more cleanly between high and low eccentricity orbits than the $j$-statistic.} 
\label{fig:inc_ecc_h_j}
\end{figure*}

The basic idea for identifying stars with high orbital eccentricities is straightforward: a radial orbit in three spacial dimensions also appears as a radial orbit in projection on the sky. This was noted by \citet{Gen03a} and revisited by \citet{Pau06} and \citet{Bar09}. These authors use the $j$ versus $p$ diagram, where
$j$ is the normalized angular momentum along the line-of-sight (positive $z$-axis),
\begin{equation}
\begin{split}
j & = \frac{j_z}{j_{\rm z(max)}} \\
&= \frac{x v_y - y v_x }{p v_p},
\end{split}
\end{equation}
and 
\begin{equation}
p = (x^2 + y^2)^{1/2}
\end{equation}
is the projected radius from the MBH. The positive $x$-axis points west and the positive $y$-axis points north. $v_x,v_y$ are the right ascension and declination velocities of a star at $(x,y)$ on the sky such that the projected velocity (i.e., proper motion) is
\begin{equation}
v_p= (v_x^2 + v_y^2)^{1/2}.
\end{equation}
The quantity $j$ is $\sim\! 1, \sim\! 0, \sim\! -1$ if the stellar orbit projected on the sky is mainly clockwise (CW) tangential, radial, or counterclockwise (CCW) tangential. \citet{Gen03a} define three $j$ ranges: CW tangential ($j \ge 0.6$), CCW tangential ($j \le -0.6$) and radial ($|j| \le 0.3$). Though a useful tool for classification, the quantity $j$ is not optimally sensitive to high-eccentricity ($j \sim 0$) orbits. This is because stars on radial orbits spend the majority of their orbital period near apoapsis with low $v_p$ with respect to the circular velocity at their projected radii $p$; this increases their value of $j$ and imparts a more tangential orbit in projection. In its place, we propose to use a new high-eccentricity statistic, $h$: $j_z$ normalized to the maximum angular momentum at projected radius $p$ (i.e. replacing $v_p$ with circular velocity at $p$):
\begin{equation}
\begin{split}
h & = \frac{j_z}{J_p} \\
&= \frac{x v_y - y v_x }{\sqrt{G \Mbh p}},
\end{split}
\end{equation}
where we use the Kepler circular velocity such that $v_{\rm circ}(p) = (G\Mbh/p)^{1/2}$. As with $j$, the quantity $h$ is $\sim\! 1, \sim\! 0, \sim\! -1$ depending on whether the stellar orbit projected on the sky is mainly CW tangential, radial, or CCW tangential, but radial orbits are now confined to low $|h|$-values. 
We show this in Figure \ref{fig:inc_ecc_h_j} where we initialize a cluster of $10^4$ isotropically arranged stars with a thermal distribution of orbital eccentricities to see how well their $j$- and $h$-values constrain their original orbital eccentricity. Each star is distributed randomly in its orbital phase and their positions and velocities are projected onto the plane of the sky to get values for $j$ and $h$. We plot the inclination of the stellar orbits\footnote{The inclination, $i$, of a stellar orbit is calculated from the angle between its angular momentum vector and the positive $z$-axis (line-of-sight). Hence $i = 0 \arcdeg$ ($i = 90 \arcdeg$) corresponds to a face-on (edge-on) orbit.} (ranging from $0 \arcdeg - 90 \arcdeg$ as the distribution in $|j|$ and $|h|$ is symmetric about this range) as a function of orbital eccentricity. The left (right) plot shows stars color-coded according to $|j|$ ($|h|$)-values.

The $h$-statistic differentiates well between tangential and radial orbits. Although low $|h|$-values can correspond to stars that have high-eccentricity and/or inclined orbits (i.e., edge-on with respect to line-of-sight), high eccentricity orbits are not contaminated by high $|h|$-values; $|h|$-values are sharply defined as a function of orbital eccentricity and inclination. 

In contrast, high $|j|$-values (yellow dots) are scattered throughout the inclination and eccentricity plane. Many high eccentricity orbits are represented by high $|j|$-values which makes them hard to isolate as a group. We refer the reader to the Appendix for statistical constraints on orbital eccentricity and inclination in different $h$-ranges, the maximum value of $|h|$ for a bound orbit and the effect of the stellar gravitational potential on its estimate.

\section{$N$-body simulations}\label{S:simulations}

We perform $N$-body simulations of two formation scenarios for the population of large-radii B-stars in the GC --- a disk origin, and a binary disruption origin. The stars in the two formation scenarios differ only in orbital angular momentum distribution. We investigate whether their original orbital eccentricities are preserved over $6 \Myr$, having been subjected to gravitational torquing from the young CW disk, and calculate the resulting $h$-values to be compared with observations. 

We use a special-purpose $N$-body integrator, which is described in detail in \citet{Mad11b}. Our integrator is based on a mixed-variable symplectic algorithm \citep{Wis91,Kin91,Sah92} and designed to accurately integrate the equation of motion of a particle in a near-Keplerian potential. We use direct $N$-body particles which move in Kepler elements along ellipses under the influence of the central object, and calculate perturbations to their orbits in Cartesian co-ordinates from surrounding $N$-body particles \citep{Dan92}. We define semi-major axis, $a$, and eccentricity, $e$, of stellar orbits with respect to a stationary MBH,
\begin{equation}
a = - \frac{G \Mbh}{2 E}, 
\end{equation}
and
\begin{equation}
e =  \left( 1 - \frac{J^2}{G \Mbh a} \right)^{1/2},
\end{equation}
where $\Mbh$ is the mass of the MBH, and $J$ and $E$ are the specific orbital angular momentum and energy of a star. The periapsis of the stellar orbits precess with retrograde motion due to Newtonian mass precession from the additional smooth potential from surrounding cluster of stars. 

The time to precess by $2\pi$ radians is
\begin{equation} \label{e:prec}
t^{\rm cl}_{\rm prec} = \pi(2 - \alpha)\dfrac{\Mbh}{N(<a)m} P(a) f(e, \alpha),
\end{equation}
where $N(<a)$ is the number of stars within a given $a$, $m$ is the mass of a single star, $f(e,\alpha)$ is a function which depends on the eccentricity of the orbit and the power-law density index $\alpha$ of the surrounding cluster of stars \citep{Iva05,Mad12}, and $P(a) = 2\pi(a^3/G\Mbh)^{1/2}$ is the orbital period of a star with semi-major axis $a$. We include the first post-Newtonian general relativistic effect, that is prograde apisidal precession with a timescale,
\begin{equation} \label{e:prec_GR}
t^{\rm GR}_{\rm prec} = \dfrac{1}{3} (1 - e^2) \dfrac{a c^2}{G\Mbh} P(a).
\end{equation}

Our simulations have four main components chosen to represent the Galactic center $\sim 6 \Myr$ in the past:
\begin{enumerate}
\item A MBH of mass $\Mbh = 4.3 \times 10^6 \Mo$ \citep{Ghe08a,Gil09b}. 

\item A smooth stellar cusp with power-law density profile $n(r) \propto r^{-\alpha}$, $\alpha = (0.5, 1.75)$, normalized with a mass of $1.5 \times 10^6 M_{\odot}$ within 1 pc \citep{Sch07, Tri08, Sch09}. We use a smooth gravitational potential for the cusp in our simulations as this greatly decreases the required computation time. Traditional two-body gravitational relaxation has little impact on the stellar orbits as its characteristic  timescale is $\mathcal{O}(1 \Gyr)$ \citep{Mer10,Bar12}. We model two-body relaxation and resonant relaxation (which occurs on a shorter timescale), using the ARMA code described in detail in \citet{Mad11b} to derive initial conditions for our simulations (see Figure \ref{fig:arma_bd}). \citet{Mad09} and \citet{Gua12} find that in the case of coherently-eccentric disks, significant angular momentum changes due to self-gravity of the disk occurs on timescales $\lesssim 1 \Myr$.  

\begin{figure}[t!]
  \centering
    \includegraphics[angle = -90, scale=0.34]{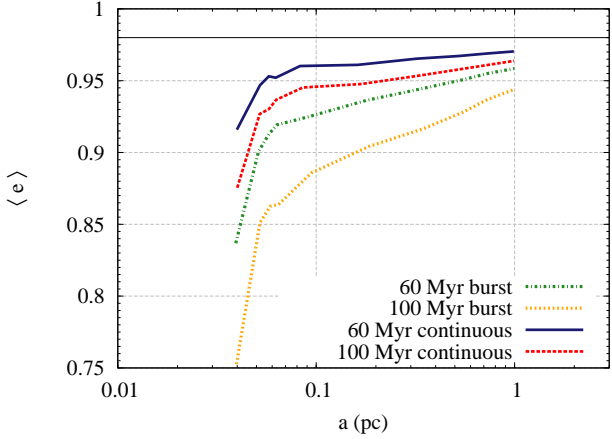}
\caption{Mean B-star orbital eccentricities as a function of semi-major axis in the {\it binary disruption} scenario. Stars begin with initial orbital eccentricity $e = 0.98$ and evolve due to stochastic relaxation over $60$ and $100 \Myr$. In the {\it burst} scenario, stars are initialized at $t = 0$; in the {\it continuous} scenario, they are initialized randomly between $t = 0, t_{\rm max}$. For this plot we use our ARMA code \citep{Mad11b} with $M(<1\pc) = 1.5 \times 10^6 \Mo$ and $\alpha = 1.75$.}
\label{fig:arma_bd}
\end{figure}

\item An eccentric stellar disk, $e = (0.3, 0.6$), representing the young CW disk with surface density profile $\Sigma(a) \propto a^{-2}$. It consists of $N$-body particles with equal masses of $100 \Mo$, total mass $M_{\rm CW} = (1,2,4) \times 10^4 \Mo$ and semi-major axes $0.03 \pc \!\le\! a \!\le\! 0.5 \pc$. In our basic model, the young CW disk is formed instantaneously, i.e., fully formed at $t = 0$, with an opening angle of $1 \arcdeg$. We also run a number of simulations wherein we model its formation using a ``switch-on'' multiplicative function for the disk mass, such that the mass of a single star is
\begin{equation} \label{e:switch}
m(t) = \dfrac{M_{\rm disk}}{N_{\rm disk}} \tanh \left(\dfrac{t-t_0}{\tau}\right),
\end{equation}
where $t_0$ is $\mbox{-} 1$ years (so that the young CW disk has mass at t = 0) and the growth timescale $\tau$ is $ 1 \times 10^5$ years \citep{BoR08}. Secular gravitational interactions with the young CW disk will change the angular momenta of the B-stars. We anticipate the greatest orbital eccentricity change for stars at similar radii to the inner edge of the disk \--- the torques are much greater at these radii, as 
\begin{equation} \label{e:tau}
\tau \sim \left( \dfrac{G M_{\rm disk} e_{\rm disk} e}{a} \right) \delta \phi,
\end{equation}
where $\tau$ is specific torque on a stellar orbit, $e_{\rm disk}$ is a typical orbital eccentricity of a star in the disk and $\delta \phi$ is the angle between them.

\begin{figure}[t!]
  \centering
    \includegraphics[angle = -90, scale=0.34]{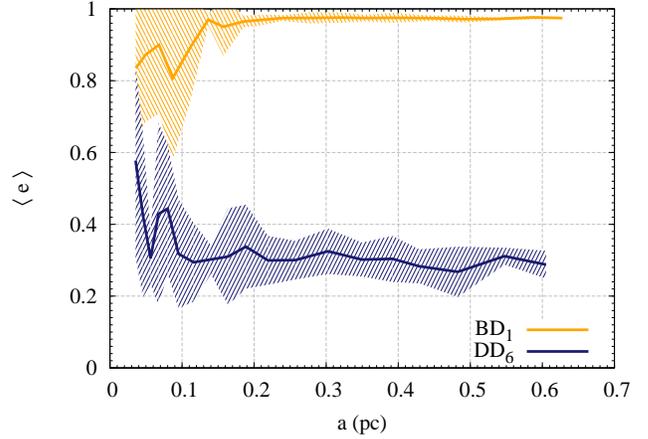}
\caption{Mean orbital eccentricity and one standard deviation of B-stars after $6 \Myr$ of evolution in the {\it dissolved disk} DD$_6$ scenario and {\it binary disruption} BD$_1$ scenario as a function of semi-major axis $a$. Stars at large $a$ do not evolve far from their initial eccentricity values.}
\label{fig:eccVa}
\end{figure}

\item A population of B-stars with semi-major axes between $0.03 \pc \!\le\! a \!\le\! 0.7 \pc$, consisting of 100 $N$-body particles with equal masses of $100 \Mo$. Though the individual masses of the stars are high with respect to real B-star masses, the orbits will respond to the gravitational potential of the young CW disk in the same way, regardless of their mass,  due to the equivalence principle. We run convergence tests to confirm this; see appendix. The stars are initialized with a surface density profile $\Sigma(a) \propto a^{-2}$ while their angular momentum distribution depends on the scenario we are simulating; we describe them both here. 

{\it Binary Disruption (BD) Scenario}:  In this scenario, the B-stars form in binaries outside the central parsec. Due to enhanced relaxation from massive perturbers \citep{Per07}, they are propelled onto near-radial orbits where they are disrupted by the MBH via Hills' mechanism and become bound to the MBH as their partners are ejected into the halo at high velocities. They have a spatially isotropic distribution and are initialized on Kepler orbits with very high eccentricities, $e = 1 - r_t/a \sim 1- (m_{\rm bin}/\Mbh)^{1/3}$ \citep[see, e.g.,][]{Pfa05}, where $r_t =  a_{\rm bin} (\Mbh/m_{\rm bin})^{1/3}$ is the tidal radius of the MBH, and $m_{\rm bin}$ and $a_{\rm bin}$ are the mass and semi-major axis of the binary.

We use our ARMA code \citep{Mad11b} to simulate the evolution of orbital eccentricities for B-stars in the {\it binary disruption} scenario under the dynamical influence of two-body and resonant relaxation. We confirm the result by PG10 that stellar orbits remain at very high eccentricities outside $\sim 0.1 \pc$. We find higher mean orbital eccentricity values at all radii however, for both steep ($\alpha = 1.75$) and shallow ($\alpha = 0.5$) cusp profiles; see Figure \ref{fig:arma_bd} in which we plot mean orbital eccentricities as a function of semi-major axis  at $60$ and $100 \Myr$. As PG10 use full $N$-body simulations, they do not include the entire stellar cusp and hence precession rates for stars are lower than in our simulations, contributing to a higher resonant relaxation rate and hence larger orbital eccentricity changes. We simulate both a {\it burst} scenario in which all B-stars begin at $t = 0$, and a {\it continuous} scenario in which they are randomly initialized between $t = 0$ and $t = t_{\rm max}$; the latter best reflects binary disruptions due to massive perturbers but the former can be directly compared with the simulations of PG10. We use a fit of the resulting orbital eccentricity distribution after $100 \Myr$ in the {\it continuous} case as our initial conditions for the {\it binary disruption} scenario. 

\begin{center}
\begin{table*}[th!]
\caption{\label{tab:sim_params} Model parameters, mean and standard deviation of $|h|$-values of stars in $N$-body simulations, with combined results of ten random viewing directions, and standard error on the mean, $s_e$.} 
{\small
\hfill{}
\begin{tabular}{cccccccccccccc}
\hline 
       \noalign{\smallskip}
{\footnotesize Simulation}$^{a}$  & $\langle e \rangle^{b}$ & $\alpha^{c}$ & $\langle e \rangle_{\rm CW }^{d}$ & $M_{\rm CW }^{e}$ &  $sw^{f}$  & $\theta^{g}$ & $\langle i \rangle^{h}$  & \multicolumn{3}{c}{ $p^{i} \geq 7 \arcsec$} &  \multicolumn{3}{c}{ $p^{i} \geq 10 \arcsec$} \tabularnewline
        \noalign{\smallskip}
\hline
        \noalign{\smallskip}
&&&&&&& & $\langle |h| \rangle^{j}$ &  $\sigma_{\langle |h| \rangle}^{k}$ & $s_e^{l}$& $\langle |h| \rangle^{j}$ & $\sigma_{\langle |h| \rangle}^{k}$ & $s_e^{l}$\\
\hline
        \noalign{\smallskip}
BD$_1$ & $0.97$ & $0.5$ & $0.6 $ c & 1 & $0$ & - & - & 0.095 & 0.063 & 0.006 & 0.096 & 0.060 & 0.006 \tabularnewline
BD$_2$ & $0.97$ & $0.5$ & $0.6 $ c & 1 & $1$ & - & -  & 0.098 & 0.063 & 0.007 & 0.097 & 0.060 & 0.006 \tabularnewline
BD$_3$ & $0.97$ & $0.5$ & $0.6 $ & 1 & $0$ & - & -  & 0.097 & 0.061 & 0.010 & 0.095 & 0.056 & 0.011 \tabularnewline
BD$_4$ & $0.97$ & $0.5$ & $0.3$ c & 1 & $0$ & - & -  & 0.099 & 0.058 & 0.012 & 0.096 & 0.055 & 0.010 \tabularnewline
BD$_5$ & $0.93$ & $1.75$ & $0.6$ c & 1 & $0$ & -  & - & 0.130 & 0.068 & 0.010 & 0.125 & 0.065 & 0.009 \tabularnewline
BD$_6$ & $0.93$ & $1.75$ & $0.3$ c & 1 & $0$ & - & -  & 0.127 & 0.072 & 0.014 & 0.123 & 0.068 & 0.011 \tabularnewline
BD$_7$ & $0.97$ & $0.5$ & $0.6 $ c & 2 & $0$ & - & - &  0.114  &  0.077  &  0.007  &  0.102  &  0.059  &  0.005  \tabularnewline
BD$_8$ & $0.97$ & $0.5$ & $0.6 $ c &  4 &$0$ & - & - &  0.141  &  0.102  &  0.009  &  0.129  &  0.095  &  0.011  \tabularnewline
        \noalign{\smallskip}
DD$_{1}$ & $0.71$ & $0.5$ & $0.6$ c & 1 & $0$ & $40$  & 80 & 0.325 & 0.236 & 0.048 & 0.313 & 0.247 & 0.051 \tabularnewline
DD$_{2}$ & $0.66$ & $0.5$ & $0.6$ c &  1 &$0$ & $40$  & 25 & 0.352 & 0.244 & 0.033 & 0.349 & 0.248 & 0.034 \tabularnewline
DD$_{3}$ & $0.68$ & $0.5$ & $0.6$ c &  1 &$0$ & $40$  & 121  & 0.283 & 0.227 & 0.038 & 0.250 & 0.188 & 0.029 \tabularnewline
DD$_{4}$ & $0.66$ & $0.5$ & $0.6$ c &  1 &$0$ & $40$  & 163  & 0.348 & 0.201 & 0.043 & 0.350 & 0.203 & 0.045 \tabularnewline
DD$_{5}$ & $0.67$ & $0.5$ & $0.6$ &  1 &$0$ & $40$  & 80  & 0.301 & 0.210 & 0.034 & 0.262 & 0.191 & 0.035 \tabularnewline
DD$_{6}$ & $0.30$ & $0.5$ & $0.6$ c &  1 &$0$ & $40$  & 80  & 0.573 & 0.279 & 0.049 & 0.563 & 0.270 & 0.067 \tabularnewline
DD$_{7}$ & $0.69$ & $1.75$ & $0.6$ c & 1 & $0$ & $40$  & 80  & 0.277 & 0.212 & 0.035 & 0.235 & 0.178 & 0.035 \tabularnewline
DD$_{8}$ & $0.67$ & $0.5$ & $0.6$ c & 1 & $0$ & $60$  & 80 & 0.311 & 0.235 & 0.021 & 0.290 & 0.230 & 0.031 \tabularnewline
DD$_{9}$ & $0.67$ & $0.5$ & $0.6$ c & 1 & $1$ & $40$  & 80  & 0.315 & 0.263 & 0.028 & 0.251 & 0.215 & 0.038 \tabularnewline
DD$_{10}$ & $0.62$ & $0.5$ & $0.3$ c &  1 &$0$ & $40$  & 80  & 0.382 & 0.246 & 0.028 & 0.364 & 0.235 & 0.033 \tabularnewline
        \noalign{\smallskip}
\hline 
\multicolumn{8}{l}{{\scriptsize {\bf Notes.}}}\tabularnewline
\multicolumn{8}{l}{{\scriptsize $^{a}$ BD: binary disruption. DD: dissolved disk.}}\tabularnewline
\multicolumn{8}{l}{{\scriptsize $^{b}$ Mean initial orbital eccentricity of B-stars across all radii.}}\tabularnewline
\multicolumn{8}{l}{{\scriptsize $^{c}$ Index of power-law density cusp profile.}}\tabularnewline
\multicolumn{8}{l}{{\scriptsize $^{d}$ Mean initial orbital eccentricity of young CW stars. c: coherent-eccentricity disk}}\tabularnewline
\multicolumn{8}{l}{{\scriptsize $^{e}$ Mass of young CW disk in units of $10^4 \Mo$}}\tabularnewline
\multicolumn{8}{l}{{\scriptsize $^{f}$ Switch-on function on/off (1/0); see Equation \ref{e:switch}.}}\tabularnewline
\multicolumn{8}{l}{{\scriptsize $^{g}$ Initial opening angle of B-star disk in arcdeg.}}\tabularnewline
\multicolumn{8}{l}{{\scriptsize $^{h}$ Mean initial inclination of B-star disk with respect to young CW disk in arcdeg.}}\tabularnewline
\multicolumn{8}{l}{{\scriptsize $^{i}$ Projected radius.}}\tabularnewline
\multicolumn{8}{l}{{\scriptsize $^{j}$ Mean value of $|h|$ for B-stars with $p \geq 7(10) \arcsec$ at end of simulation.}}\tabularnewline
\multicolumn{8}{l}{{\scriptsize $^{k}$ Standard deviation on $|h|$ for B-stars with $p \geq 7(10) \arcsec$ at end of simulation.}}\tabularnewline
\multicolumn{8}{l}{{\scriptsize $^{l}$ Standard error on the mean of $|h|$ for B-stars with $p \geq 7 \arcsec$ at end of simulation.}}\tabularnewline
       \noalign{\smallskip}
\end{tabular}}
\hfill{}
\end{table*}
\end{center}

 \vspace{-7ex} 

{\it Dissolved Disk (DD) Scenario}: In this scenario, the B-stars form in-situ around the MBH in a nuclear stellar disk during an earlier episode ($\sim 100 \Myr$) of gas infall and fragmentation: we call this the {\it dissolved disk} scenario as the disk structure should have `puffed up' due to gravitational interactions between stars, in particular due to vector resonant relaxation \citep{Koc11} and the gravitational influence of the circum-nuclear disk \citep{Sub09}. We distribute the B-stars in a stellar disk with their initial eccentricities, disk opening angle and inclination with respect to the young CW disk varying with simulation number. Inclinations are selected such that co-rotating and counter-rotating cases with both large and small angles between the two disks are explored. The basic model, DD$_{1}$, draws eccentricities from a thermal distribution. 

\end{enumerate}

Our simulations explore a wide range of parameters relevant for the GC as listed in Table \ref{tab:sim_params}. An important variation in the models is the power-law stellar density index of which we choose two values \--- a steep stellar cusp $\alpha\!=\!1.75$, and a shallow stellar cusp $\alpha\!=\!0.5$. This variation reveals itself in the precession rate of stars at different radii, and hence the persistence of stellar torques. Another variation is the mean eccentricity of the B-star and the young CW disk orbits, which affects their eccentricity evolution through the strength of the torques between the two groups. We run simulations with both coherently-eccentric (or lopsided) young CW disks, in which the eccentricity vectors of the stellar orbits initially overlap, and non-coherent ones. 

\begin{figure*}[t]
  \centering
  \subfigure{
    \includegraphics[angle = -90, scale=0.28]{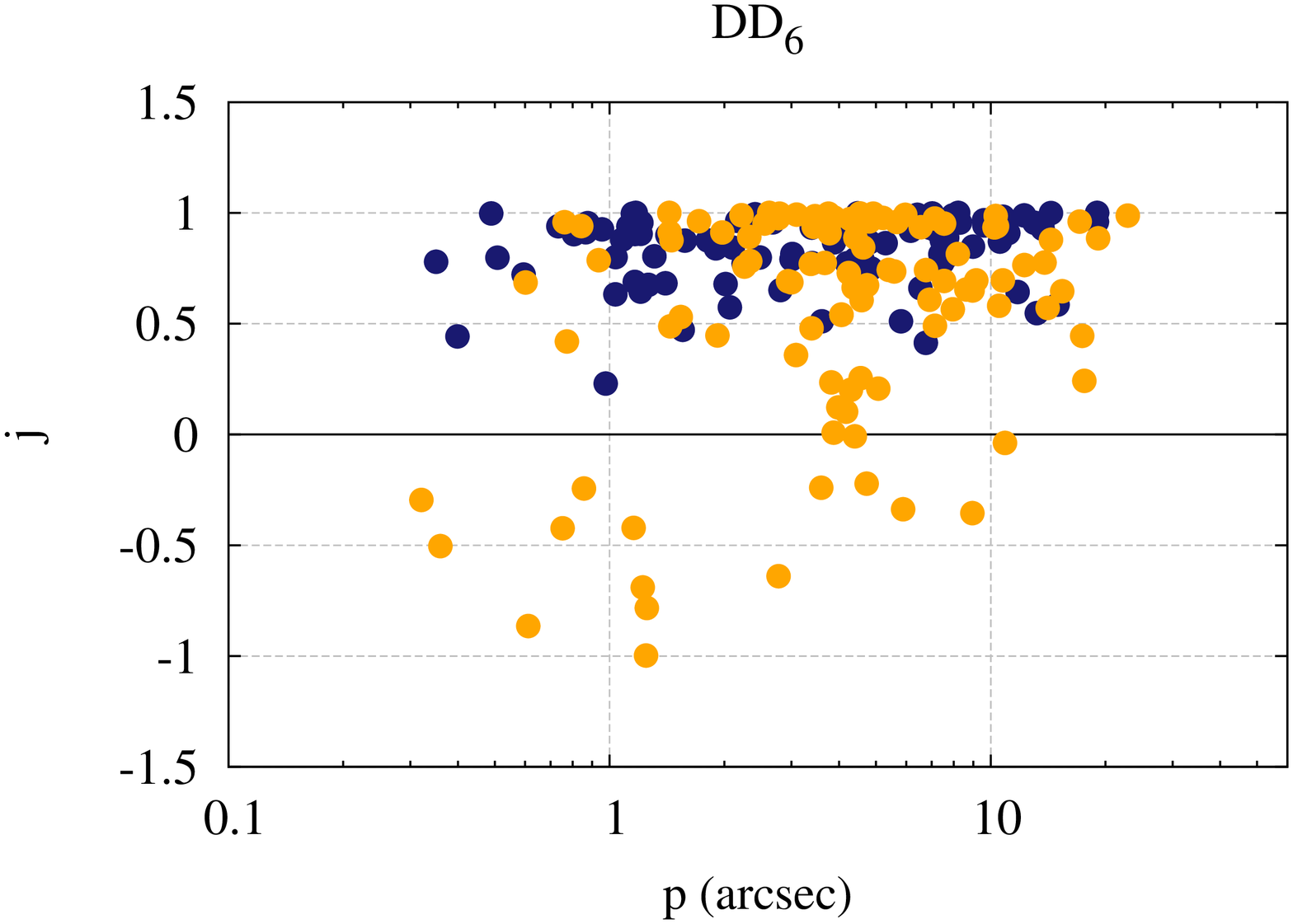} 
  }
  \subfigure{
      \includegraphics[angle = -90, scale=0.28]{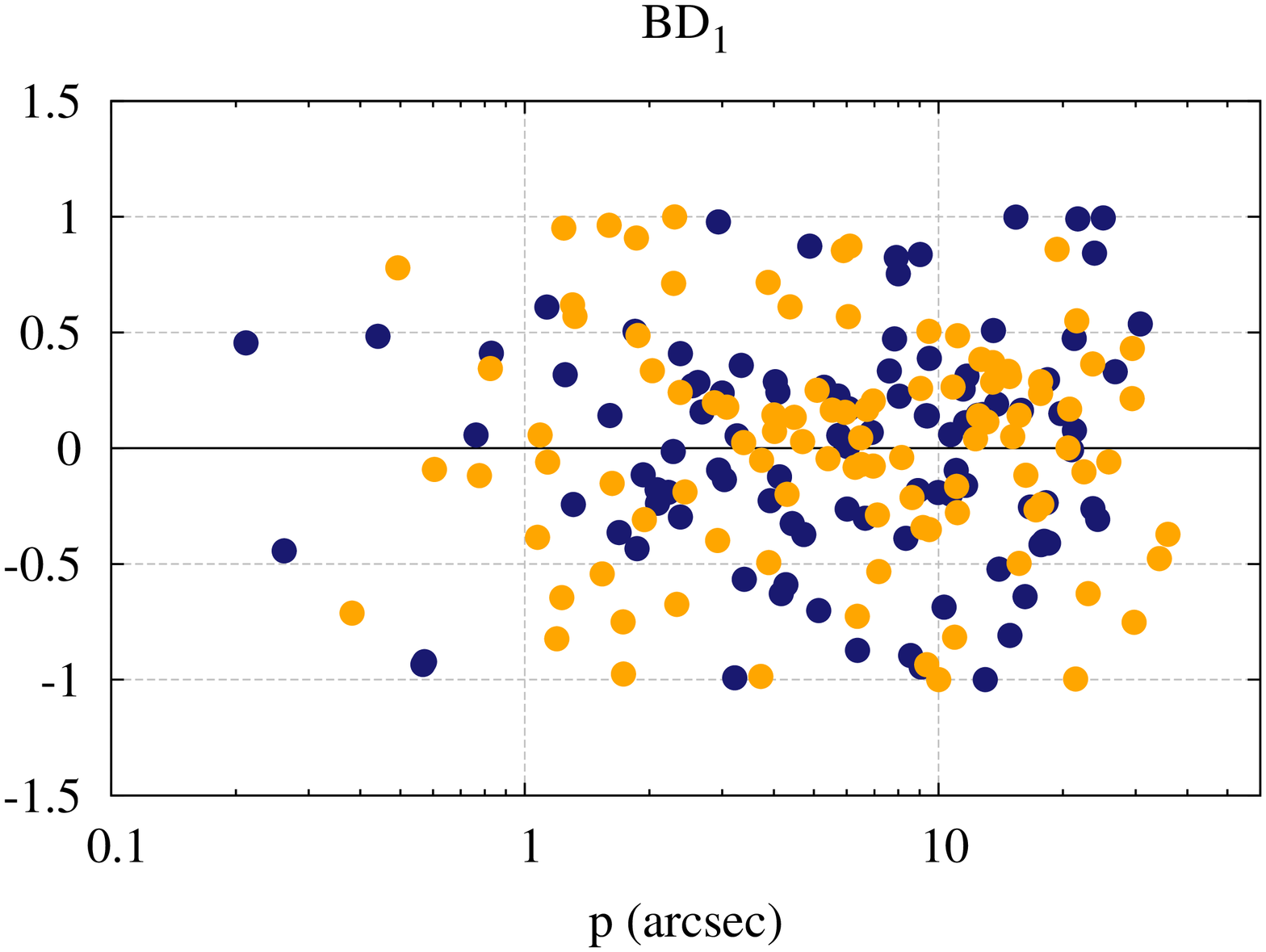} 
  }
  \subfigure{
            \includegraphics[angle = -90, scale=0.28]{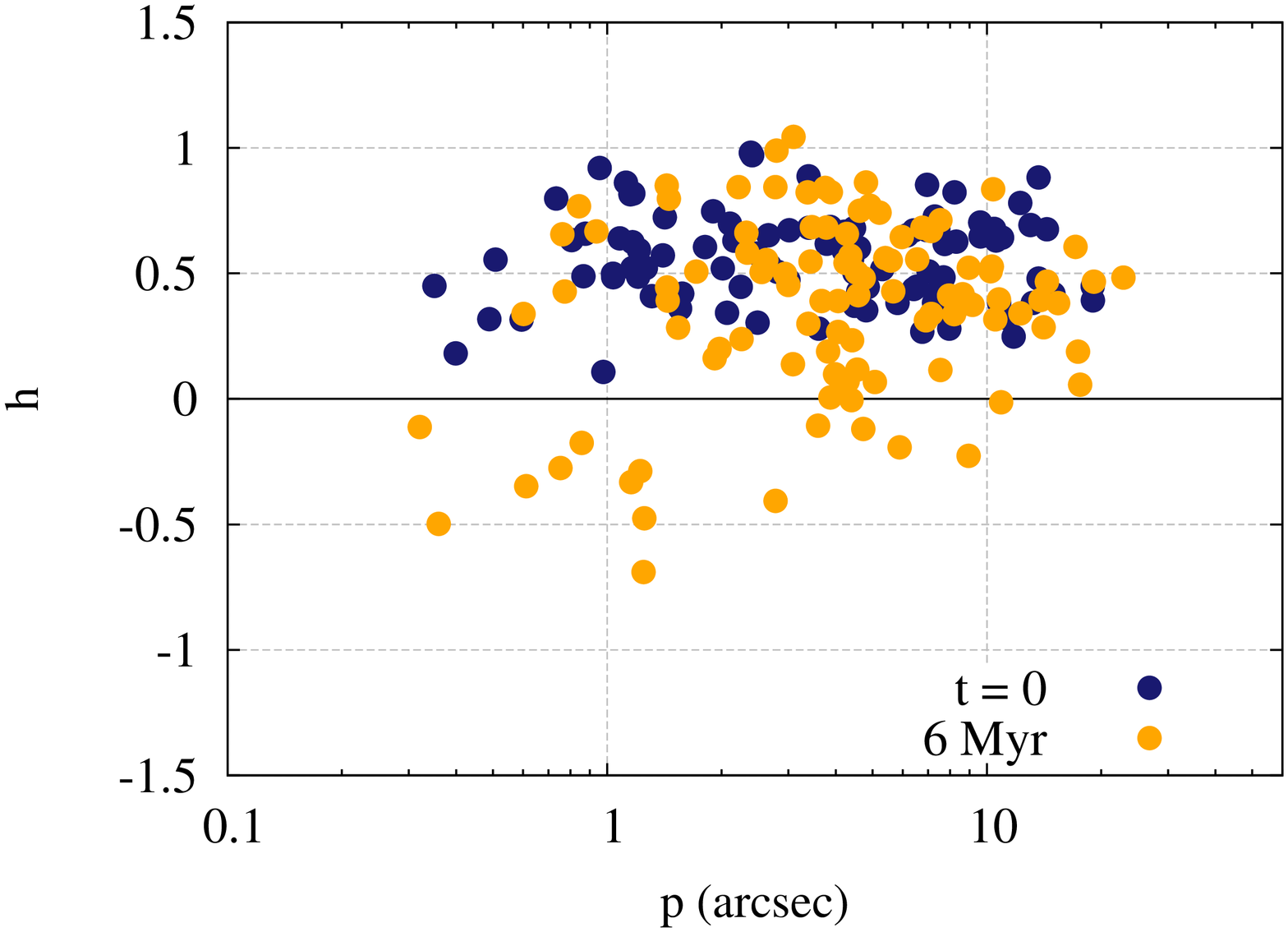} 
  }
  \subfigure{
      \includegraphics[angle = -90, scale=0.28]{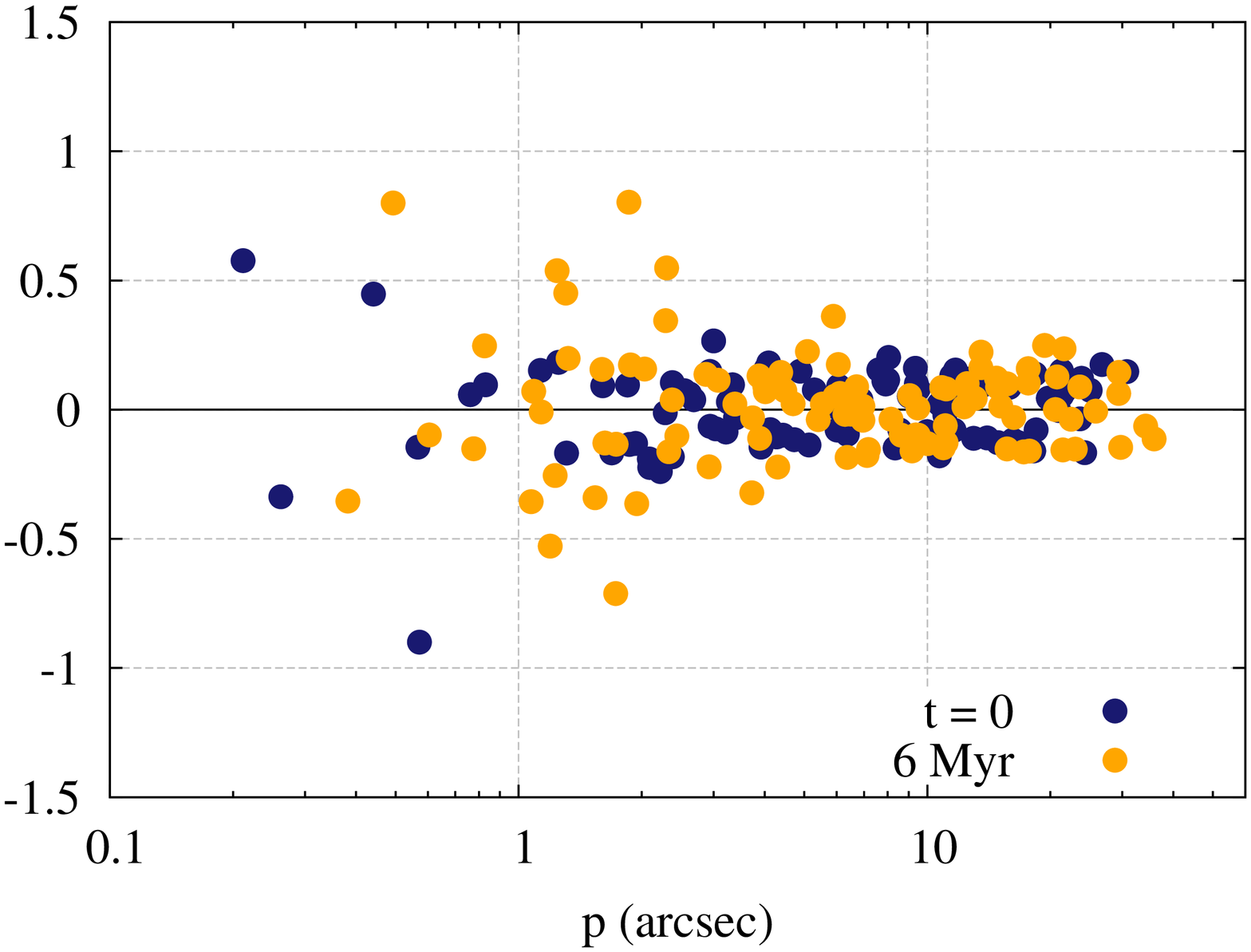} 
  }
\caption{(Top) Distribution of projected, normalized orbital angular momentum on the sky, $j$, for stars in the {\it dissolved disk} DD$_6$ scenario (left) and {\it binary disruption} BD$_1$ scenario (right) at $t = 0, 6 \Myr$. In plotting the parameter $j$ we lose evidence of high eccentricity orbits as stars on near-radial orbits spend most of their orbital period near apoapsis and their $v_p$ value will be lower than the circular velocity at their projected radius $p$.
(Bottom) Distribution of $h$ for stars in the DD$_6$ scenario (left) and BD$_1$ scenario (right) at $t = 0, 6 \Myr$. This new statistic highlights high eccentricity orbits by focusing them at zero.}\label{fig:jh}
\end{figure*}

\begin{figure*}[ht!]
  \centering
      \subfigure{
      \includegraphics[trim=0cm 0cm 0cm 0cm, clip=true, angle = -90, scale=0.32]{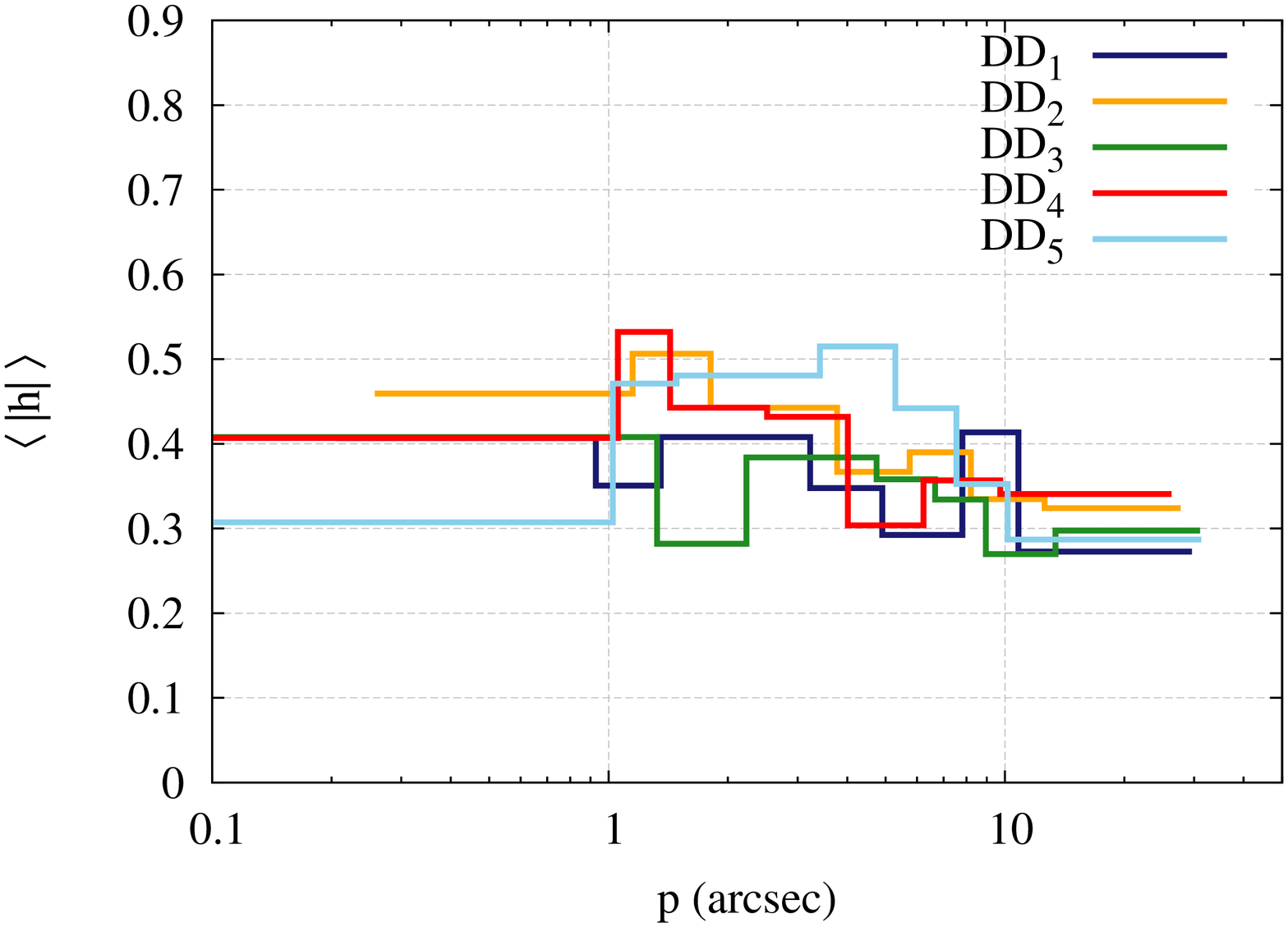}
}
  \subfigure{
    \includegraphics[trim=0cm 0.cm 0cm 0cm, clip=true, angle = -90, scale=0.32]{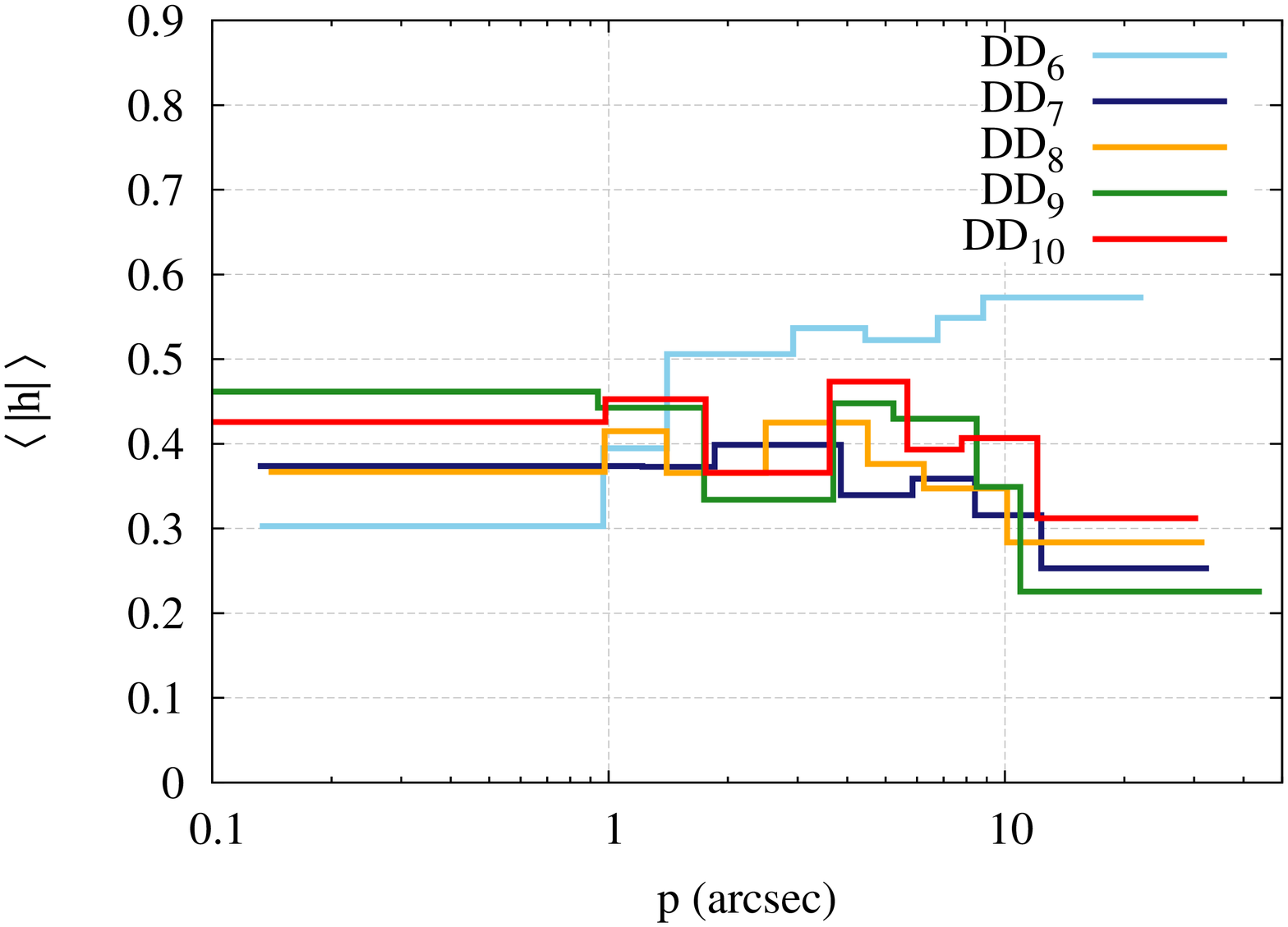} 
  }
  \subfigure{
    \includegraphics[trim=0cm 0cm 0cm 0cm, clip=true, angle = -90, scale=0.32]{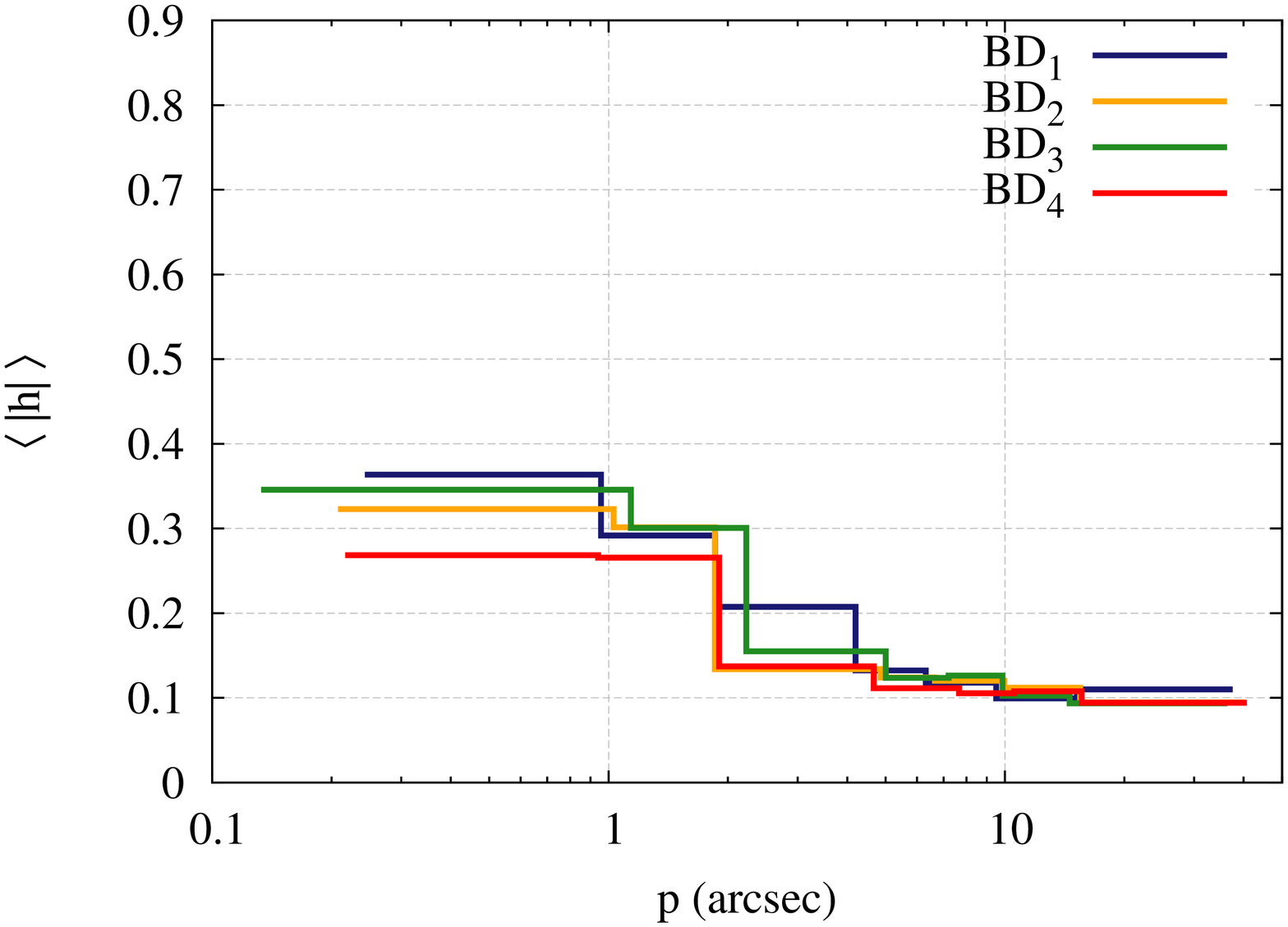} 
  }
    \subfigure{
    \includegraphics[trim=0cm 0cm 0cm 0cm, clip=true, angle = -90, scale=0.32]{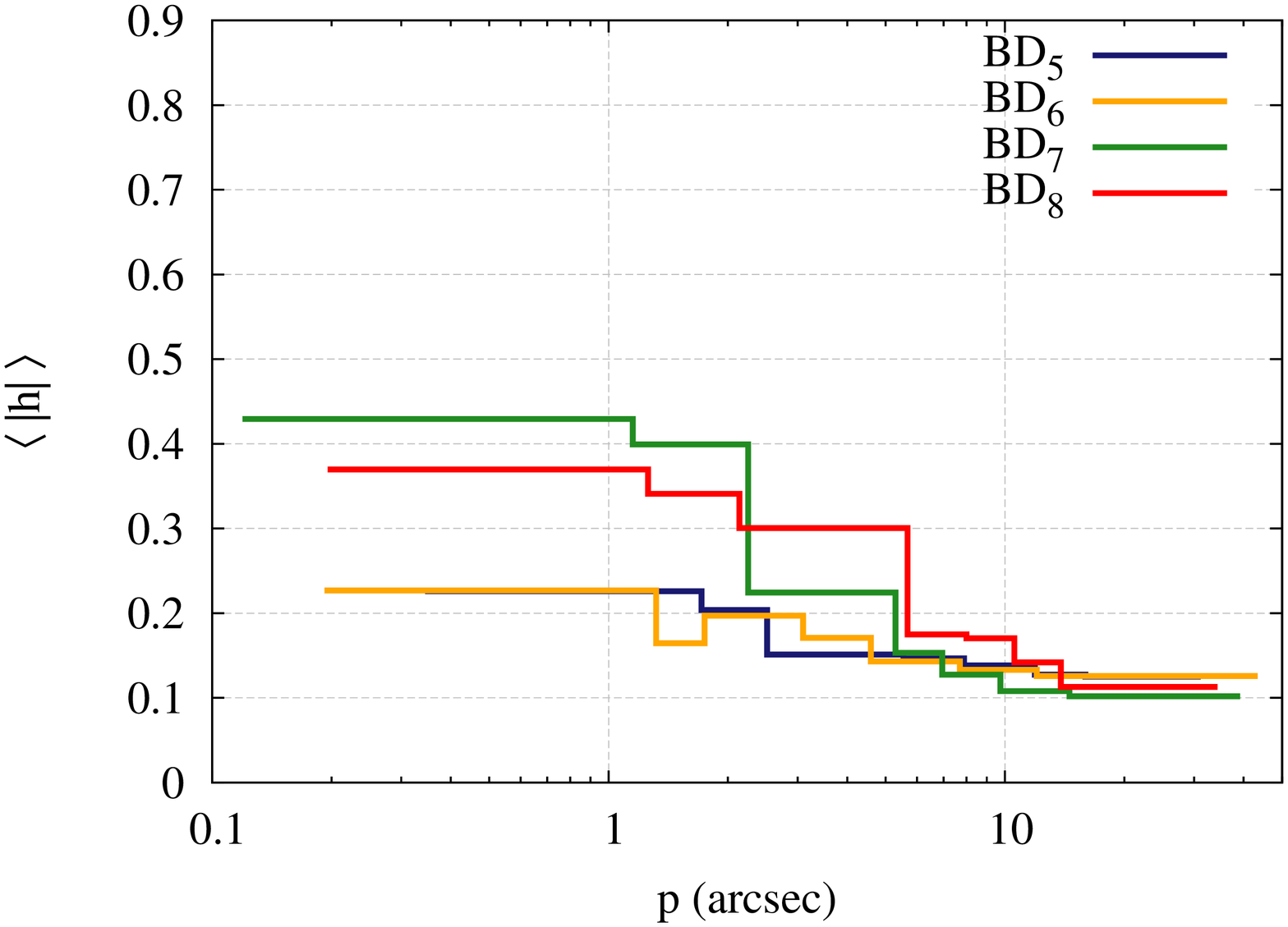} 
  }
  \caption{Mean $|h|$-values as a function of projected distance, $p$, for the {\it dissolved disk} scenario and {\it binary disruption} scenario. All variants of the {\it binary disruption} scenario produce low $|h|$-values at large projected radii, the most distinguishable feature between the two scenarios.}
\label{fig:h_mean}
\end{figure*}

\begin{figure*}[t!]
\begin{minipage}[b]{0.47\linewidth}
\centering   
 \includegraphics[trim=0cm 0cm 0.cm 0cm, clip=true, angle = -90, scale=0.38]{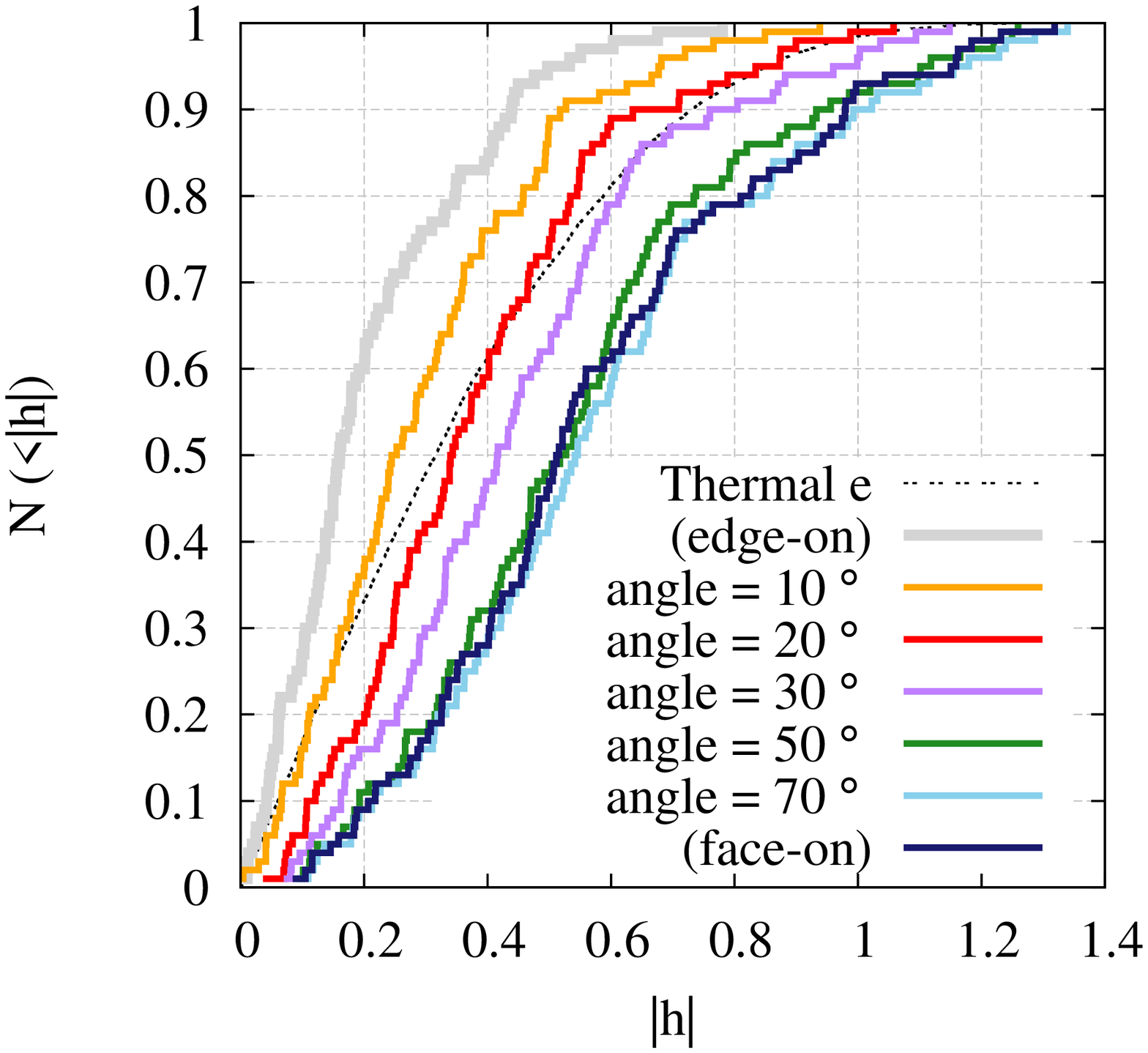}
\end{minipage}
\begin{minipage}[b]{0.47\linewidth}
\centering
    \includegraphics[trim=0cm 0cm 0.cm 0cm, clip=true, angle = -90, scale=0.38]{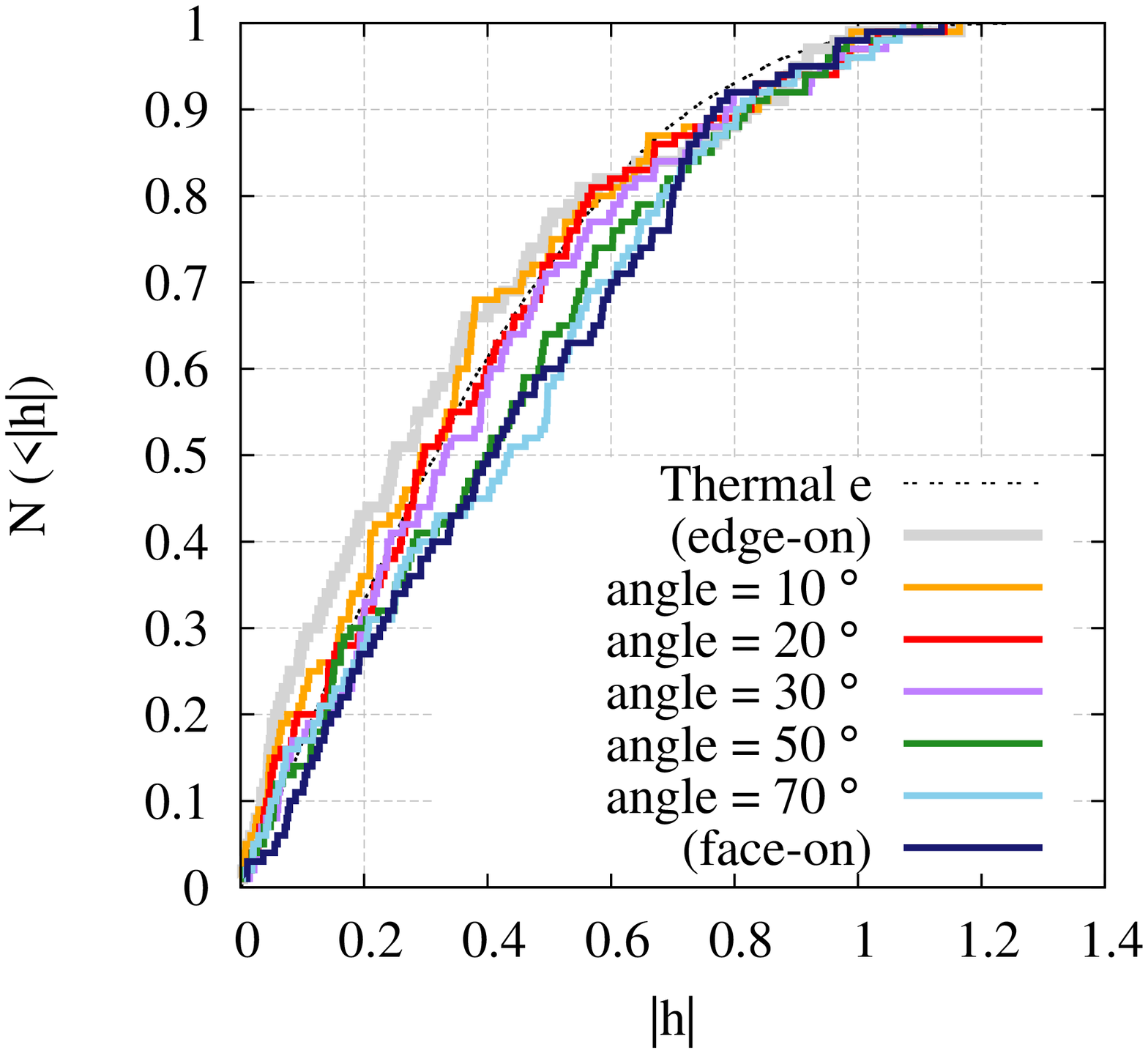}
\end{minipage}
\caption{Cumulative distribution function of $|h|$ as a function of viewing angle with respect to the B-star disk edge in the {\it dissolved disk} simulation DD$_1$ (viewing angle of $0 \arcdeg$ is edge-on, $90\arcdeg$ is face-on). Left: $t  = 0 \Myr$. Right: $t = 6 \Myr$.} 
		\label{fig_DD_inc_cum}
\end{figure*}

\subsection{Evolution of orbital eccentricities}

We follow the change in orbital angular momentum in the B-stars over $6 \Myr$ in response to the young CW disk. The B-stars with small semi-major axes not only experience a greater torque, $\tau \propto 1/a$, but have lower angular momentum, $J$, and hence the relative change in angular momentum is high. This brings about a rapid change in their orbital eccentricities. In contrast, B-stars with large semi-major axes retain memory of their initial orbital eccentricities; these stars can best constrain their formation scenario. This is as PG10 found for resonant relaxation, but the change in eccentricities of B-stars over $6 \Myr$ under the persistent torques of the young CW disk proves to be more rapid than resonant relaxation over $100 \Myr$. We show this effect in Figure \ref{fig:eccVa}, plotting the eccentricity, $e$, of B-star orbits in two simulations (DD$_6$, BD$_1$) as a function of semi-major axis, $a$. The {\it binary disruption} scenario produces the most dramatic signature as stars with large semi-major axes retain their high orbital eccentricities. In the {\it dissolved disk} scenario, the B-stars with large semi-major axes have similar eccentricities as their initial input values.

\subsection{Simulated $h$-values and difference in initial conditions}

We contrast the $j$ and $h$ versus $p$ diagrams for both scenarios using an arbitrary viewing angle in Figure \ref{fig:jh}. The advantage of the $h$-statistic is not obvious in the {\it dissolved disk} scenario (left) as there are few very high eccentricity orbits. However, in the {\it binary disruption} scenario (right), the B-stars at large radii retain their high eccentricities over the $6 \Myr$ simulation, and their $h$ values are centered about zero.

We plot the mean value of $|h|$ as a function of projected radius $p$ after $6 \Myr$ in Figure \ref{fig:h_mean}. The mean value of $|h|$ is derived from combining $h$-values from three different viewing directions for each simulation. The most distinguishing feature between the two scenarios is the disparate $|h|$-signature at large projected radii. The simulations that model the young CW disk formation by employing the switch-on function of Eq. \ref{e:switch} show no significant differences with respect to the basic models. 

We find an inverse relation in the {\it binary disruption} simulations between the value of the power-law index of the stellar cusp, $\alpha$, and the slope of the distribution of $\langle |h| \rangle$-values with distance from the MBH. Steep stellar cusps result in flatter $\langle |h| \rangle$-$r$ distributions and vice versa. This is due to the coherence time over which stellar torques can act on the individual B-star orbits. Simulations BD$_{1-4}$ which have a shallow stellar cusp ($\alpha = 0.5$) result in a steep $\langle |h| \rangle$-distribution across projected radii. The B-star orbits at small radii precess relatively slowly and hence secular changes in $J$ are efficient at changing their eccentricities. This results in large $\langle |h| \rangle$-values at small radii.  B-star orbits at large radii precess relatively quickly and secular changes in $J$ are less efficient as the coherence time is short. These B-stars retain their high orbital eccentricities, and have low $\langle |h| \rangle$-values. Simulations BD$_{5-6}$ which have a steep stellar cusp ($\alpha = 1.75$) result in a flatter $\langle |h| \rangle$-$r$ distribution. The stellar orbits at small radii precess relatively quickly and hence secular changes in $J$ are inefficient relative to that experienced by stars in simulations BD$_{1-4}$. This results in lower $\langle |h| \rangle$-values at small radii. The stellar orbits at large radii precess relatively slowly and secular changes in $J$ are more efficient as the coherence time is longer. $\langle |h| \rangle$-values at large radii are consequently larger than for BD$_{1-4}$. One can in principle constrain the mass distribution of the underlying stellar cusp using observations of stars which have been disrupted from a binary by the MBH from the relation between the stellar cusp profile and $\langle |h| \rangle$-values as a function of distance from the MBH. 

Simulations BD$_{7-8}$ (larger mass young CW disk) show the most evolution in B-star eccentricities. $\langle |h| \rangle$-values are high relative to the simulations with $M_{\rm CW } = 10^4\Mo$ but still lower than in the {\it dissolved disk} simulations.

The DD$_6$ simulation, in which the B-stars are initialized in an $e = 0.3$ disk, and DD$_{10}$ with an $e = 0.3$ young CW disk, result in the lowest eccentricities of the {\it dissolved disk} simulations and hence the highest $\langle |h| \rangle$-values. DD$_7$ which has a steep stellar cusp ($\alpha = 1.75$) shows the most evolution in B-star eccentricity at large radii, reaching the lowest $\langle |h| \rangle$-values of the {\it dissolved disk} simulations. In the DD$_{2}$ simulation, where the initial mean inclination angle between the two disks is $\langle i \rangle = 25 \arcdeg$, the B-star angular momentum vectors overlap with those of the young CW disk by the end of the simulation. The same overlap is observed in the DD$_{1}$ simulation ($\langle i \rangle = 80 \arcdeg$) but to a lesser extent. Simulations DD$_{3,4}$, in which the B-star and young CW disks are counter-rotating with respect to one another, $\langle i \rangle > 90 \arcdeg$, show a substantial number of stars from each disk with reversed signs of angular momentum but little overlap. The counter-rotating instability \citep[e.g.][]{Tou12} is suppressed due to precession resulting from the presence of a stellar cusp.

The distribution in angular momentum vectors of the young CW disk stars can become very spread out, in contrast to the observed dispersion angle \citep{Pau06,Bel06,Lu09,Bar09}. This is due to secular gravitational torquing of the stellar orbits between the two disks. We find that small angles between the disks cause the most spreading as the torques are stronger when they are closer together (see Equation \ref{e:tau}). Without the second disk, the orbits of the young CW disk are less spread out in angular momentum. A steep density profile in the stellar cusp also hinders spreading as the rapid orbital precession time decreases the timescale over which orbits can feel coherent torques. The inclination angle between the B-star disk and the young CW disk affects the resulting $\langle |h| \rangle$-values; the DD$_{3}$ simulation with $\langle i \rangle = 121 \arcdeg$ shows the lowest $\langle |h| \rangle$-values at large radii.

For each simulation, we chose ten random viewing directions to calculate a mean $|h|$-value for each star. In Table \ref{tab:sim_params} we show $\langle |h| \rangle$ and one standard deviation, $\sigma_h$, for all stars with projected radii $p \geq 7 \arcsec$ and $p \geq 10 \arcsec$. The {\it binary disruption} simulations produce low $\langle |h| \rangle$-values with a range ($0.095 - 0.141$) for $p \geq 7 \arcsec$ and ($0.095 - 0.129$) for $p \geq 10 \arcsec$ with $\sigma_h \sim 0.06$ on most values. \\ We also show the standard error on the mean, $s_e$, of $\langle |h| \rangle$ for the ten different viewing directions. {\it Binary disruption} simulations have small $s_e$ values, reflecting the fact that the B-stars are isotropically distributed and as a consequence the different viewing directions produce similar results. 

The {\it dissolved disk} simulations produce higher mean $\langle |h| \rangle$-values with a range ($0.277 - 0.573$) for $p \geq 7 \arcsec$ and ($0.251 - 0.563$) for $p \geq 10 \arcsec$ with $\sigma_h \sim 0.2$ on most values. {\it Dissolved disk} simulations have large $s_e$ values as the different viewing directions produce different results, depending on the angle between the viewing direction and the plane of the B-star disk. To further demonstrate this, in Figure \ref{fig_DD_inc_cum} we plot the cumulative $|h|$-distribution function as a function of viewing angle with respect to the B-star disk in {\it dissolved disk} simulation, DD$_1$. The left plot shows data from the start of simulation in which the disk has an opening angle of $30 \arcdeg$. Larger viewing angles with respect to the disk plane produce larger $|h|$-values. The right plot shows the data after $6 \Myr$. Here the viewing directions produce more similar results to each other as the disk opening angle has increased and the orbits are more isotropically distributed. The disk structure is still non-isotropic however; the $\langle |h| \rangle$-values increase from $0\arcdeg - 90\arcdeg$.

\section{Observational data}\label{S:obs}

\subsection{Observations and data reduction}

The observational data comprise 207 spectroscopically identified early-type stars within $25\arcsec$ ($\sim 1\pc$) from SgrA*. The data set incorporates those used in the studies of \cite{Bar09}, \cite{Bar10}, \cite{Pfu11} and additional stars found in new fields which will be reported in Fritz et al. (in prep). We refer the reader to the \cite{Bar10} for a detailed description of the data reduction process. 
 
 The corresponding observations have taken place between 2002 and 2010 at the VLT in Cerro Paranal Chile (ESO programs 075.B-0547, 076.B-0259, 077.B-0503, 179.B-0261 and 183.B-0100). The imaging data were obtained with the adaptive optics camera NACO \citep{Rou03, Har03}. The photometric reference images were taken on the 29th of April 2006 and on the 31st of March 2010. Most of the images were obtained, using the $K'$-band filter ($2.17\, \rm \mu m$) together with the 27\,$\rm{mas/pixel}$ camera of NACO. Each image was processed in the same way, using sky-subtraction, bad-pixel and flat-field correction as described in \cite{Tri08}. The spectroscopic data were obtained with the adaptive optics assisted integral field spectrograph SINFONI \citep{Eis03, Bon04}. The data output of SINFONI consists of cubes with two spatial axes and one spectral axis. Depending on the plate scale, an individual cube covers $3.2\arcsec \times3.2\arcsec$ or $8\arcsec \times8\arcsec$; the spectral resolution varies between 2000 and 4000 depending on the chosen bandpass and the field-of-view. \\ A major challenge for the identification of stars in the Galactic Center is stellar crowding. The bulk of the resolved stellar population are low-mass giants later than K0III. Due to the extreme extinction towards the Galactic Center ($A_K\approx2.7$), foreground stars are easily excluded by their blue color. The patchiness of the extinction, however, prevents a photometric distinction between evolved giants ($\rm T\sim 4000\, \rm K$) and young early-type stars ($\rm T>15,000\,\rm K$; spectral types between B9V up to WR$/$O) in the cluster. $K$-band spectra of the stars classify them as late-type if CO absorption features (2.29 -2.40$\rm \,\mu m$) are present, or early-type if Br$\gamma$ (2.166$\rm \,\mu m$) or HeI (2.058$\rm \,\mu m$) absorption lines are present. In this way, the combination of imaging and spectroscopy allows a clear identification of the young and massive early-type stars amongst the equally bright but much more numerous old low mass giants. 

\begin{figure}[t!]
\vspace{-0pt}
  \centering
        \includegraphics[trim=0cm 0cm 0.cm 0cm, clip=true, angle = -90, width=0.48\textwidth]{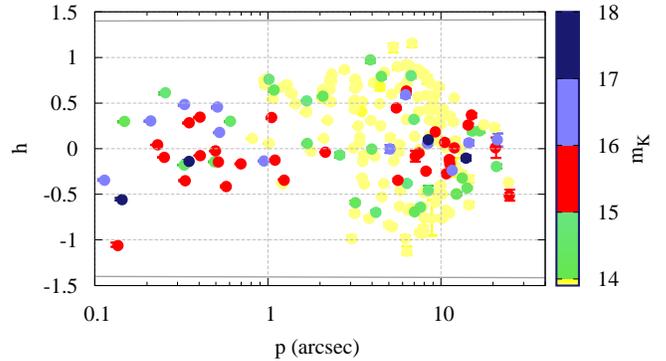}
    		\vspace{-0 cm}
		\caption{Observed $h$-values of stars with error bars as a function of projected radius, $p$. The maximum value of $h = \sqrt{2}$ for a bound orbit is plotted. The color of each point corresponds to the $K$-magnitude of the star.} 
		\label{fig_h_plane_magbins}
\end{figure}

\begin{figure}[t!]
\vspace{-0pt}
  \centering
    \includegraphics[trim=0cm 0cm 0.cm 0cm, clip=true, angle = -90, width=0.54\textwidth]{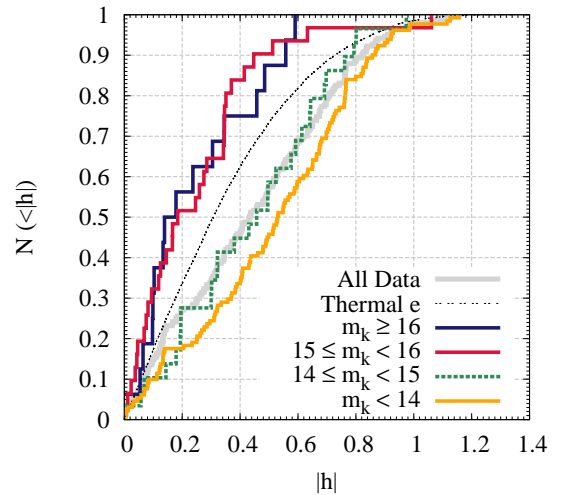}
    		\vspace{-0 pt}
		\caption{Cumulative $|h|$-distribution function for observational data (binned in $K$-magnitude, $m_K$) and that for a simulated thermal eccentricity distribution.} 
		\label{fig_cum_data}
\end{figure}

\begin{figure}[t!]
\vspace{-0pt}
  \centering
    \includegraphics[trim=0cm 0cm 0.cm 0cm, clip=true, angle = -90, width=0.48\textwidth]{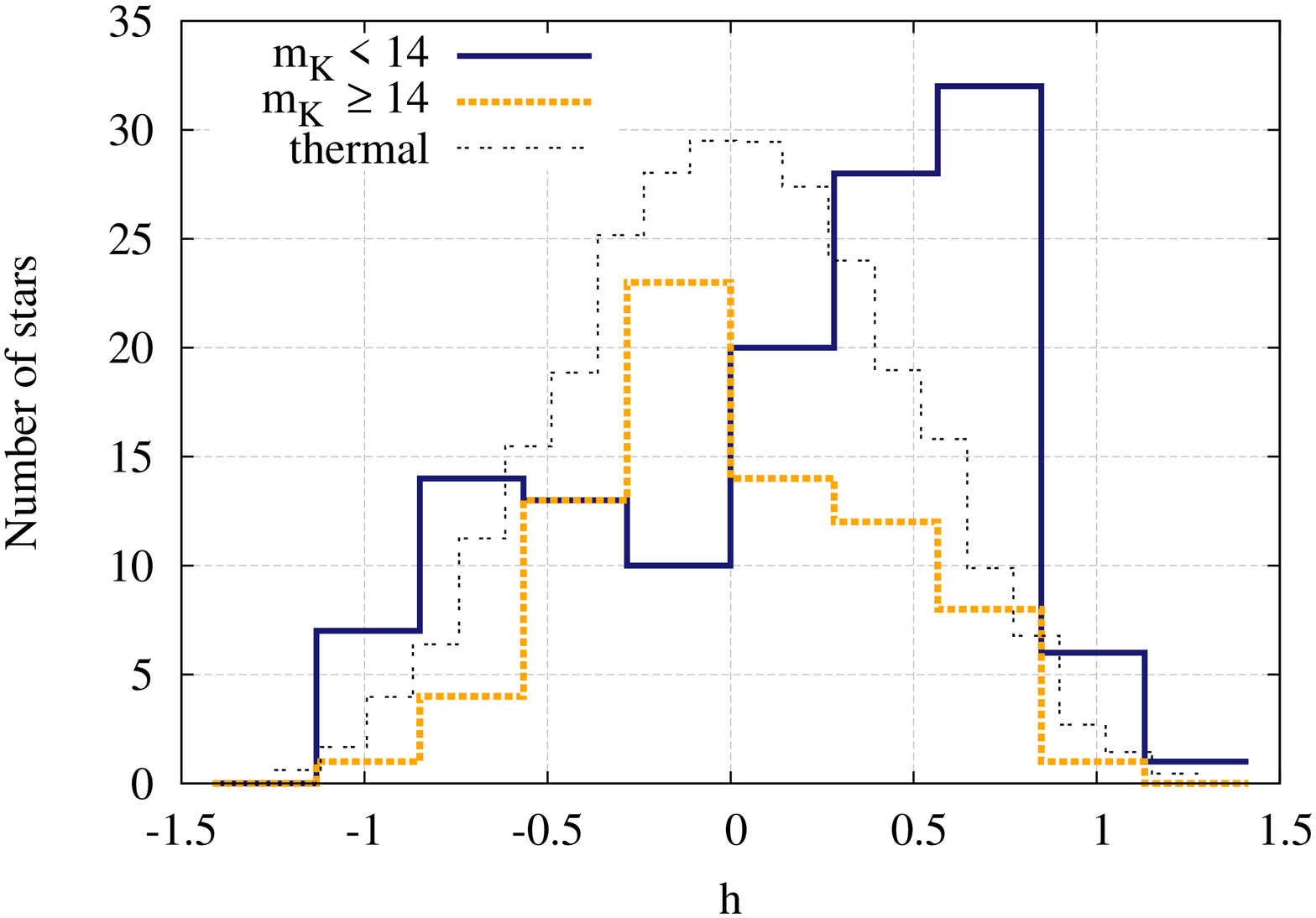}
    		\vspace{-0 pt}
		\caption{Histogram of $h$-values for observational data (binned in $K$-magnitude, $m_K < 14$, $m_K \ge 14$) and that for a simulated thermal eccentricity distribution.} 
		\label{figHistAllData}
\end{figure}

\begin{figure}[t!]
\vspace{-0pt}
  \centering
    \includegraphics[trim=0cm 0cm 0.cm 0cm, clip=true, angle = -90, width=0.48\textwidth]{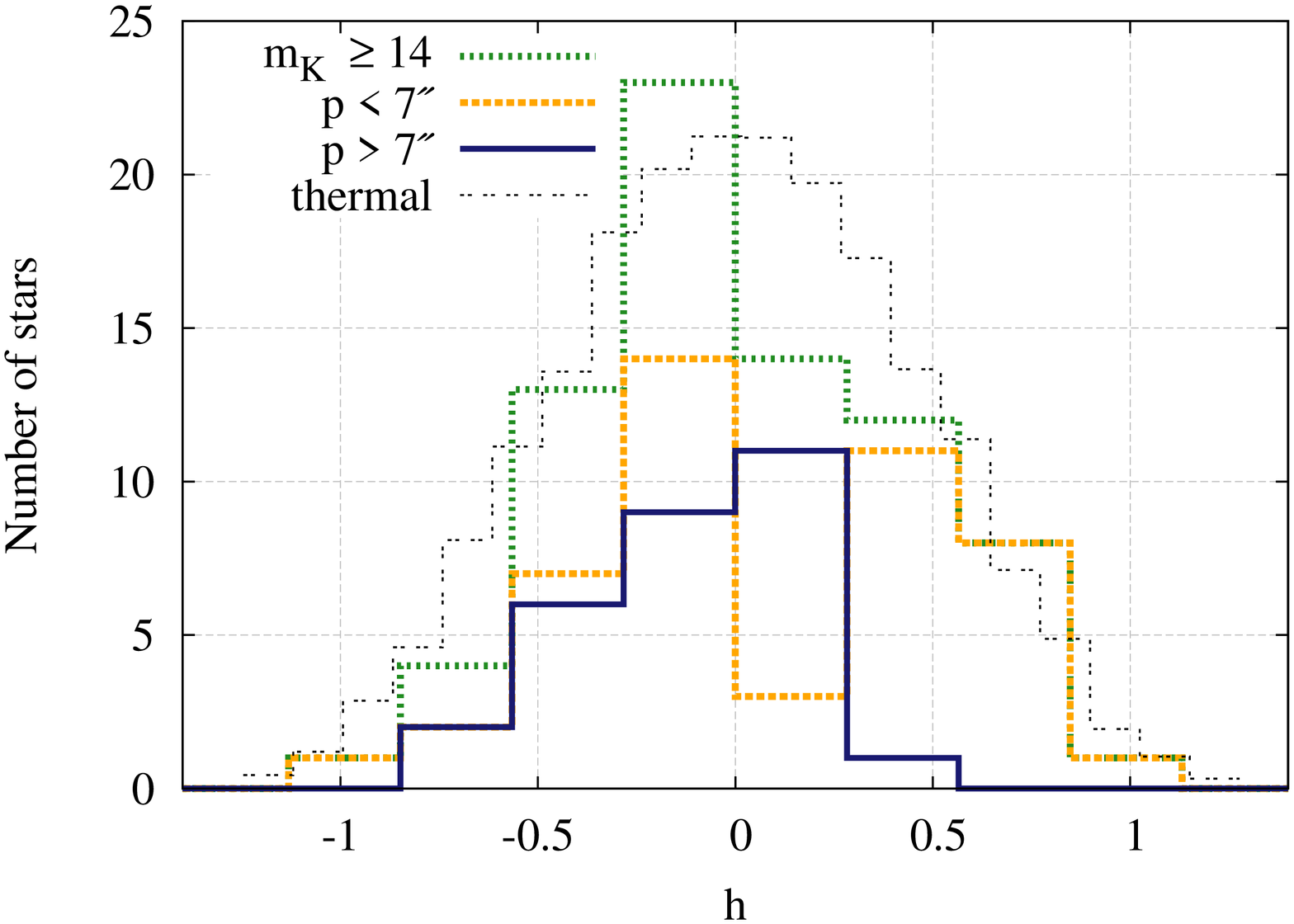}
    		\vspace{-0 pt}
		\caption{Histogram of $h$-values for observational data with $m_K \ge 14$ binned in projected radius, $p > 7\arcsec$, $p < 7\arcsec$, and that for a simulated thermal eccentricity distribution.} 
		\label{figHistpgtr14}
\end{figure}

\begin{figure}[t!]
\vspace{-0pt}
  \centering
    \includegraphics[trim=0cm 0cm 0.cm 0cm, clip=true, angle = -90, width=0.48\textwidth]{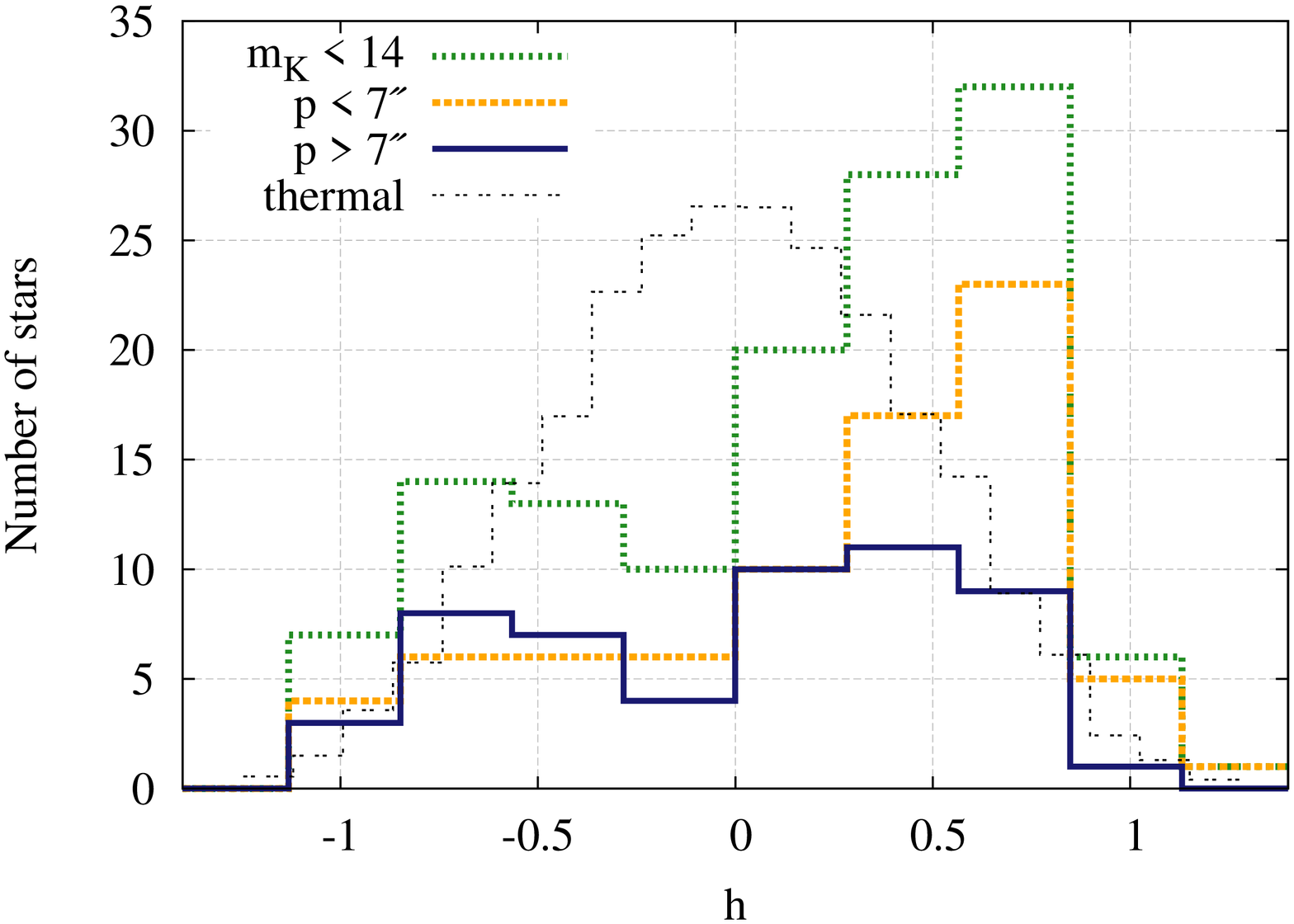}
    		\vspace{-0 pt}
		\caption{Histogram of $h$-values for observational data with $m_K < 14$ binned in projected radius, $p > 7\arcsec$, $p < 7\arcsec$, and that for a simulated thermal eccentricity distribution.} 
		\label{figHistpless14}
\end{figure}

\begin{figure}[t!]
\vspace{-0pt}
  \centering
    \includegraphics[trim=0cm 0cm 0.cm 0cm, clip=true, angle = -90, width=0.48\textwidth]{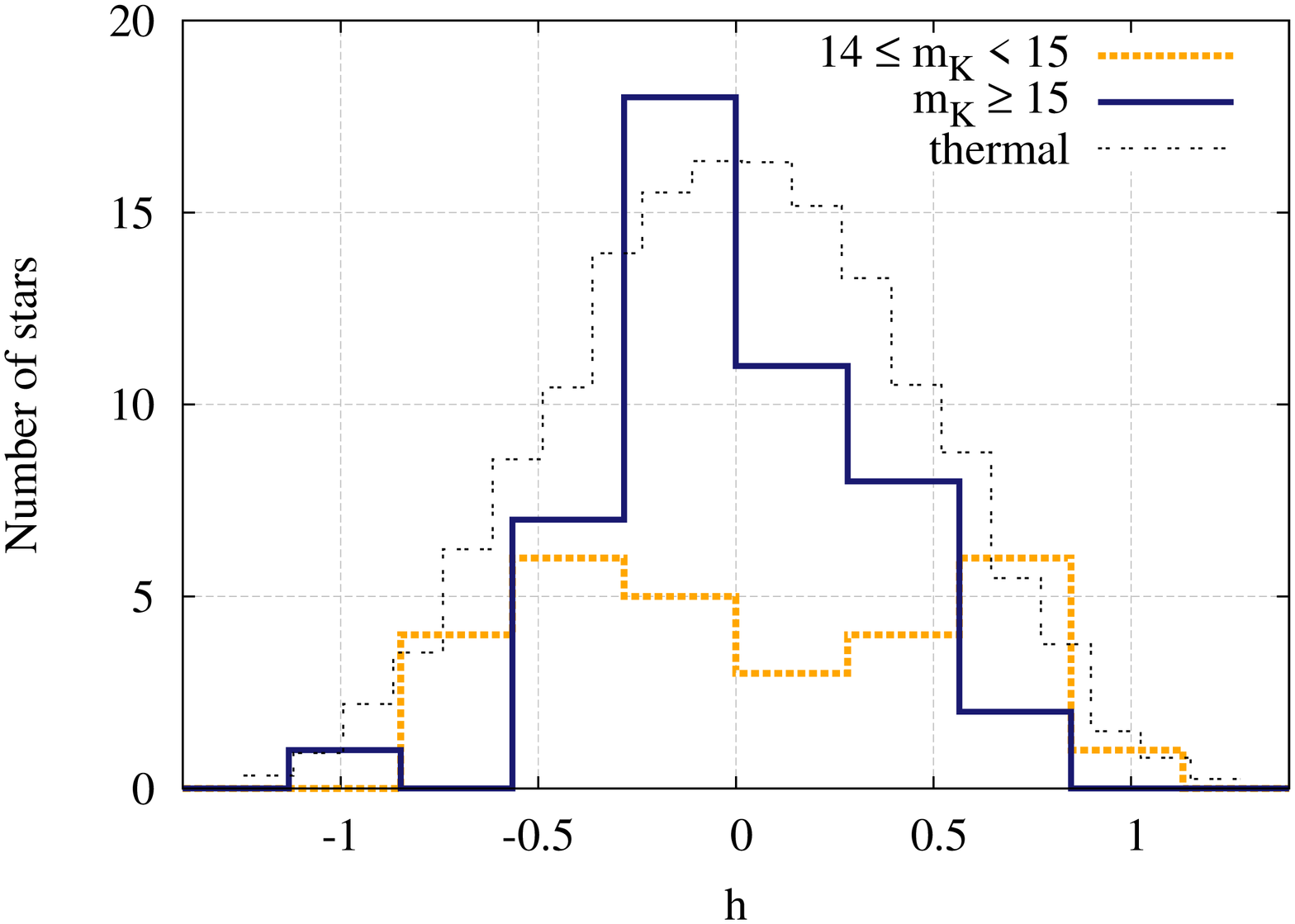}
    		\vspace{-0 pt}
		\caption{Histogram of $h$-values for observational data binned in $K$-magnitude, $14 \le m_K < 15$, $m_K \ge 15$, and that for a simulated thermal eccentricity distribution.} 
		\label{figHist1415}
\end{figure}

\begin{figure}[t!]
\vspace{-0pt}
  \centering
    \includegraphics[trim=0cm 0cm 0.cm 0cm, clip=true, angle = -90, width=0.48\textwidth]{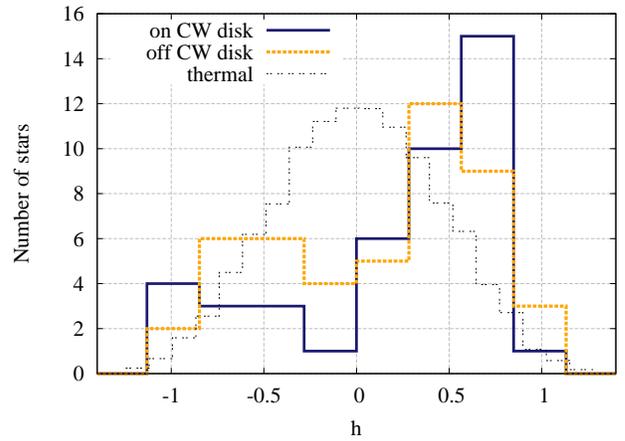}
    		\vspace{-0 pt}
		\caption{Histogram of $h$-values for observational data ($m_K < 14$) for stars analyzed to be on or off the young CW disk, and that for a simulated thermal eccentricity distribution.} 
		\label{figHistOnOffDisk}
\end{figure}
\begin{figure}[t!]
\vspace{-0pt}
  \centering
    \includegraphics[trim=0cm 0cm 0.cm 0cm, clip=true, angle = -90, width=0.54\textwidth]{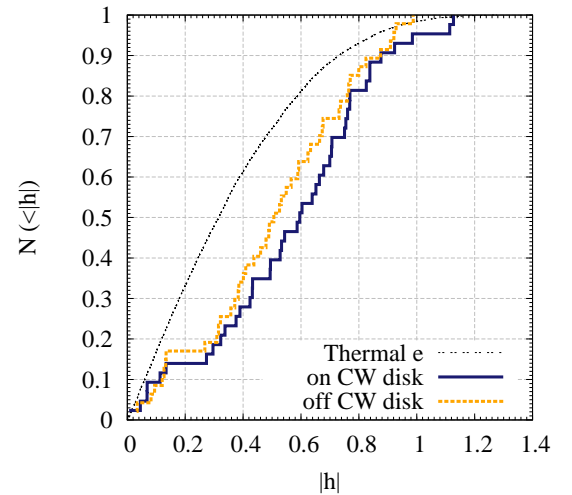}
    		\vspace{-0 pt}
		\caption{Cumulative $|h|$-distribution function for the observational data ($m_K < 14$), both on and off the young CW disk, and that for a simulated thermal eccentricity distribution.} 
		\label{fig_cum_data_CWdisk}
\end{figure}

\begin{figure*}[t!]
\begin{minipage}[b]{0.47\linewidth}
\centering
    \includegraphics[angle = -90, scale=0.38]{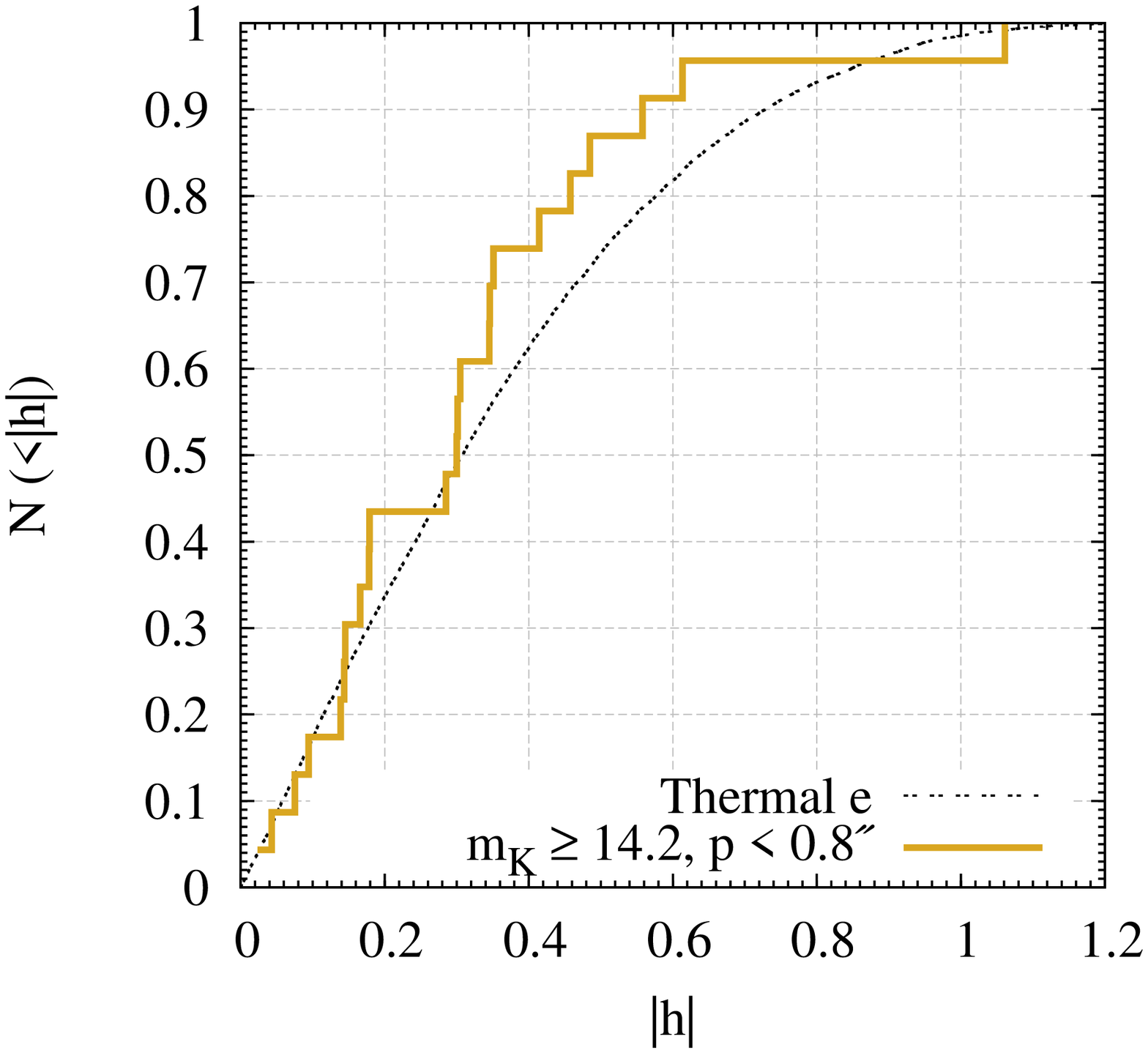}
\end{minipage}
\begin{minipage}[b]{0.47\linewidth}
\centering
    \includegraphics[angle = -90, scale=0.38]{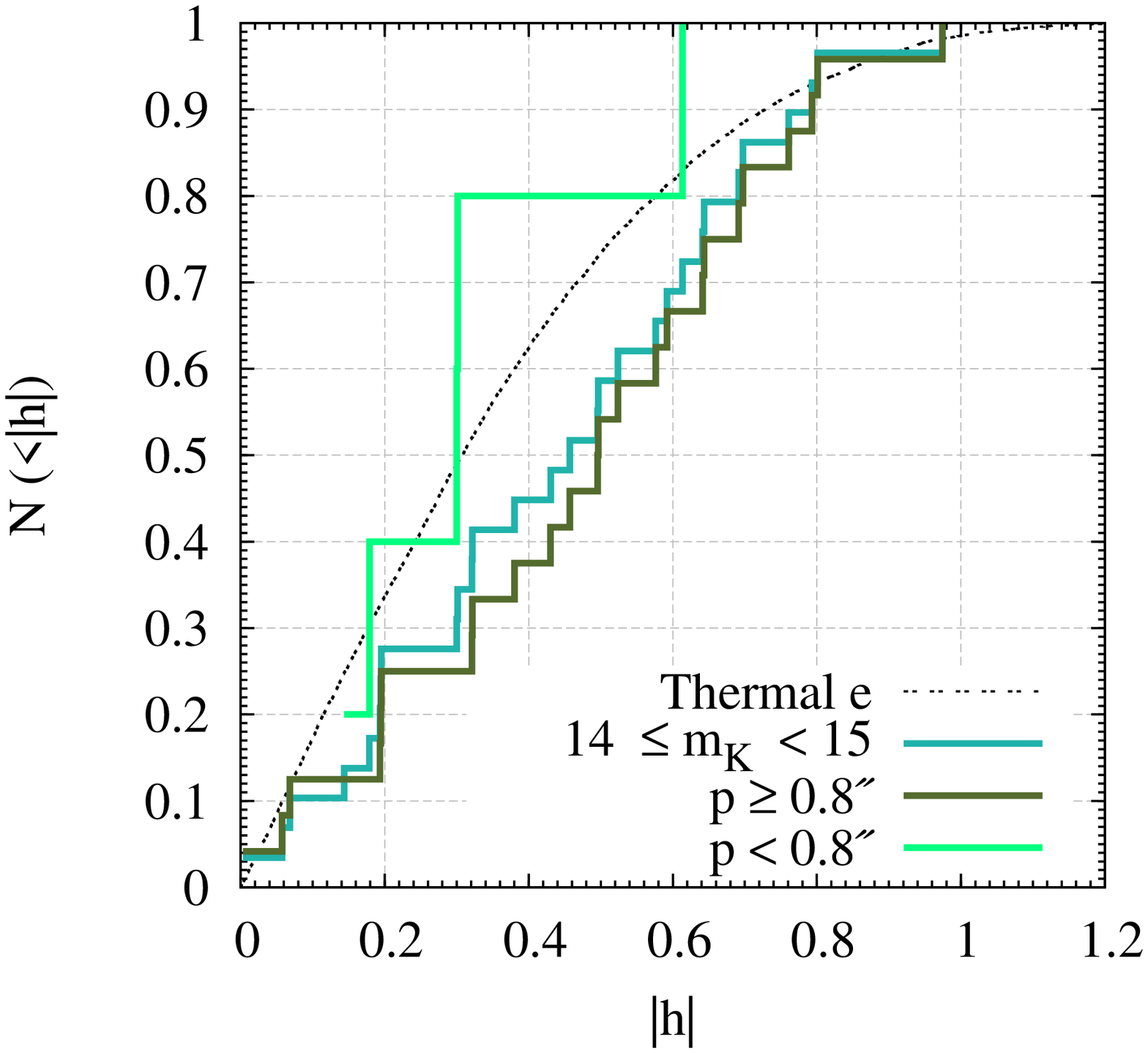}
\end{minipage}

\vspace{0.1 cm}

\begin{minipage}[b]{0.47\linewidth}
\centering
    \includegraphics[angle = -90, scale=0.38]{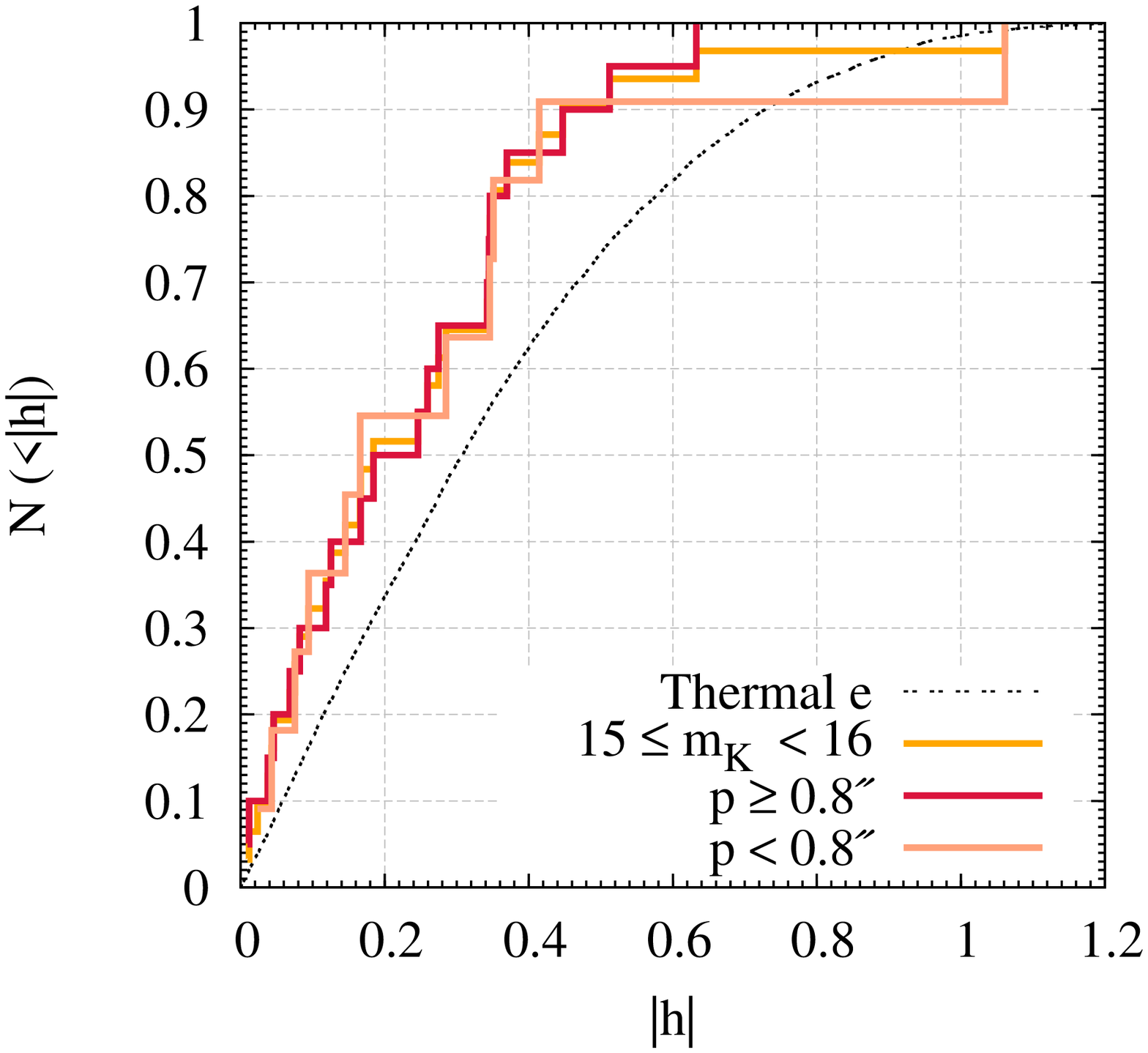}
\end{minipage}
\begin{minipage}[b]{0.47\linewidth}
\centering
    \includegraphics[angle = -90, scale=0.38]{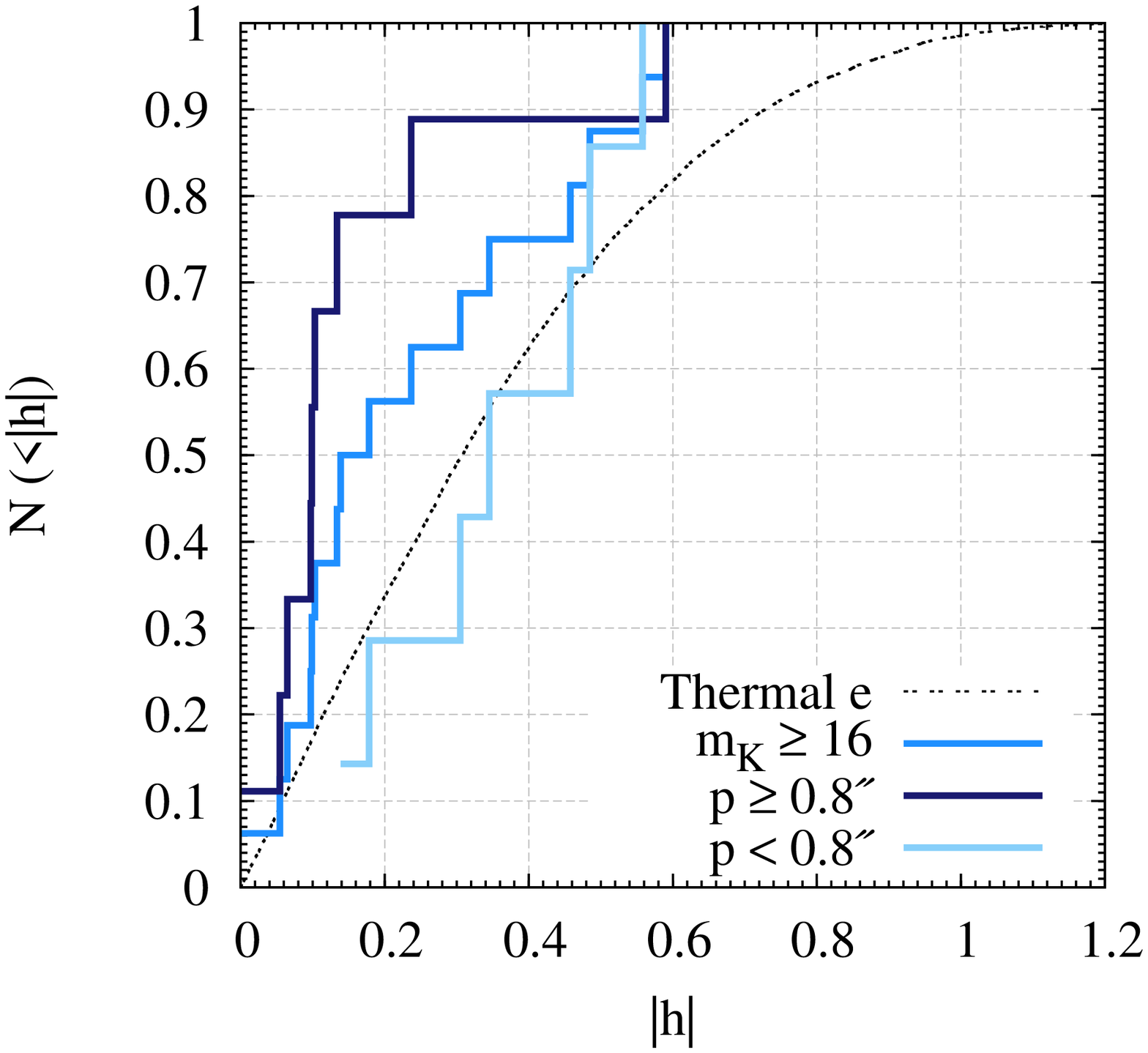}
\end{minipage}
\caption{Cumulative $|h|$-distribution function for observational data in different $K$-magnitude ranges, with a cut at projected radius $p = 0.8 \arcsec$. {\it Top left:} The `S-stars' with $p < 0.8 \arcsec$, $14.2 \leq m_K < 17.5$, are slightly more eccentric than a thermal distribution. {\it Top right:} Stars between $14 \leq m_K < 15$ appear more eccentric (or more `edge-on') at smaller projected radii, $p < 0.8 \arcsec$ than at large ($p_{\rm ks} = 0.16$). {\it Bottom left:} Stars between $15 \leq m_K < 16$ show no significant variation in $|h|$-distribution across $p = 0.8 \arcsec$. {\it Bottom right:} Stars with $m_K \geq 16$ are more eccentric, or more `edge-on', at $p \geq 0.8 \arcsec$ ($p_{\rm ks} = 0.017$).}\label{fig:p_gtr_8}
\end{figure*}

\subsection{Astrometry}
Similar to the method of \cite{Tri08} the stars in the individual images were detected using the algorithm FIND \citep{Ste87}. The individual positions in each image were retrieved using Gaussian fits with formal fit errors of the order $\approx300 \, \rm \mu as$. About 560 bright isolated stars in the field served as an astrometric reference frame. The proper motions were  computed by fitting a linear function to the astrometric star positions with time. Typical errors were $\rm \approx0.13\,mas$/$\yr$ ($\rm 5\,km/s$).

\subsection{Stellar masses and main-sequence lifetimes}
The early-type stars in our sample contain main-sequence B-stars up to evolved WR/Ofpn stars. An individual classification of the spectral type however is non-trivial with $K$-band spectra alone. One difficulty is the absence of a simple correlation between Br$\gamma$ (HeI) strength and spectral type. Only the most massive WR$/$O stars can be recognized due to their strong wind emission lines, but the large mass-loss rates make a current mass estimate even more difficult and these stars must be modeled in detail to derive temperatures and masses \citep[e.g.,][]{Mar07}. Only stars more massive than $\sim20-25 \Mo$ reach the WR phase however \citep{Mae04}. The mass of the less luminous B-stars can be deduced from their observed luminosity \citep[using e.g.,][isochrones]{Ber94}. Although the spectral information alone provides not a unique classification, the known absolute magnitude of the stars allows to constrain their masses quite accurately. Unlike any other OB clusters, the distance to the Galactic Center is known to better than 5\% ($\rm 8.33\pm0.35\,kpc$). Together with precise extinction measurements \citep[][$A_K\approx2.7$]{Fri11}, the absolute magnitudes of the stars are known to $\approx$ 0.2\,mag. We use magnitude, infrared color, temperature and mass calibrations from \cite{Cox00}, and ages from \cite{Sal06}. An $m_K=14.1$ star at the distance of the GC corresponds to a B0V dwarf with an initial mass of $M_{\rm MS}=17\,M_{\odot}$ and a main-sequence lifetime of $t_{\rm MS}=8\rm \Myr$, which is of the same order as the young CW disk age. Fainter B-dwarfs with $m_K=15.5$ ($M_{\rm MS}=11\,M_{\odot}$, $t_{\rm MS}=25\rm \Myr$) and $m_K=16.5$ ($M_{\rm MS}=6\,M_{\odot}$, $t_{\rm MS}=120\rm \Myr$) can be significantly older than the young CW disk.

\begin{figure}[ht]
\vspace{-0pt}
  \centering
    \includegraphics[trim=0cm 0cm 0.cm 0cm, clip=true, angle = -90, width=0.47\textwidth]{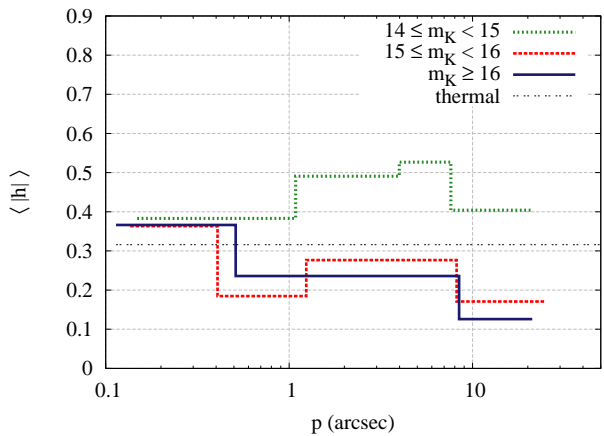}
    		\vspace{-0 pt}
		\caption{Histogram of observed $|h|$-values as a function of $p$. The median value of $|h| = 0.32$ is plotted for a thermal eccentricity distribution for comparison.} 
		\label{fig_hist_fn_p}
\end{figure}

\begin{figure*}[th!]
\begin{minipage}[b]{0.47\linewidth}
\centering   
    \includegraphics[trim=0cm 0cm 0.cm 0cm, clip=true, angle = -90, scale=0.35]{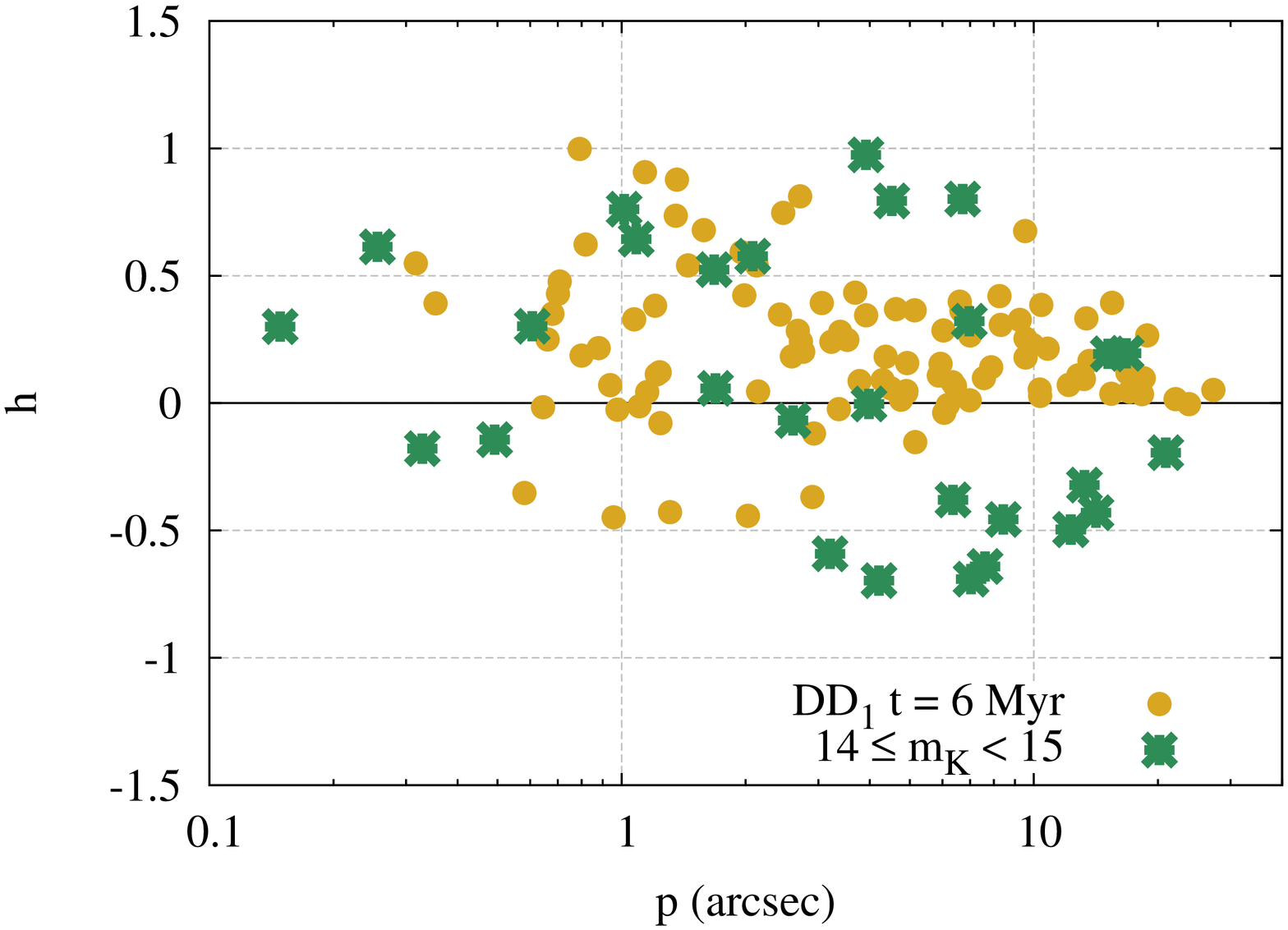}
\end{minipage}
\begin{minipage}[b]{0.47\linewidth}
\centering
    \includegraphics[trim=0cm 0cm 0.cm 0cm, clip=true, angle = -90, scale=0.35]{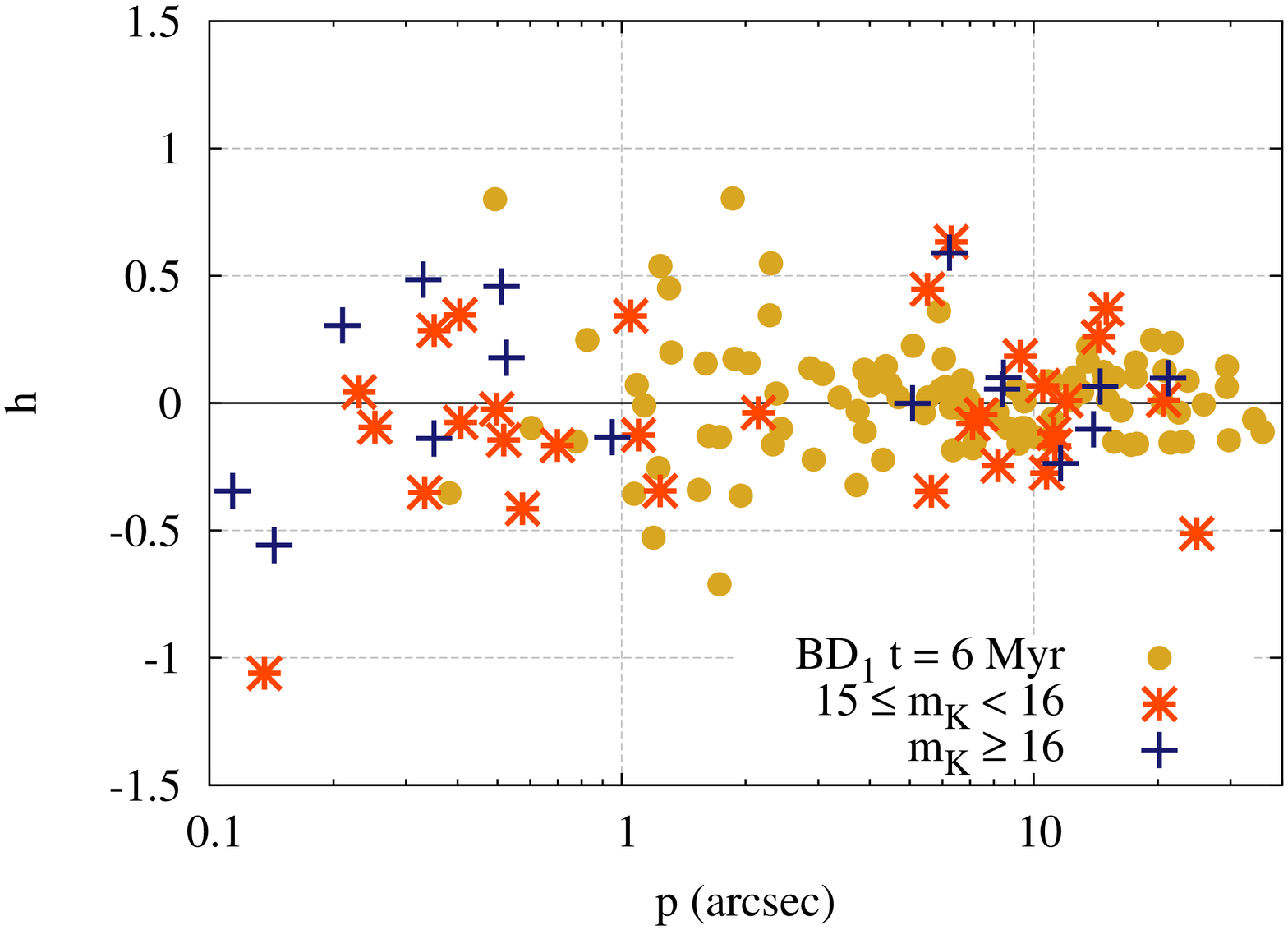}
\end{minipage}
\caption{Plot of measured $h$-values of stars as a function of projected radius, $p$, for the {\it dissolved disk} simulation DD$_1$ (left) and the {\it binary disruption} simulation BD$_1$ (right). Also plotted are $h$-values for stars in different magnitude ranges, $14 \leq m_K < 15$  (left), $15 \leq m_K < 16$ and $m_K \geq 16$ (right).} 
		\label{fig_h_plane_comp_sims_data}
		\hspace{1cm}
\end{figure*}

\begin{center}
\begin{table*}[th!]
\caption{\label{tab:properMotions} Proper motions of young ($m_{\rm K} > 14$) stars.} 
{\small
\hfill{}
\begin{tabular}{ccccccccccc}
\hline 
       \noalign{\smallskip}
{\footnotesize Star} & $m_{\rm K}$ & x & y & p & ex & ey & vx & vy & evx & evy \tabularnewline
& (mag) & (arcsec) & (arcsec) &  (arcsec) & (arcsec) & (arcsec) &  (mas $\peryr$) & (mas $\peryr$)  & (mas $\peryr$) & (mas $\peryr$)  \tabularnewline
        \noalign{\smallskip}
\hline 
       \noalign{\smallskip}
 1 & 16.7283 & 0.0463 & -0.1040 & 0.1139 & 0.0009 & 0.0008 & -7.2066 & -27.1863 & 0.4076 & 0.3534  \tabularnewline 
 2 & 15.8604 & -0.0673 & 0.1182 & 0.1360 & 0.0015 & 0.0006 & 42.2079 & 26.0934 & 0.4189 & 0.1749  \tabularnewline 
 3 & 17.4537 & 0.1403 & -0.0311 & 0.1437 & 0.0020 & 0.0017 & -2.3484 & -25.4647 & 0.7190 & 0.6107  \tabularnewline 
 4 & 14.2067 & 0.0281 & 0.1458 & 0.1485 & 0.0002 & 0.0006 & -8.8333 & 25.1138 & 0.0793 & 0.2235  \tabularnewline 
 5 & 16.9512 & -0.1984 & -0.0702 & 0.2105 & 0.0004 & 0.0006 & -17.2400 & -18.2476 & 0.1970 & 0.2380  \tabularnewline 
 6 & 15.6632 & 0.1873 & 0.1345 & 0.2306 & 0.0005 & 0.0004 & 23.1651 & 18.5350 & 0.1129 & 0.0978  \tabularnewline 
 7 & 15.7113 & 0.2264 & 0.1111 & 0.2522 & 0.0007 & 0.0005 & -5.2460 & -6.1800 & 0.2829 & 0.1908  \tabularnewline 
 8 & 14.7296 & 0.0231 & -0.2547 & 0.2558 & 0.0001 & 0.0001 & 22.6434 & -18.2185 & 0.0470 & 0.0564  \tabularnewline 
 9 & 14.5520 & 0.3052 & 0.1216 & 0.3285 & 0.0002 & 0.0001 & 12.8644 & -0.6664 & 0.0561 & 0.0324  \tabularnewline 
10 & 16.5369 & 0.1906 & -0.2697 & 0.3302 & 0.0002 & 0.0003 & 22.7268 & -6.9537 & 0.1068 & 0.1335  \tabularnewline 
        \noalign{\smallskip}
\hline 
\multicolumn{11}{l}{{\scriptsize {\bf Notes.}}}\tabularnewline
\multicolumn{11}{l}{{\scriptsize Table \ref{tab:properMotions} is published in its entirety in the online journal. A portion is shown here for guidance regarding its form and content.}}\tabularnewline
       \noalign{\smallskip}
\end{tabular}}
\hfill{}
\end{table*}
\end{center}

\subsection{$h$-values in different $K$-magnitude ranges}

In Figure \ref{fig_h_plane_magbins} we plot $h$-values of stars as a function of projected radius, $p$, for different $K$-magnitude ranges. Error bars are determined through error propagation from position and proper motion uncertainties given in Table \ref{tab:properMotions}, and an MBH mass uncertainty of $0.36 \times 10^6 \Mo$ \citep{Gil09b}. Stars in our sample have, on average, lower $|h|$-values as their $K$-magnitude increases. In Figure \ref{fig_cum_data} we plot the cumulative $|h|$-distribution function for stars in these different $K$-magnitude ranges. Stars with $m_K \geq 15$ have lower $|h|$-values than those with $m_K < 15$, and if isotropically distributed, form a population more eccentric than a thermal distribution, $N(e) de \sim e de$. We compare different distributions to each other using the one-dimensional two-sample Kolmogorov-Smirnov (KS) test, under the null hypothesis that the samples are drawn from the same distribution. We refer to the test statistic as $p_{\rm ks}$ to distinguish it from the projected radius of a star, $p$. We reject the null hypothesis if $p_{\rm ks}$ is smaller than or equal to the significance level, $\alpha = 0.05$. A KS test between stars with $14 \leq m_K < 15$ and those with $m_K \geq 15$ yields $p_{\rm ks} = 0.007$, which suggests they are not drawn from the same population, and between stars with $15 \leq m_K < 16$ and $m_K \geq 16$ yields $p_{\rm ks} = 0.965$.
  
In Figures \ref{figHistAllData} - \ref{figHist1415} we plot histograms of $h$-values for observational data, binned in $K$-magnitude and radius, compared to that for a simulated thermal eccentricity distribution. The $m_K \ge 14$ population contains fractionally more CCW orbits and low $|h|$-values than the $m_K < 14$ population. Stars with $14 \leq m_K < 15$ have a flatter $h$-distribution than those with $m_K \ge 15$ and are missing the central low-$|h|$ peak. We plot $h$-values for stars with projected radii less than and greater than $7\arcsec$ as the azimuthal coverage drops at his radius (corresponds to middle black circle in Figure 1 of \citet{Bar10}). The $m_K \ge 14$ stars with $p > 7\arcsec$ have low $|h|$-values; this is not seen in the $m_K < 14$ population. The are more $m_K \ge 14$ stars with $p > 7\arcsec$ on CCW orbits than CW orbits but this result is not statistically significant. 

In Figure \ref{figHistOnOffDisk} we plot a histogram of $h$-values for $m_K < 14$ stars on and off the young CW disk using a re-analysis of \citet{Bar10} data (this re-analysis revealed 43 stars to be on the CW disk, instead of 45). The off-disk population have slightly lower $|h|$-values than the on-disk population, indicative of a more eccentric or inclined population. As the CW disk is more edge-on than face-on, a population of stars with the same eccentricities but more isotropically distributed should have higher $|h|$-values. This suggests the off-disk population is more eccentric. They also have more $h < 0$ values or CCW orbits. In Figure \ref{fig_cum_data_CWdisk} we plot the cumulative $|h|$-distribution function for the same data. The relatively high $|h|$-values of the young CW disk stars suggests, given their inclination, that they are actually quite a low-eccentricity population.

\subsection{$|h|$-values as function of projected radius, $p$}

In Figure \ref{fig:p_gtr_8} we plot the cumulative $|h|$-distribution function for data in different $K$-magnitude ranges, binned according their projected radius, either less or greater than $p = 0.8 \arcsec$. All stars in our sample with $p < 0.8 \arcsec$ have $m_K \geq 14.2$; these are the `S-stars'. They have a $|h|$-distribution that, if isotropically distributed, is just slightly more eccentric than a thermal eccentricity distribution. A KS-test between the two populations (S-stars and thermal distribution) yields $p_{\rm ks} = 0.48$. Stars with $m_K \geq 16$, if isotropically distributed, form a more eccentric group at large radii, $p \geq 0.8 \arcsec$, than at small radii ($p_{\rm ks} = 0.017$ between the two populations). Stars between $14 \leq m_K < 15$ appear more eccentric at smaller projected radii, $p < 0.8 \arcsec$, but the two groups do not differ in a statistically significant way ($p_{\rm ks} = 0.16$). In Figure \ref{fig_hist_fn_p} we show a histogram of $|h|$-values as a function of projected distance for stars in different magnitude ranges. While the binning is arbitrary, this visually confirms that stars with $m_K \geq 15$ have lower $|h|$-values  at larger projected radii (i.e., become more eccentric/inclined to $xy$-plane), while the opposite is true for stars with $14 \leq m_K < 15$. In Table \ref{tab:p_gtr_7} we compare the mean $|h|$-values for stars with $p \geq 7 \arcsec, 10 \arcsec$. Stars with higher $K$-magnitude have lower mean $|h|$-values at large radii. 

\begin{table}[h!]
\caption{Mean and standard deviation of $|h|$-values for observational data binned in $K$-magnitude} 
\begin{center}
\begin{tabular}{ccccccc}
 Sample &  \multicolumn{3}{c}{ $p^{a} \geq 7 \arcsec$} &  \multicolumn{3}{c}{$p^{a} \geq 10 \arcsec$} \tabularnewline
        \noalign{\smallskip}
\hline
        \noalign{\smallskip}
 & Num & $\langle |h| \rangle^{b}$ & $\sigma_{|h|}^{c}$ &  Num  & $\langle |h| \rangle^{b}$ &  $\sigma_{|h|}^{c}$ \tabularnewline
        \noalign{\smallskip}
\hline
all & 82 & 0.390 & 0.261 & 39 & 0.300 & 0.204 \tabularnewline
$14 \leq m_K < 15$ & 10 & 0.412 & 0.182 & 7 & 0.333 &0.142 \tabularnewline
$15 \leq m_K < 16$ & 13 & 0.181  &  0.149 & 9 & 0.200 & 0.170 \tabularnewline
$m_K \geq 15$ & 19 &0.158 & 0.131 & 13 & 0.176  & 0.148 \tabularnewline
$m_K \geq 16$ & 6 & 0.109 & 0.066 & 4 & 0.126 & 0.076 \tabularnewline
        \noalign{\smallskip}
\hline 
\multicolumn{5}{l}{{\scriptsize {\bf Notes.}}} \tabularnewline
\multicolumn{5}{l}{{\scriptsize $^{a}$ Projected radius.}} \tabularnewline
\multicolumn{5}{l}{{\scriptsize $^{b}$ Mean value of $|h|$ for stars with $p \geq 7(10) \arcsec$.}} \tabularnewline
\multicolumn{5}{l}{{\scriptsize $^{c}$ Standard deviation on $|h|$ for stars with $p \geq 7(10) \arcsec$.}} \tabularnewline
       \noalign{\smallskip}
\end{tabular}
\end{center}
\label{tab:p_gtr_7}
\end{table} 

\begin{figure}[ht!]
\vspace{-0pt}
  \centering
    \includegraphics[trim=0cm 0cm 0.cm 0cm, clip=true, angle = -90, width=0.46\textwidth]{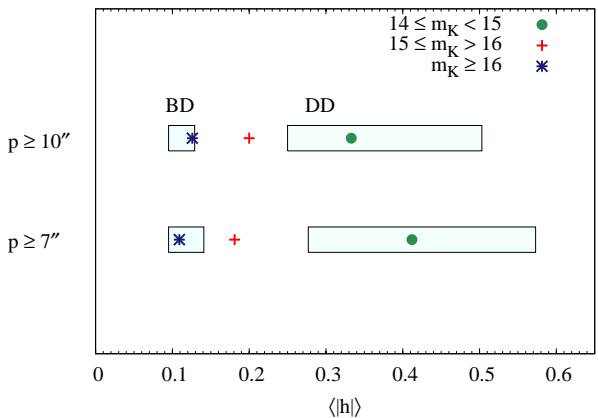}
    		\vspace{-0 pt}
		\caption{Comparison of the range in $\langle |h| \rangle$-values for both the {\it binary disruption} (BD: blue box) and the {\it dissolved disk} simulations (DD: green box), for stars with $p \geq 7 \arcsec$ and $p \geq 10\arcsec$, and for observational data binned in $K$-magnitude. The observational data within $14 \leq m_K < 15$ have $\langle |h| \rangle$-values that are compatible with the {\it dissolved disk} scenario range, not the {\it binary disruption} scenario.  The observational data with $m_K \geq 16$ have much lower $\langle |h| \rangle$-values that are compatible with the {\it binary disruption} scenario.} 
		\label{fig_h_grt_7_10}
\end{figure}

\begin{figure*}[th!]
\begin{minipage}[b]{0.5\linewidth}
\centering   
    \includegraphics[trim=0cm 0cm 0.cm 0cm, clip=true, angle = -90,  scale = 0.4]{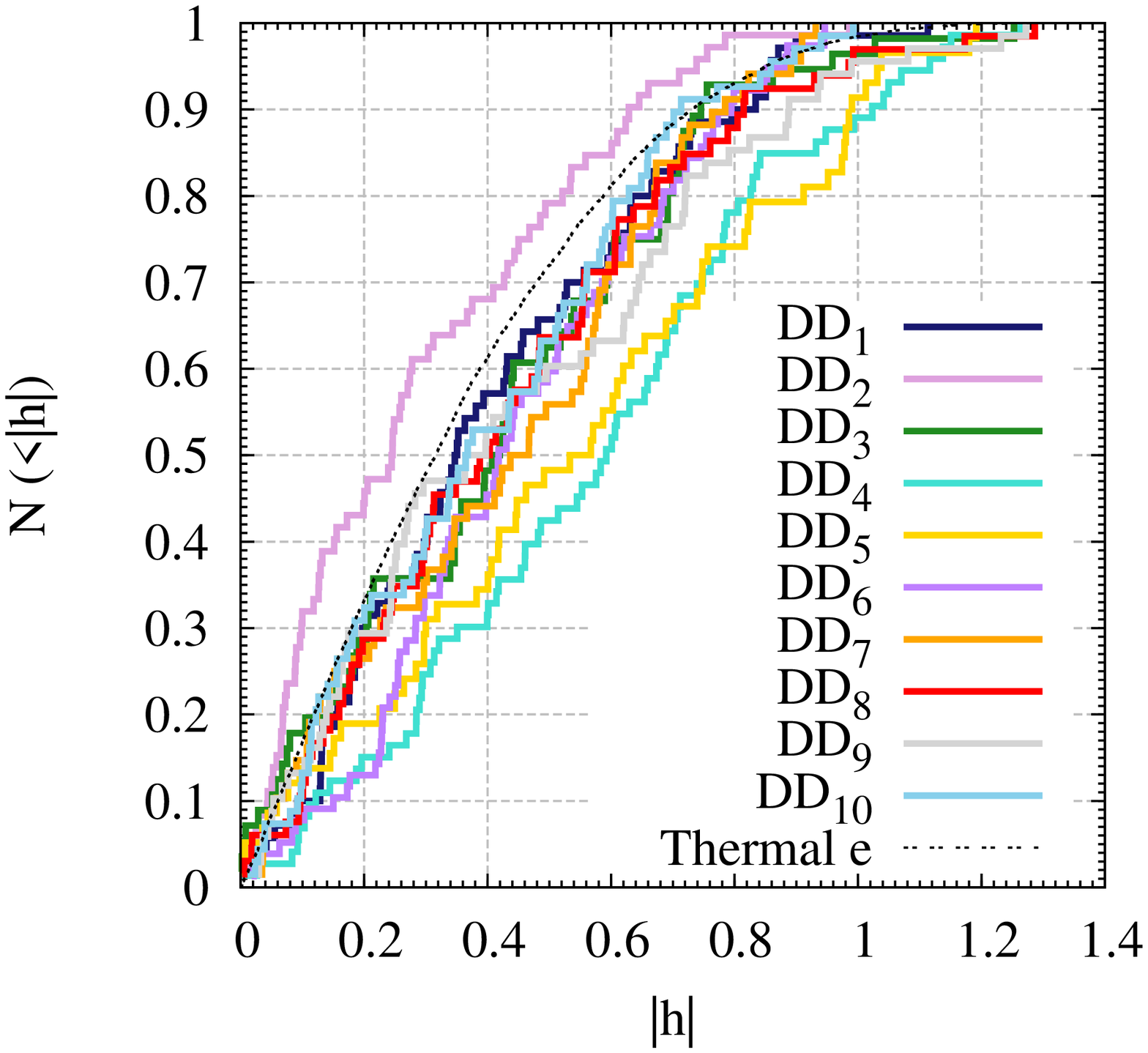}

    		\vspace{-0 pt}
				\end{minipage}
\begin{minipage}[b]{0.5\linewidth}
\centering 
 \includegraphics[trim=0cm 0cm 0.cm 0cm, clip=true, angle = -90, scale = 0.4]{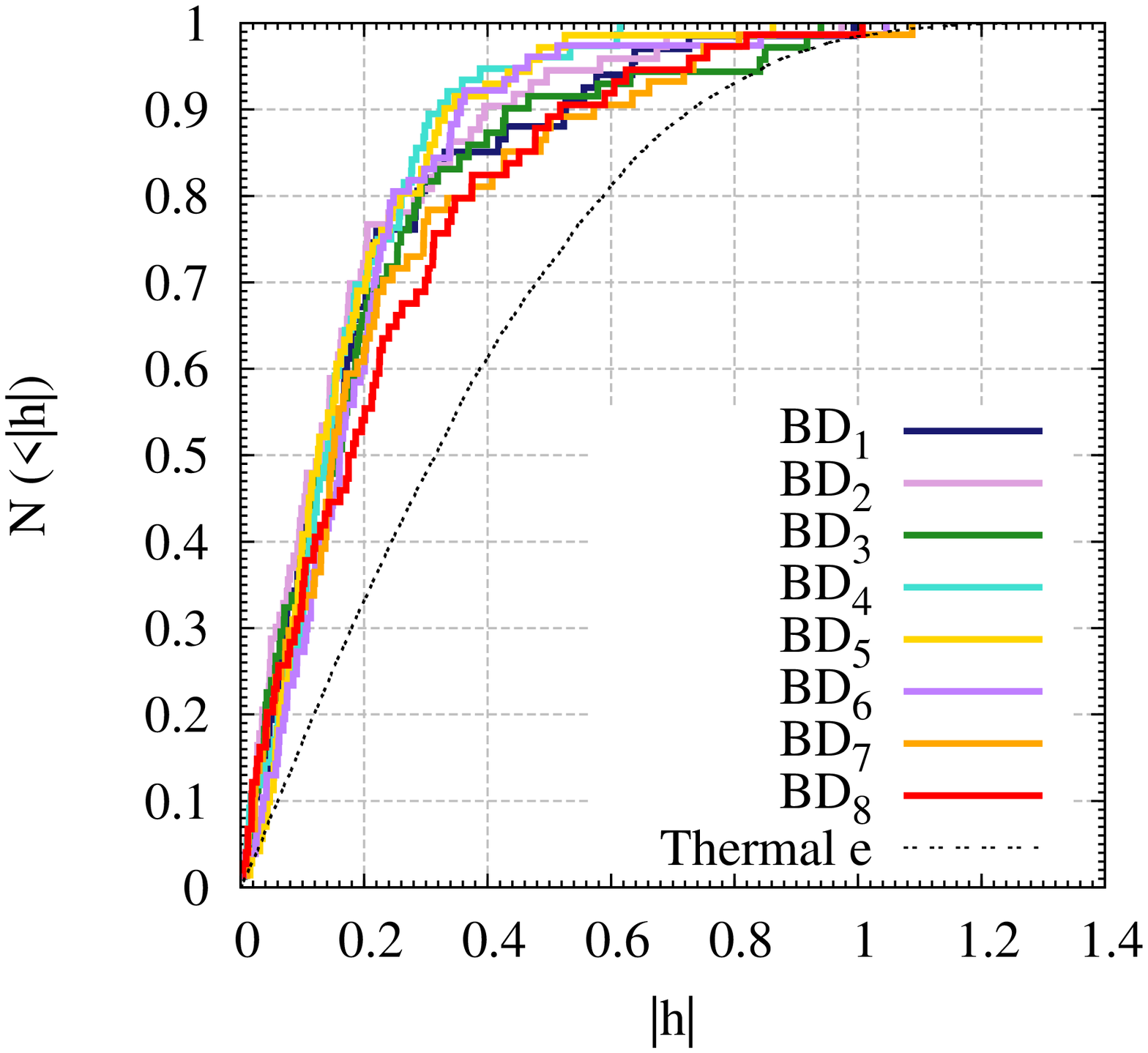}
    		\vspace{-0 pt}
\end{minipage}
\caption{Cumulative $|h|$-distributions of B-stars in  {\it dissolved disk} simulations (left) and {\it binary disruption} simulations (right). $|h|$-values are drawn from a random viewing direction and sampled using the observational completeness correction. } 
\label{fig:cum_h_bd_sims_cc}
\end{figure*}

\begin{figure}[ht!]
\vspace{-0pt}
  \centering
    \includegraphics[trim=0cm 0cm 0.cm 0cm, clip=true, angle = -90, width=0.46\textwidth]{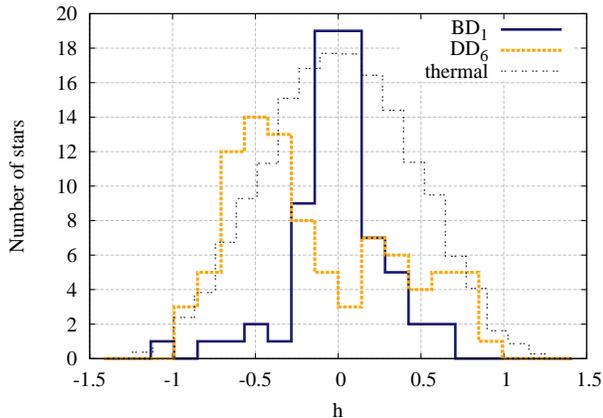}
    		\vspace{-0 pt}
		\caption{Histogram of $h$-distributions of B-stars in DD$_6$ and BD$_1$ simulations. $h$-values are drawn from a random viewing direction and sampled using the observational completeness correction.} 
		\label{fig_HISTBD1DD6}
\end{figure}

\section{Comparison between observations and simulations}\label{S:comp}

We compare the observational data with simulation results from two formation scenarios of B-stars in the Galactic center: the {\it binary disruption} scenario in which stars begin their lives near the MBH on high-eccentricity ($e \simeq 0.97$) orbits, and the {\it dissolved disk} scenario in which they form in a nuclear stellar disk of lower eccentricity ($0.3 \lesssim e \lesssim 0.7$). In Figure \ref{fig_h_plane_comp_sims_data} we show $h$-values of both simulated stars (at the end of the simulation, $t = 6 \Myr$) and data in different magnitude ranges, as a function of their projected radii, $p$. The higher magnitude stars, $m_K \geq 15$, qualitatively match the {\it binary disruption} simulations well while the lower magnitude stars, $14 \leq m_K < 15$, match better the {\it dissolved disk} simulations. 

We compare the range in $\langle |h| \rangle$-values for stars at large projected radii in our simulations to the single values taken from observational data binned in $K$-magnitude (see Table \ref{tab:p_gtr_7}), in Figure \ref{fig_h_grt_7_10}. For stars with $p \geq 7 \arcsec$ and $14 \leq m_K < 15$, $\langle |h| \rangle = 0.412$, which places this population well outside the range found in the {\it binary disruption} simulations ($0.095 - 0.141$), whereas for $m_K \geq 16$, $\langle |h| \rangle = 0.109$, which lies inside the correct range for {\it binary disruption} simulations, not for {\it dissolved disk} simulations, for which $0.277 \leq \langle |h| \rangle \leq 0.573$. 

\subsection{KS-testing, correcting for incompleteness and incorporating observational errors}

\begin{table}[h!]
\caption{Fractional Completeness as a function of radius and magnitude} 
\begin{center}
\begin{tabular}{cccc}
 Radial Bin &  $m_K < 14$ & $14 \leq m_K < 15$ & $15 \leq m_k < 17$ \tabularnewline
 (arcsec) & (mag) & (mag) & (mag)  \tabularnewline
        \noalign{\smallskip}
\hline
0 - 5 & 0.564400 & 0.454493 & 0.155868 \tabularnewline
5 - 10 & 0.390942 & 0.237714 & 0.144753 \tabularnewline
10 - 15 & 0.249561 & 0.106667 & 0.066129 \tabularnewline
15 - 20 & 0.164209 & 0.043871 & 0.026194 \tabularnewline
        \noalign{\smallskip}
\hline 
\end{tabular}
\end{center}
\label{tab:completenessCorrection}
\end{table} 
     
To include the dependence of $|h|$-values on projected radius, we compare our simulations with the data using radial completeness corrections derived from from observations in \citet{Bar10}. We show the corrections used in radial and magnitude bins in Table \ref{tab:completenessCorrection}. We calculate projected $|h|$ and $p$ values of the stars in our simulations and then sample according to the completeness at their projected radii and $K$-magnitude. For illustration, we plot the cumulative $|h|$-distributions of all the {\it dissolved disk} simulations and {\it binary disruption} simulations in Figure \ref{fig:cum_h_bd_sims_cc}. $|h|$-values are sampled from a single random viewing direction, using the observational completeness correction for stars with $14 \leq m_K < 15$ in the {\it dissolved disk} simulations, and for stars with $15 \leq m_K < 17$ in the {\it binary disruption} simulations. We note that the results from the {\it dissolved disk} simulations are highly-dependent on the chosen viewing angle, in contrast to the {\it binary disruption} simulations. In Figure \ref{fig_HISTBD1DD6} we plot a histogram of the $h$-distributions of B-stars in DD$_6$ and BD$_1$ simulations. 

\begin{figure*}[h]
\begin{minipage}[b]{0.5\linewidth}
\centering   
  \includegraphics[trim=0cm 0cm 0.cm 0cm, clip=true , angle = 0, scale = 0.42]{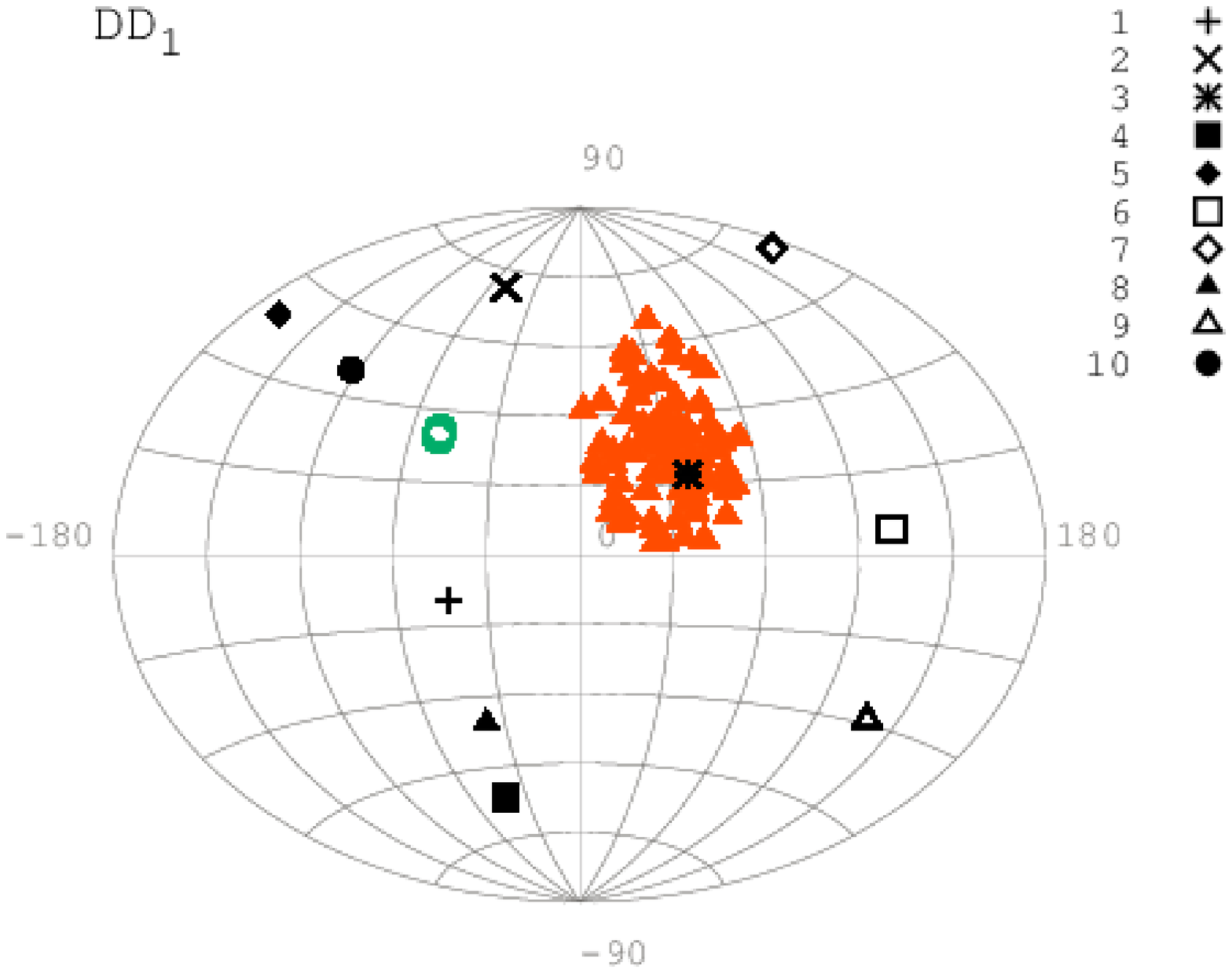}
    		\vspace{-15ex}
				\end{minipage}
\begin{minipage}[b]{0.5\linewidth}
\centering
  \includegraphics[trim=0cm 0cm 0.cm 0cm, clip=true, angle = 0,  scale = 0.42]{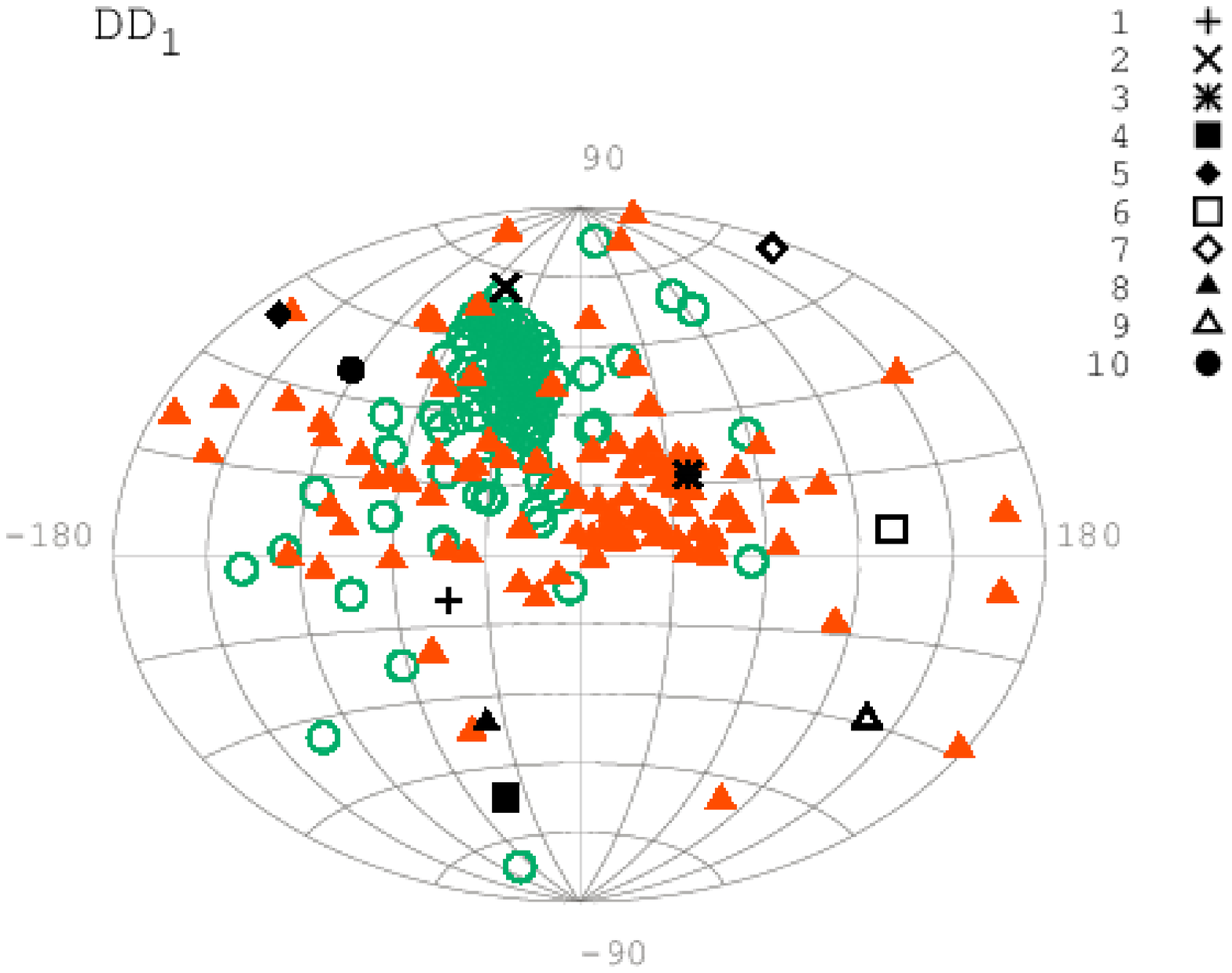}
    		\vspace{-15ex}
\end{minipage}
\vspace{0.1 cm}
\begin{minipage}[b]{0.5\linewidth}
\centering
  \includegraphics[trim=0cm 0cm 0.cm 0cm, clip=true , angle = 0, scale = 0.42]{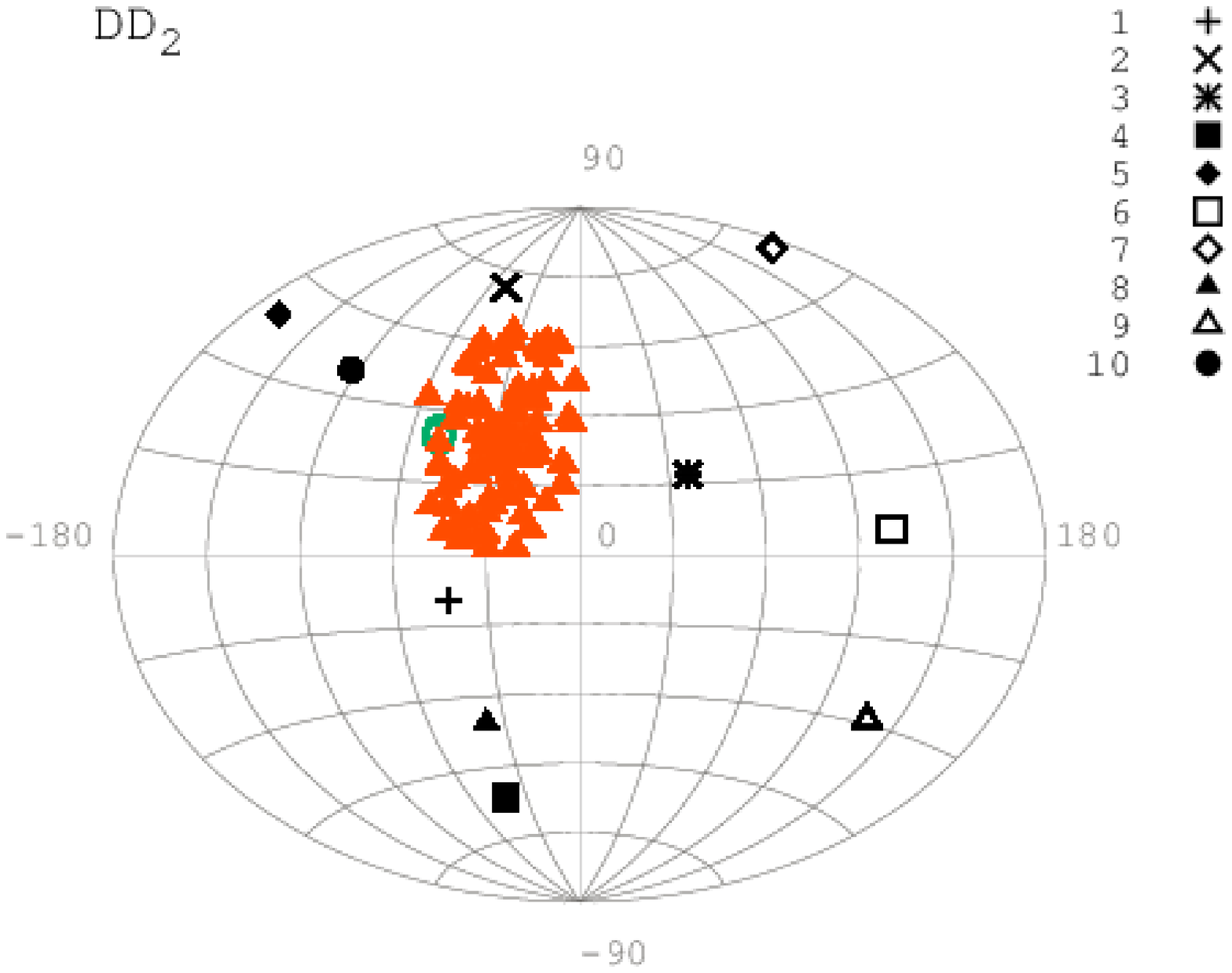}
    		\vspace{-15ex}
\end{minipage}
\begin{minipage}[b]{0.5\linewidth}
\centering
  \includegraphics[trim=0cm 0cm 0.cm 0cm, clip=true, angle = 0,  scale = 0.42]{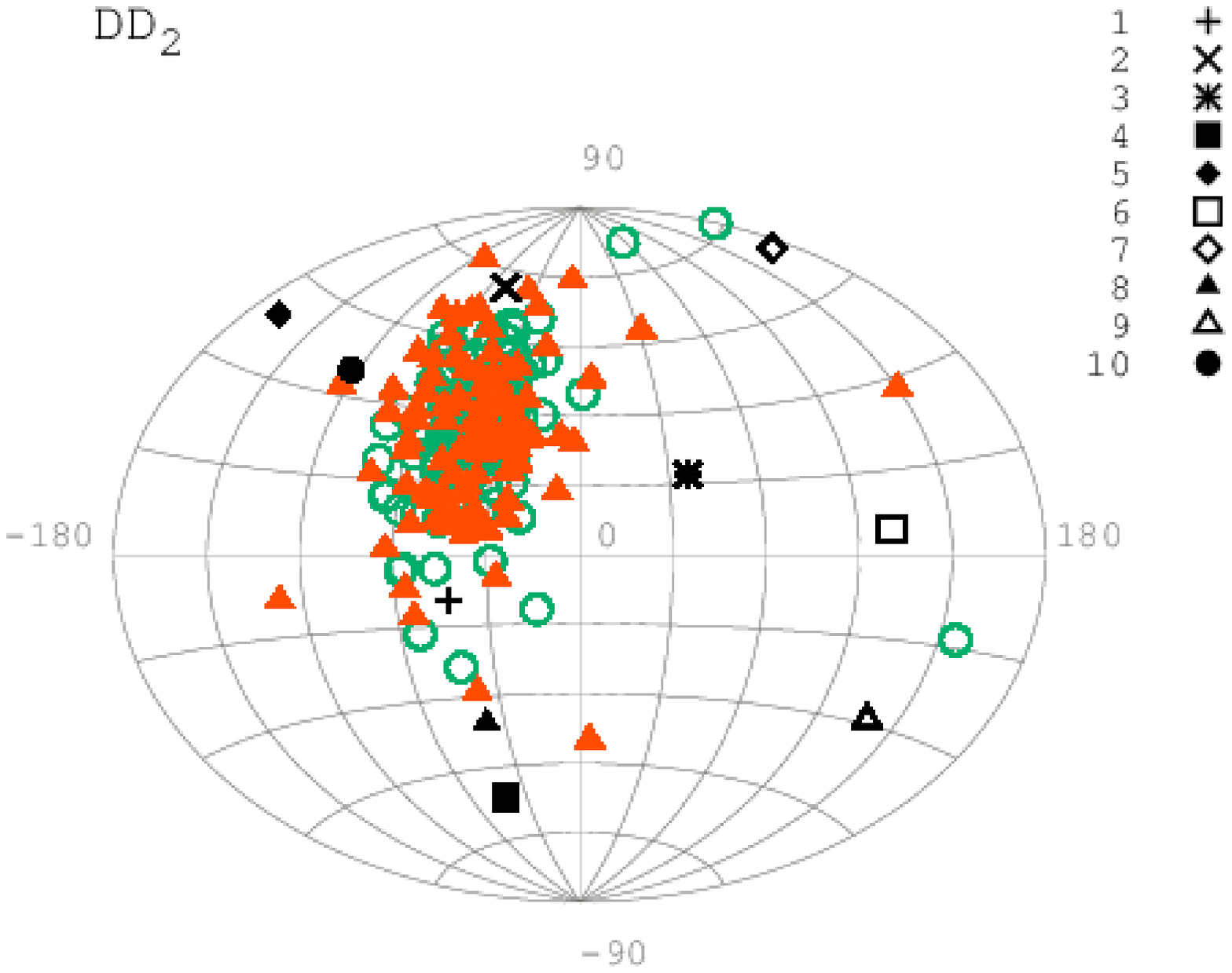}
    		\vspace{-15ex}
\end{minipage}
\begin{minipage}[b]{0.5\linewidth}
\centering   
  \includegraphics[trim=0cm 0cm 0.cm 0cm, clip=true , angle = 0, scale = 0.42]{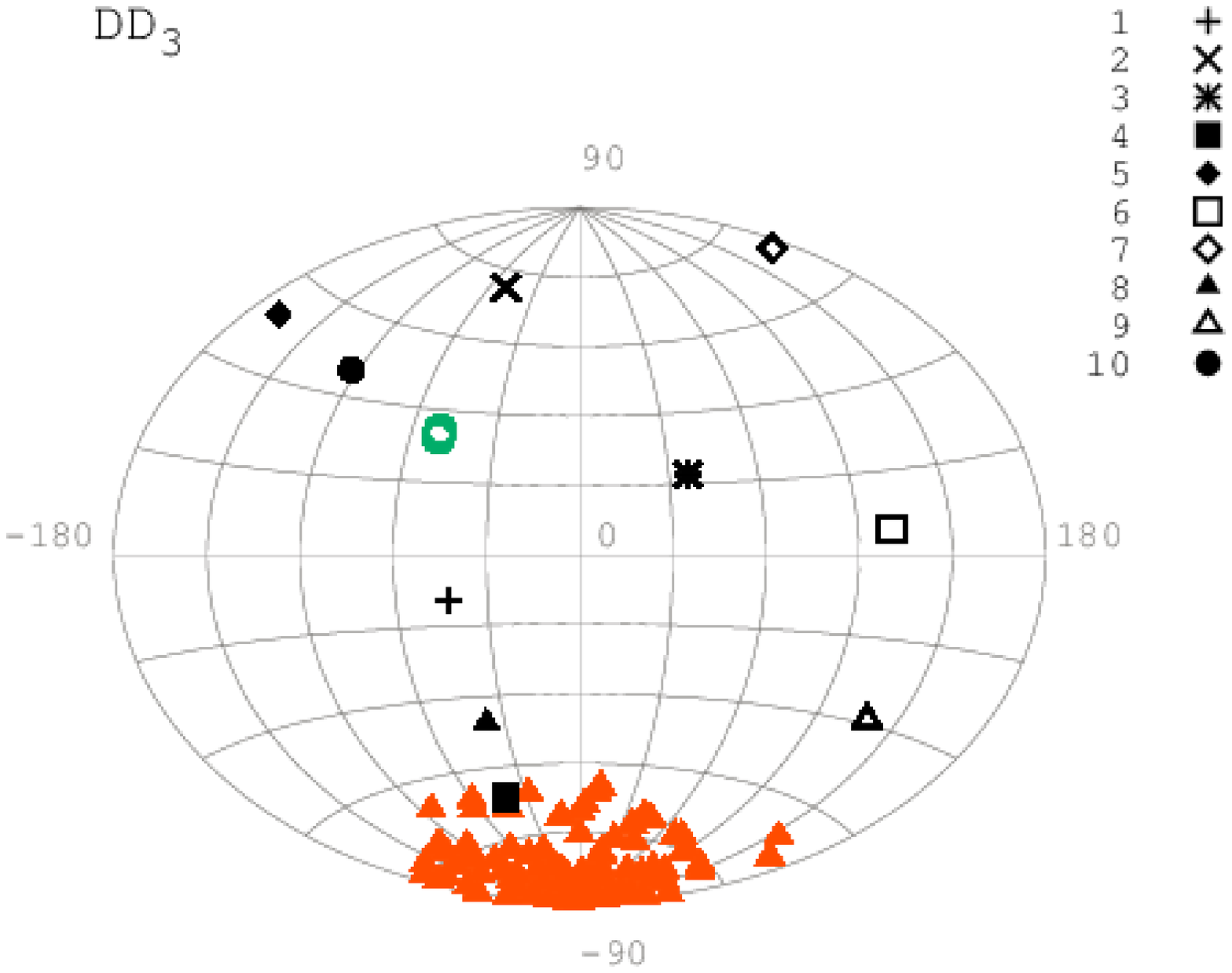}
    		\vspace{-15ex}
				\end{minipage}
\begin{minipage}[b]{0.5\linewidth}
\centering
  \includegraphics[trim=0cm 0cm 0.cm 0cm, clip=true, angle = 0,  scale = 0.42]{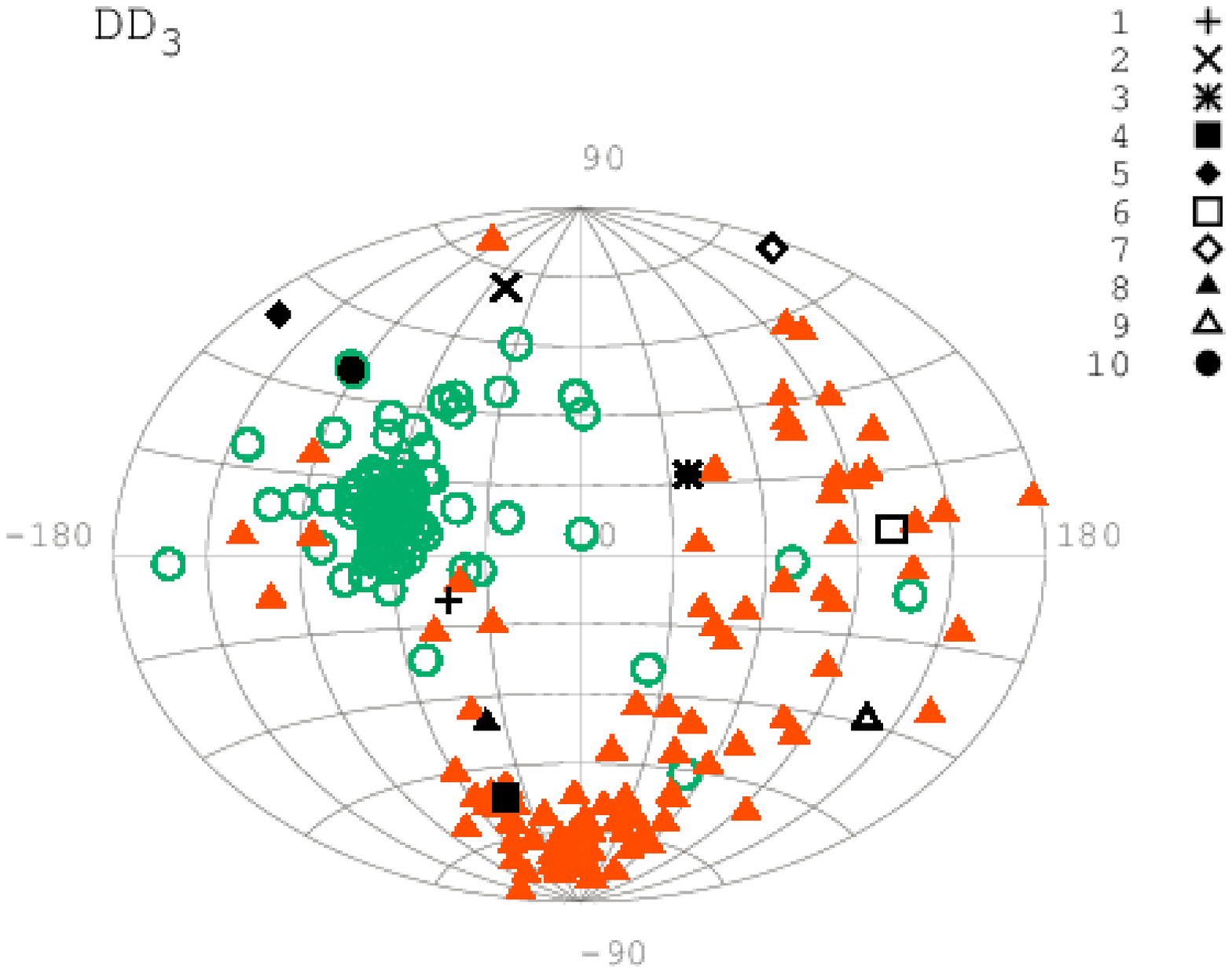}
    		\vspace{-15ex}
\end{minipage}
\vspace{0.1 cm}
\begin{minipage}[b]{0.5\linewidth}
\centering
  \includegraphics[trim=0cm 0cm 0.cm 0cm, clip=true , angle = 0, scale = 0.42]{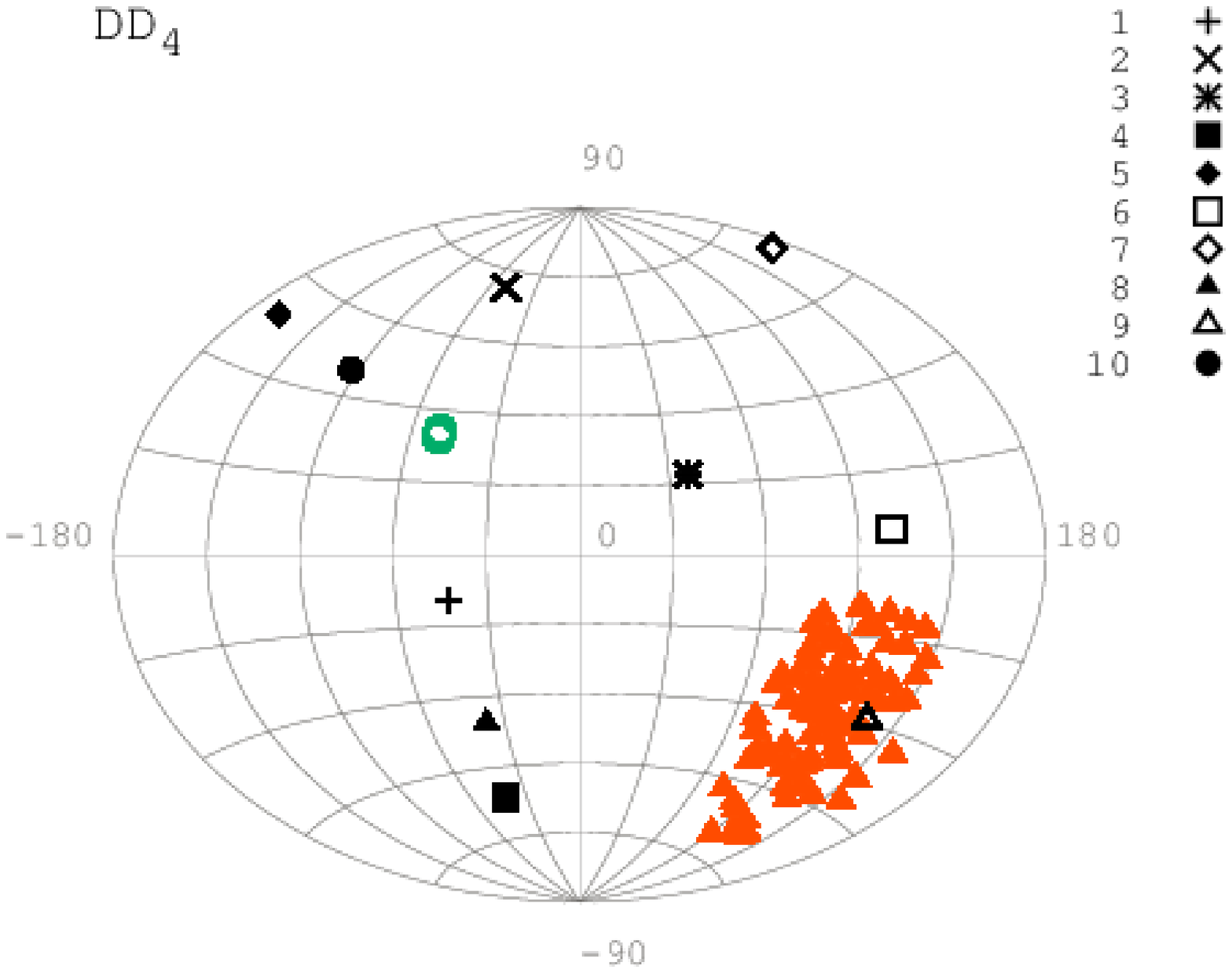}
    		\vspace{-15ex}
\end{minipage}
\begin{minipage}[b]{0.5\linewidth}
\centering
  \includegraphics[trim=0cm 0cm 0.cm 0cm, clip=true, angle = 0,  scale = 0.42]{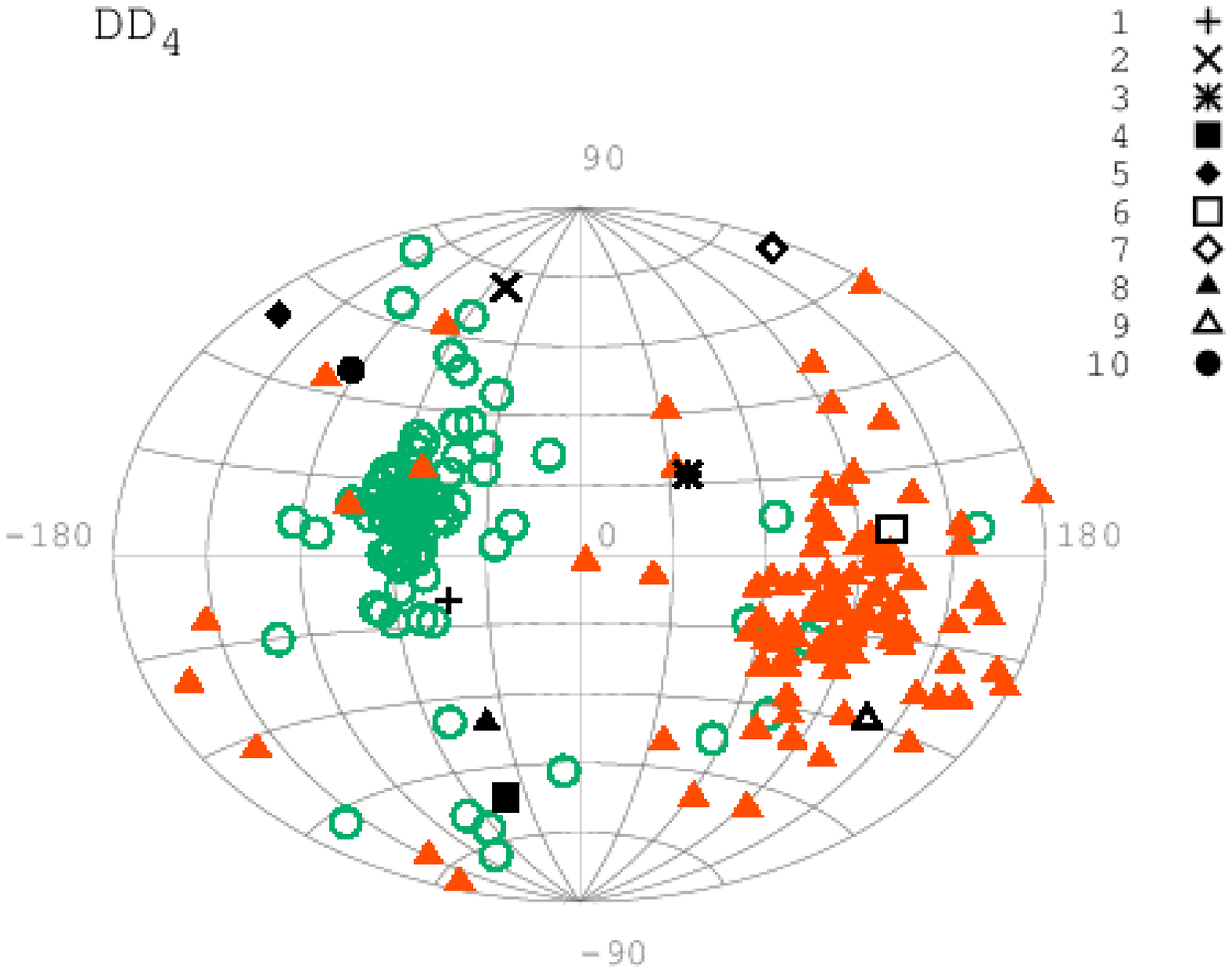}
    		\vspace{-15ex}
\end{minipage}
\caption{Aitoff projections of the angular momentum vectors of young CW disk stars (green circles) and B-stars (red triangles) at $t = 0$ (left) and $t = 6 \Myr$ (right) in the {\it dissolved disk} simulations DD$_1$, DD$_2$, DD$_3$ and DD$_4$. Ten randomly-selected viewing directions, used to calculate $|h|$-values for Figure \ref{fig:scatter_ks_sims}, are indicated with black symbols.} 
\label{fig:aitoff}
\end{figure*}

To incorporate the errors in the observed $h$-values, we take each star in a magnitude-selected range and Monte Carlo sample 100 times from a Gaussian distribution with mean $h$ and standard deviation $e_h$. We run a KS-test between the resulting cumulative distribution functions and those from completeness-corrected simulations. For the {\it dissolved disk} simulations we select ten random viewing directions with respect to the B-star disk in the simulation and generate new $|h|$-values before sampling according to $p$-values. Hence each {\it dissolved disk} simulation will have $10 \times 100$ $p_{\rm ks}$ values. In Figure \ref{fig:aitoff} we plot the angular momentum vectors of young CW disk stars and B-stars in an Aitoff projection in the {\it dissolved disk} simulations DD$_1$, DD$_2$, DD$_3$ and DD$_4$, indicating the ten randomly-selected viewing directions.

\begin{figure*}[t!]
\begin{minipage}[b]{0.5\linewidth}
\centering   
  \includegraphics[trim=0cm 0cm 0.cm 0cm, clip=true, angle = -90, scale = 0.4]{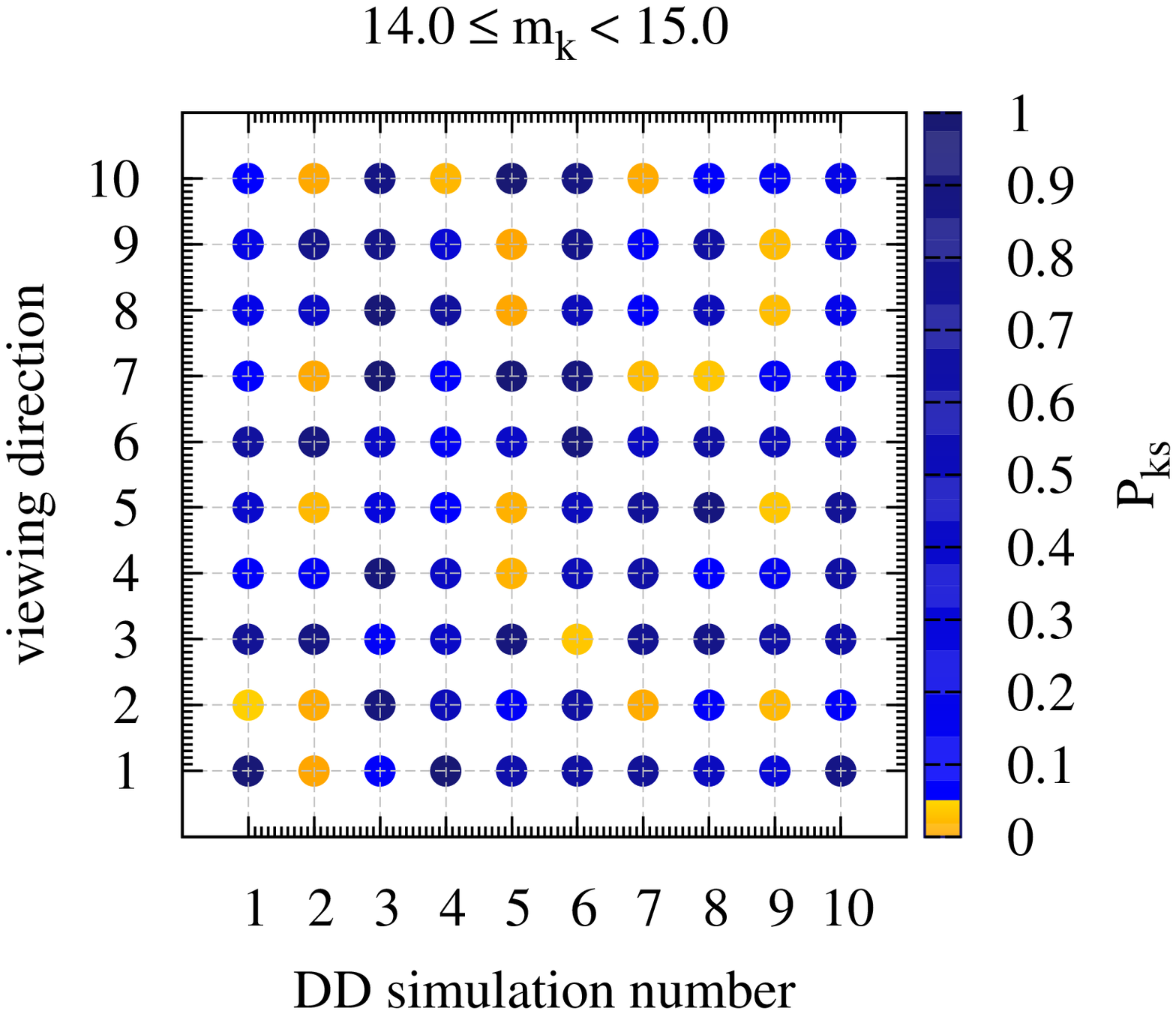}
    		\vspace{-0 pt}
				\end{minipage}
\begin{minipage}[b]{0.5\linewidth}
\centering
  \includegraphics[trim=0cm 0cm 0.cm 0cm, clip=true, angle = -90,  scale = 0.4]{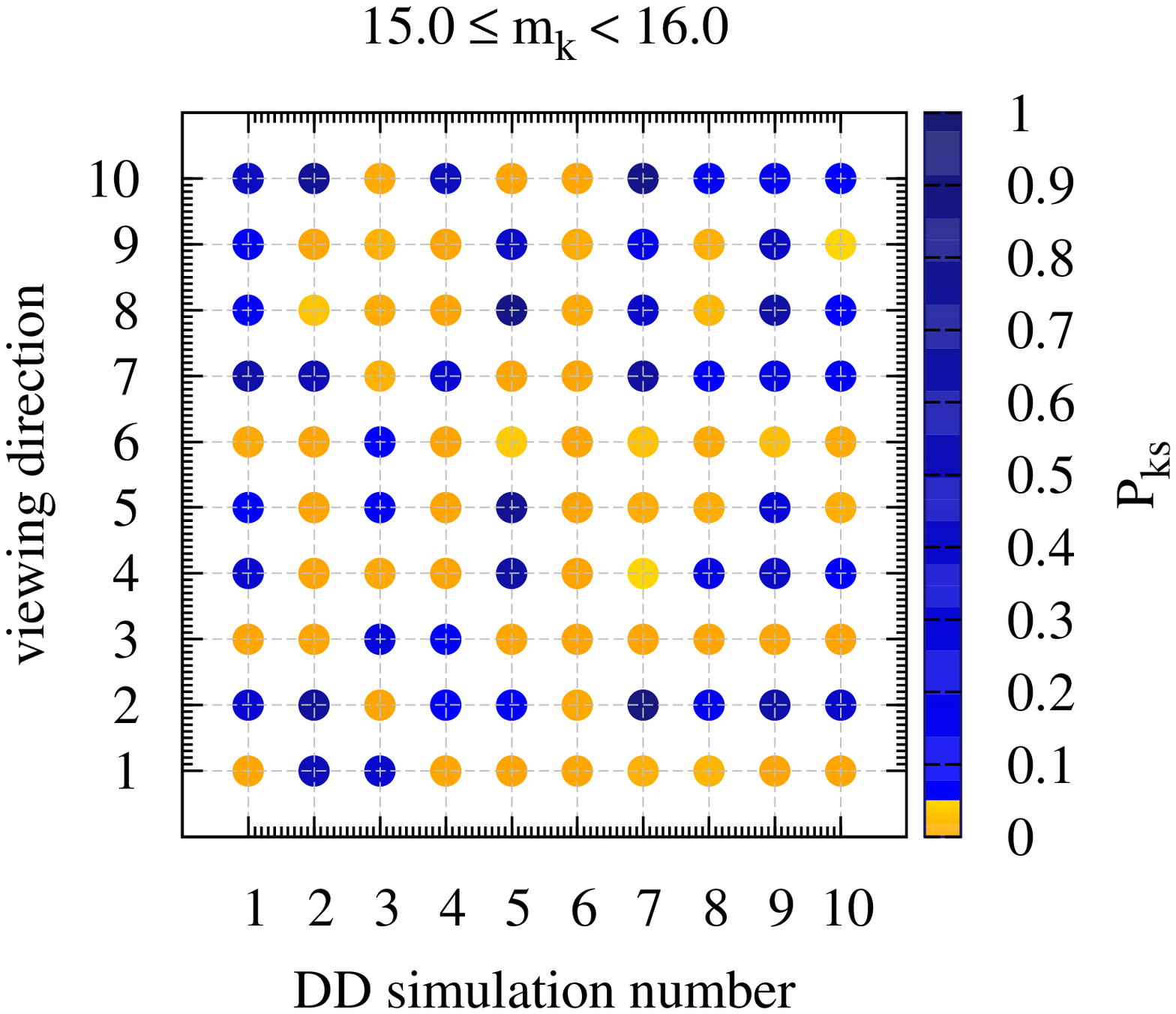}
    		\vspace{-0 pt}
\end{minipage}
\begin{minipage}[b]{0.5\linewidth}
\centering
  \includegraphics[trim=0cm 0cm 0.cm 0cm, clip=true, angle = -90,  scale = 0.4]{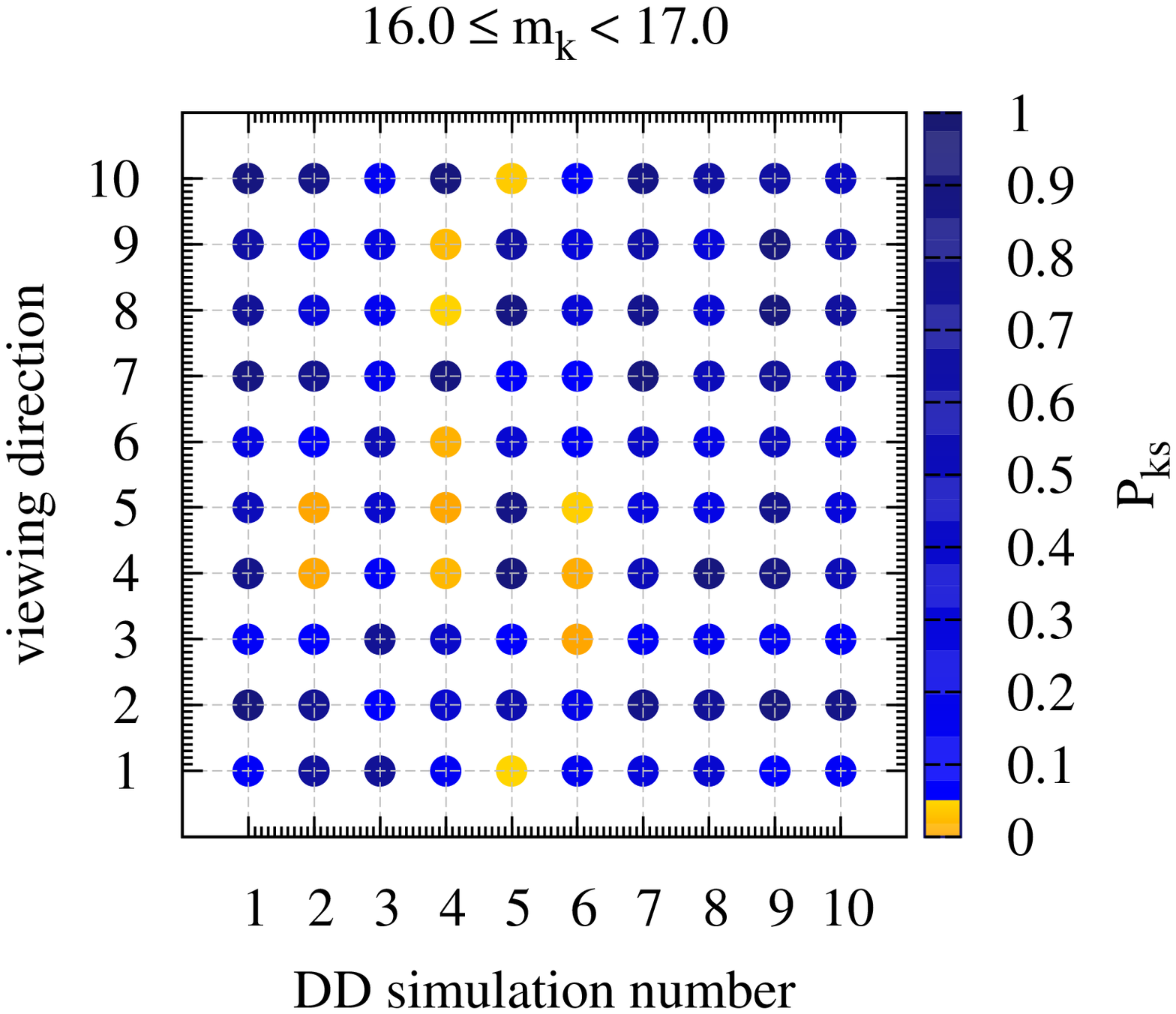}
    		\vspace{-0 pt}
		\end{minipage}
\begin{minipage}[b]{0.5\linewidth}
\centering
  \includegraphics[trim=0cm 0cm 0.cm 0cm, clip=true, angle = -90,  scale = 0.4]{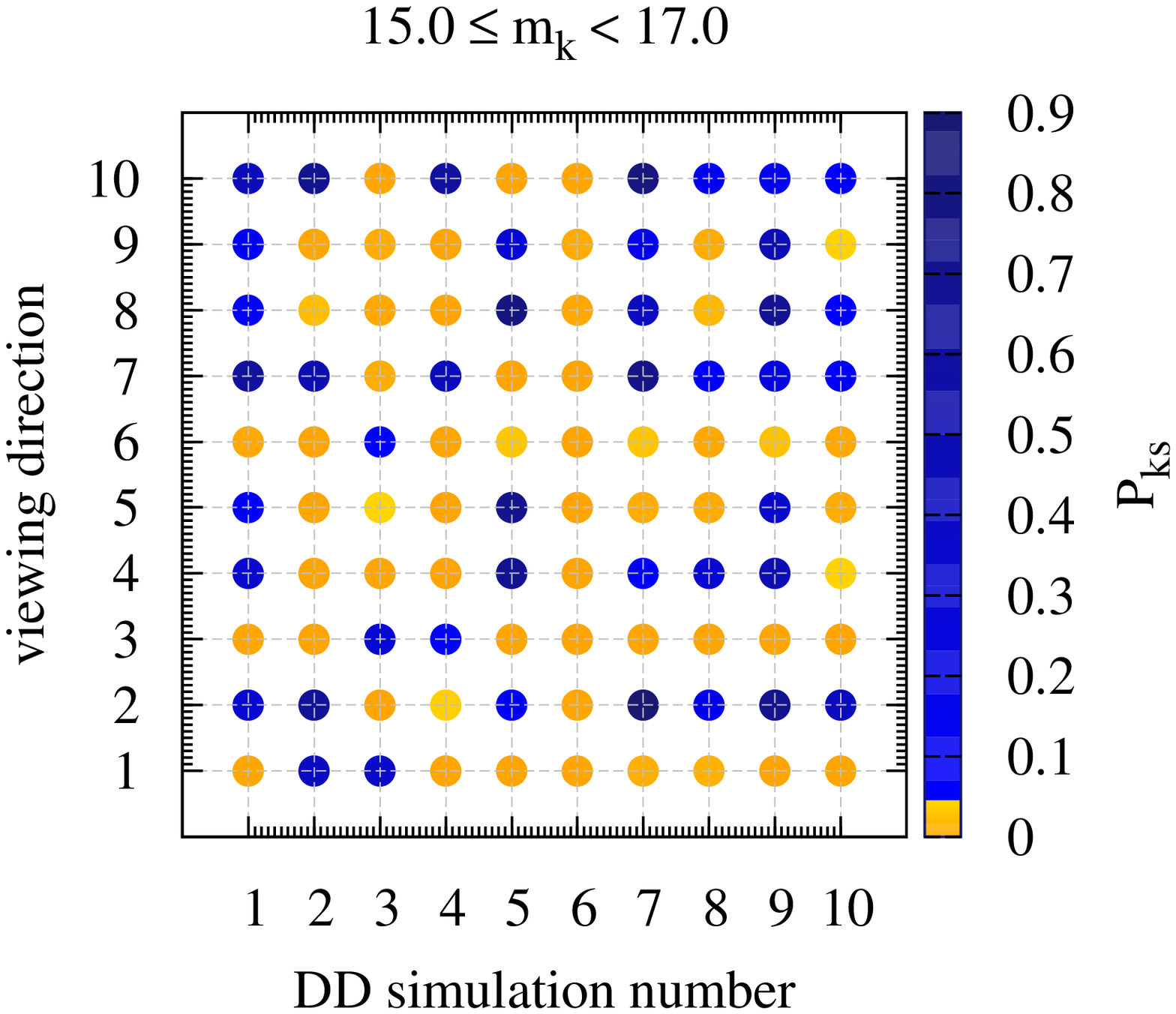}
    		\vspace{-0 pt}
\end{minipage}
\begin{minipage}[b]{0.5\linewidth}
\centering
  \includegraphics[trim=0cm 0cm 0.cm 0cm, clip=true, angle = -90,  scale = 0.4]{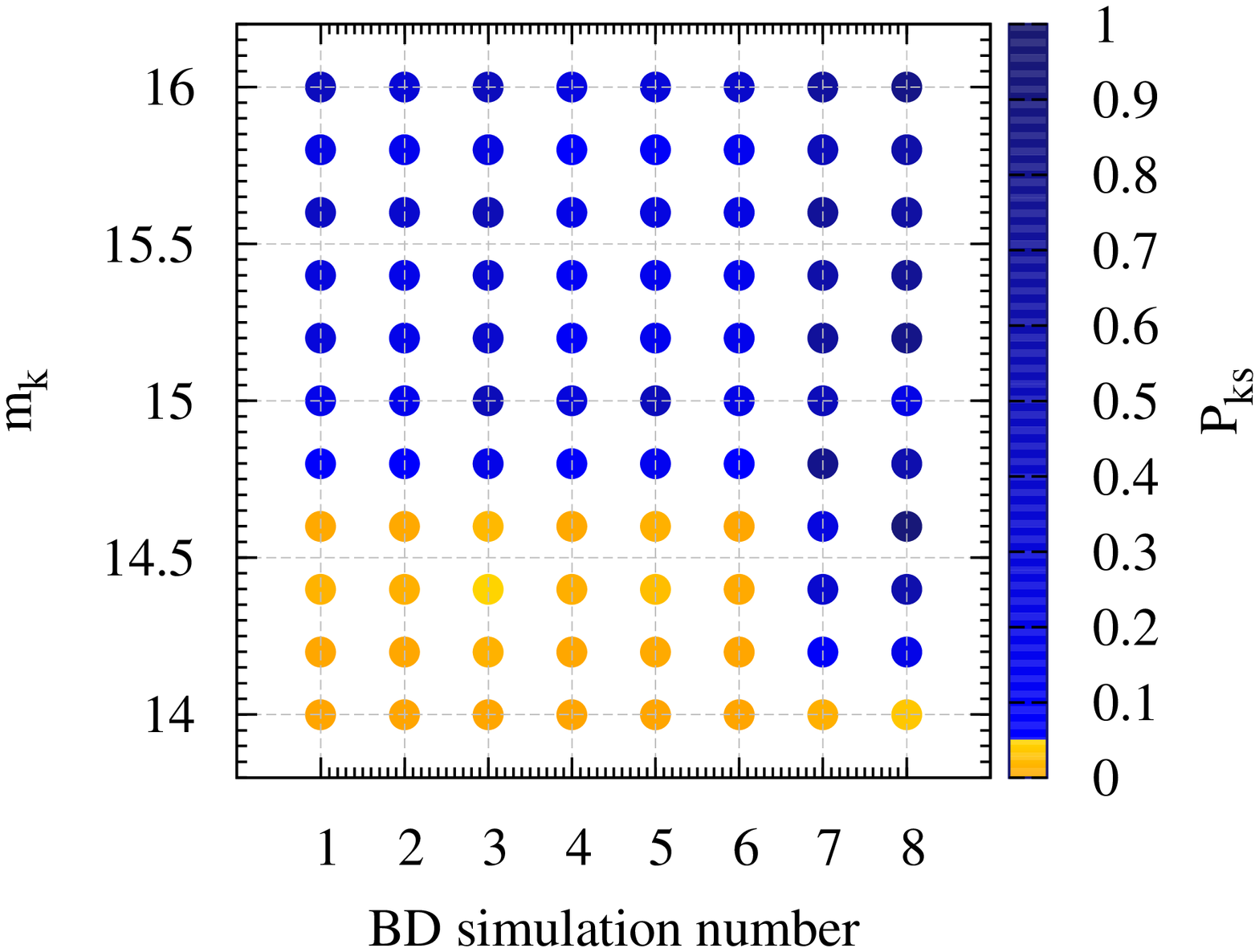}
    		\vspace{-0 pt}
\end{minipage}
\caption{Scatter plot of $p_{\rm ks}$ values between data (average of 100 realizations taking errors into account) and simulations. {\it Top left:} $p_{\rm ks}$ values between observational data with $14 \leq m_K < 15$ and {\it dissolved disk} (DD) simulations as a function of viewing direction with respect to the disk plane.  {\it Top right:} Same but for magnitude cut $15 \leq m_K < 16$. {\it Middle left:} Same but for magnitude cut $16 \leq m_K < 17$. {\it Middle right:} Same but for magnitude cut $15 \leq m_K < 17$. {\it Bottom left:} $p_{\rm ks}$ values between {\it binary disruption} (BD) simulations and observations as a function of $m_K$ selection. The $y$-axis shows the minimum $m_K$ value for each selection, the maximum value being one higher in magnitude ($\Delta m_K = 1$).} 
\label{fig:scatter_ks_sims}
\end{figure*}

In Figure \ref{fig:scatter_ks_sims} we plot the average $p_{\rm ks}$ value between the completeness corrected simulation (listed on $x$-axis) and stars from a particular magnitude bin (title) as a function of the viewing direction. The $p_{\rm ks}$ values are color-coded into two ranges for simplicity: orange for $p_{\rm ks} < 0.05$, blue for  $p_{\rm ks} \ge 0.05$, where we reject the null hypothesis that two samples are drawn from the same distribution if $p_{\rm ks} < 0.05$. The first four plots show that, for most viewing directions, the {\it dissolved disk} simulations are compatible with observations in the lowest magnitude range, $14 \leq m_K < 15$. Stars in the range $16 \leq m_K < 17$ are also mostly compatible with {\it dissolved disk} simulations. There are however few stars in this range, and both simulations and observations are close to a thermal distribution over all semi-major axes, though they are not so similar at large $p$. The fifth plot shows results for {\it binary disruption} simulations. The viewing direction is unimportant as the B-stars are distributed isotropically. In this plot the $y$-axis shows the lower value on the range of the magnitude cut taken for the observations. The range spans one magnitude in total, $\Delta m_K = 1$ (i.e., $m_K = 14$ spans the range $14 \leq m_K < 15$). This plot shows that lower magnitude stars and the binary disruption scenario are incompatible. A change occurs in the simulations BD$_{1-6}$ (young CW disk mass of $10^4 \Mo$) for the range $14.8 \leq m_K < 15.8$ and above. Here $p_{\rm ks} \geq 0.05$ for all simulations. This transition is also seen in the {\it dissolved disk} simulations but in the opposite direction: mostly  $p_{\rm ks} \geq 0.05$ for $14 \leq m_K < 14.8$, and $p_{\rm ks} < 0.05$ for $m_K \ge 14.8$.

\clearpage 

\begin{table}[h!]
\caption{Example results of 1D two-sample KS testing between completeness corrected simulations (random viewing direction) and observational data.} 
\begin{center}
\begin{tabular}{cc}
Samples & $p_{\rm ks}$ \\
\hline
$14 \leq m_K < 15$, BD$_8$ &  0.039 \\
$14 \leq m_K < 15$, DD$_1$ & 0.953  \\
\hline
$15 \leq m_K < 16$, BD$_8$ & 0.253 \\
$15 \leq m_K < 16$, DD$_1$ & 0.001 \\
\hline
$m_K \geq 16$, BD$_8$ & 0.997  \\
$m_K \geq 16$, DD$_1$ & 0.094  
\end{tabular}
\end{center}
\label{tab:pks}
\end{table}

In Table \ref{tab:pks} we provide a sample of $p_{\rm ks}$-values between the data binned in $K$-magnitude and completeness-corrected simulations. The population with $14 \leq m_K < 15$ is inconsistent with the basic {\it binary disruption} scenario (BD$_8$). In contrast, a comparison with the {\it dissolved disk} scenario (DD$_1$) yields a high $p_{\rm ks}$-value. Again, {\it dissolved disk} scenarios give more ambiguous predictions as the $h$-values are dependent on the viewing direction with respect to the B-star disk. Fainter ($m_K \geq 16$) stars have lower $|h|$-values and hence are more eccentric or edge-on in inclination. A KS test between this population and BD$_8$ yields a high $p_{\rm ks}$ value $= 0.997$. In Figure \ref{fig_cum_data_with_sims_cc} we plot the cumulative $|h|$-distribution function for observational data binned in $K$-magnitude and for the simulations that match well with observations --- DD$_1$ and BD$_8$ simulations sampled with completeness corrections.

\section{Discussion}\label{S:discussion}

We present a new, directly-observable statistic, $h$, which uses the position of stars on the sky ($x$, $y$) and their proper motion ($v_x$, $v_y$) to recognize groups of high-eccentricity orbits. It is particularly useful for stars with long-period orbits for which dynamical accelerations, and hence orbital parameters, are difficult to determine. We use a Monte Carlo ARMA code and $N$-body simulations to evolve stellar orbits in two formation scenarios for the B-stars in the GC; a {\it dissolved disk} scenario based on the model proposed by \citet{Set06}, and a {\it binary disruption} scenario due to enhanced stellar relaxation from massive perturbers by \citet{Per07}. We investigate the change in the B-star orbital parameters after $6 \Myr$ of gravitational interaction with the young CW disk and compare the results to observational data using the $h$-statistic. We summarize our results here:

\begin{enumerate}

\item Although the gravitational potential of the young CW disk can effectively exert torques on the orbits of the surrounding cluster stars within a few $\Myr$, for a disk mass of $\sim 10^4 \Mo$, stars with semi-major axes greater than $\sim 0.2 \pc$ retain memory of their origin through their eccentricity distribution. The more massive the young CW disk, the greater torque it  exerts and larger the eccentricity evolution of surrounding stars over $6 \Myr$. This result does not qualitatively change if the young CW disk is younger \--- such as the $\sim4 \Myr$ as found by \citet{Do13} and \citet{Lu13} \--- but the high-eccentricity signature of the B-stars in the {\it binary disruption} scenario will be even more prominent as the stars have less time to interact with the young CW disk.

\item  Simulations in which the B-star and young CW disks have small angles with respect to one another produce a large spread in angular momentum vectors of the young CW disk stars, in contrast to the observed dispersion angle \citep{Pau06,Bel06,Lu09,Bar09}. The concentration seen in the data for the young CW disk hints at a low initial eccentricity, a cusp rather than a core in stellar density and/or no second disk structure. However it is difficult to keep a distinct concentration of angular momentum vectors and, at the same time, produce  off-disk and counter-rotating orbits with a secular mechanism. 

\item The {\it binary disruption} scenario leaves a signature of decreasing mean values, and scatter, in $|h|$ with increasing radii. The {\it dissolved disk} scenario results in a broad range of $\langle |h| \rangle$-values with a large scatter due to the dependence on viewing angle with respect to the initial disk. 

\item The B-stars in our data set have lower $|h$|-values with increasing $K$-magnitude intervals. If their orbits are isotropically distributed, this means that the lower mass, potentially much-older B-stars are more eccentric than their more massive, younger companions. If the B-stars are preferentially aligned with the young CW disk, incomplete azimuthal coverage at large radii and sampling along the disk can lower the $|h|$-distribution and mimic a high eccentricity signature. However the $m_K < 14$ stars are sampled with the same azimuthal coverage as the $m_K \ge 14$ population so this cannot explain the difference between the low and high $K$-magnitude stars. 

\item The cumulative $|h|$-distribution function for the S-stars is similar to, but slightly lower than, that of an isotropic, thermal eccentricity distribution. As we know that they are isotropically distributed, this tells us that they form a population that is slightly more eccentric than thermal. This matches the distribution found from orbital fitting of individual stars \citep{Gil09a}. 

\begin{figure}[t!]
\vspace{-0pt}
  \centering
    \includegraphics[trim=0cm 0cm 0.cm 0cm, clip=true, angle = -90, width=0.54\textwidth]{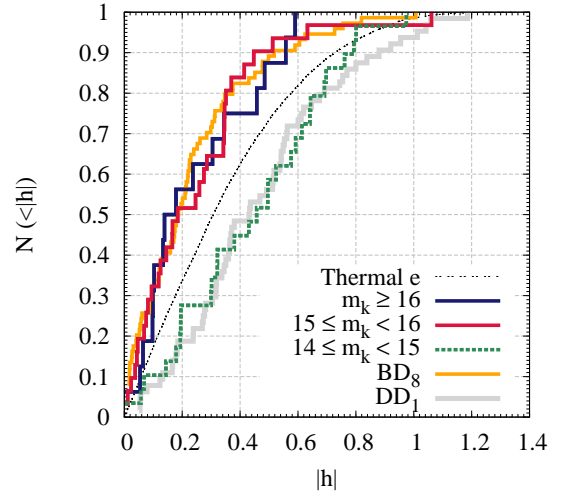}
    		\vspace{-0 pt}
		\caption{Cumulative $|h|$-distribution function for observational data (binned in $K$-magnitude), for an isotopic eccentricity distribution and for {\it dissolved disk} (DD$_1$) simulation and {\it binary disruption} (BD$_8$) simulation, both sampled using the observational completeness correction.} 
		\label{fig_cum_data_with_sims_cc}
		    		\vspace{-0.5 ex}

\end{figure}

\item Stars with $14 \leq m_K < 15$ and those with $m_K \geq 15$ have different cumulative $|h|$-distributions with a KS value of $p_{\rm KS} = 0.007$. This suggests that they are not drawn from the same ($e,i$) population. The difference is important for the interpretation of the K-band luminosity function (KLF) and initial mass function (IMF) of the young CW disk. If this dynamical information relates to a different origin mechanism, then then the KLF slope of the young CW disk must get flatter and the IMF is even more top-heavy than previously reported \citep{Bar10, Do13, Lu13}.   

\item The stars with the highest $K$-magnitudes in our sample, $m_K \geq 16$, have similar $|h|$-distributions to those with $15 \leq m_K < 16$ with a KS value of $p_{\rm KS} = 0.965$. They have lower $|h|$-values at large radii, $ p \geq 0.8 \arcsec$. If they are isotropically distributed, they are more eccentric than those located closer in projected distance to the MBH. Stars with $m_K \geq 16$ and $p \geq 0.8 \arcsec$ do not appear to be drawn from the same distribution as those with $m_K \geq 16$ and $p < 0.8 \arcsec$ ($p_{\rm KS} = 0.02$). In context of the {\it binary disruption} scenario, this can be explained by the decreasing relaxation times as orbits get closer to the MBH, which rapidly changes their low angular momenta and hence $|h|$-values. We would like to increase our sample of $m_K \geq 16$ stars to confirm this result.
 
\item In comparing the observed B-star data with simulations of the two formation scenarios, we find the following: Stars with $14 \leq m_K < 15$ have higher $|h|$-values than expected for a thermal distribution and hence fit better with the {\it dissolved disk} scenario. Given the short lifetimes for these stars ($t_{\rm MS} \lesssim 13 \Myr$), they are most likely members of the recent star formation episode which created with young CW disk. Fainter B-stars with $m_K \geq 15$, and hence longer lifetimes, have lower $|h|$-values than expected for a thermal distribution and for that reason fit better with Hills {\it binary disruption} scenario, though the data are not as eccentric (if isotropically distributed) as in most of the simulations. The best match to the data involves a steep stellar cusp such that coherence times for stellar torques are high at large radii, and/or a larger mass of the young CW disk ($\sim 4 \times 10^4 \Mo$).  

\end{enumerate}
   
An alternative scenario for the origin of the B-stars is formation in the same star formation episode that formed the young CW disk \citep{Pau06,Do13,Lu13}. To explain the low $|h|$-values of the high-magnitude stars however, there must exist a mechanism which differentiates between low- and high-mass stars in orbital eccentricity and/or inclination, and it must act on a short timescale. \citet{Ale07} show that, for an initially circular disk, energy relaxation between stars of different masses can significantly change the velocity dispersion of different populations of stars of different masses, and hence change their orbital eccentricities. We repeat their analysis and find that, though important for low and moderately eccentric disks ($e \lesssim 0.7$), this mechanism cannot account for the magnitude of the difference in $h$-values of the low and high $K$-magnitude stars.

It is important to obtain more observations of B-star positions and proper motions, particularly those at high $K$-magnitude and large projected radii, to increase the sample-size that can be used in comparison with simulations. Recent observations show that the oldest stars observable in the GC, red giants, do not form a cusp within $0.5 \pc$ of the MBH, in contradiction with theoretical predictions \citep[see e.g.,][]{Do13b}. Whether or not this result is specific to the red giant population or true also of the less luminous stellar distribution is important for NSC formation and evolution theory. If the high magnitude B-stars originate from the {\it binary disruption} scenario, the shape of the $\langle |h| \rangle$-$r$ relation can be a probe of the mass distribution of the dark stellar cluster.  \\

\acknowledgments{
Support for this work was provided by the National Aeronautics and Space Administration through Einstein Postdoctoral Fellowship Award
Number PF2-130095 issued by the Chandra X-ray Observatory Center, which is operated by the Smithsonian Astrophysical Observatory for and on behalf of the National Aeronautics Space Administration under contract NAS8-03060. Y.L. was supported by a VIDI fellowship from NWO and the Australian Research Council Future Fellowship. A.-M.M. thanks Fabio Antonini and Clovis Hopman for valuable comments on an early draft of this paper. She thanks Sylvana Yelda, Tuan Do and Leo Mayer for discussions. We thank the anonymous referee for useful comments.}

\appendix
 
 \section{Constraints on $\lowercase{h}$-statistic for bound stars}

The maximum value of $|h|$ for a bound stellar orbit is calculated by limiting the velocity of a star to the escape velocity at its projected radius. The maximum angular momentum of a stellar orbit on the plane of sky is
\begin{equation}
\begin{split}
j_z =& v_{\rm esc} . p \\
=& \sqrt{2G\Mbh p},
\end{split}
\end{equation}
and therefore,
\begin{equation}
h = \frac{j_z}{J_p} \le \sqrt{2}.
\end{equation}
 For a bound Kepler orbit, $h \le \sqrt{2}$. One can use this constraint to find unbound stars and/or stars affected by confusion with incorrect proper motion values. \\
  
A star with $1 < |h| \le \sqrt{2}$ has a $j_z$ value larger than the circular angular momentum at its projected radius. This requires that $e > 0$, but moreover that $p < a$, i.e., that the star's semi-major axis is larger than its projected radius. This star is traveling on the inner part of its orbit, closer to periapsis than apoapsis. This may provide an extra constraint on $z$, the position of a star along the line-of-sight, when estimating stellar orbital parameters \citep{Bar09, Lu09}.

\section{Effect of massive stellar potential on value of $\lowercase{h}$-statistic}
 
\begin{figure}[h!]
\vspace{-0pt}
  \centering
            \includegraphics[trim=0cm 0cm 0.cm 0cm, clip=true, angle = -90, scale=0.32]{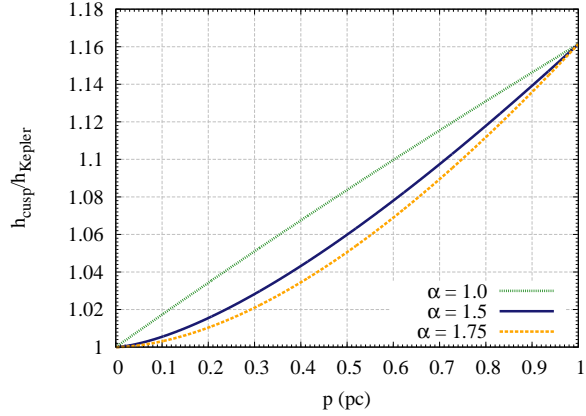}
    		\vspace{-0 pt}
		\caption{Fraction of $h$ evaluated with the mass of the stellar cusp over $h$ for a Kepler potential. The lines plotted are for a stellar mass of $1.5 \times 10^6 \Mo$ within $1 \pc$ and varying power-law indices $\alpha$ where the stellar mass density is $m(r) \propto r^{-\alpha}$.} 
		\label{fig_stellarmass}
\end{figure}

The $h$-statistic is defined for a Kepler orbit ($h = j_z/J_p$). In a real NSC, the mass contained within a stellar orbit is due to both the black hole mass and the enclosed mass of the stellar cusp. If we do not take the latter mass into account, our theoretical value of $J_p$, the maximum angular momentum at $p$, will be smaller than the true value. Hence, $|h|$ will appear artificially larger (i.e. a stellar orbit will appear less eccentric or more `face-on') than its true value. Due to the increasing enclosed stellar mass with radius, this effect will increase with radius. We quantify this in Figure \ref{fig_stellarmass} where we plot the fraction of two evaluations of $h$ (one takes stellar mass into account in calculating $J_p$, $h_{\rm cusp}$, the other does not, $h_{\rm Kepler}$) for a face-on circular orbit ($e = i = 0$). We assume a stellar mass of $1.5 \times 10^6 \Mo$ within $1 \pc$ and vary the power-law index $\alpha$, where the mass within radius $r$ scales as $m(<r) \propto r^{\alpha}$. From this simple analysis we see that a star with a face-on circular orbit at $p=0.4 \pc \sim 10 \arcsec$, will have a $h$-value that is fractionally larger by ($0.034, 0.043, 0.067$) than for a Kepler potential for this particular power-law density profile with $\alpha = (1.75, 1.5, 1.0)$. This plot can inform us on the magnitude of the expected error on $h$ due to the stellar cusp.
 
\section[]{Statistical constraints on orbital eccentricity and inclination from the $\lowercase{h}$-statistic}

\begin{figure}[h!]
\vspace{-0pt} 
  \centering
    \includegraphics[trim=0cm 0cm 0.cm 0cm, clip=true, angle = -90, width=0.58\textwidth]{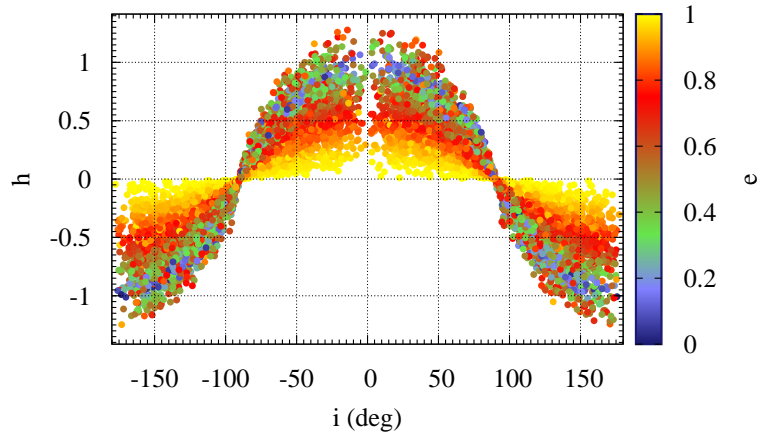}
    		\vspace{-0 pt}
		\caption{Inclination of stellar orbits as a function of $h$-value for an isotropic, thermal stellar distribution. Colors of points correspond to the stellar orbital eccentricities. $i = 0 \arcdeg$ ($i = 90 \arcdeg$) corresponds to a face-on (edge-on) orbit.} 
		\label{fig_inc}
\end{figure}

There is a degeneracy in the value of $h$ with respect to orbital eccentricity and inclination. However, if we assume a known distribution for one of the parameters we can place constraints on the mapping of the other to $h$. For example, we can take a cut in inclination (see Figure \ref{fig_inc}) and see how the range in $h$-values map to orbital eccentricity. 

\begin{figure*}[ht]
  \centering
    \includegraphics[trim=0cm 0cm 0cm 8cm, clip=true, angle = -90, scale=0.56]{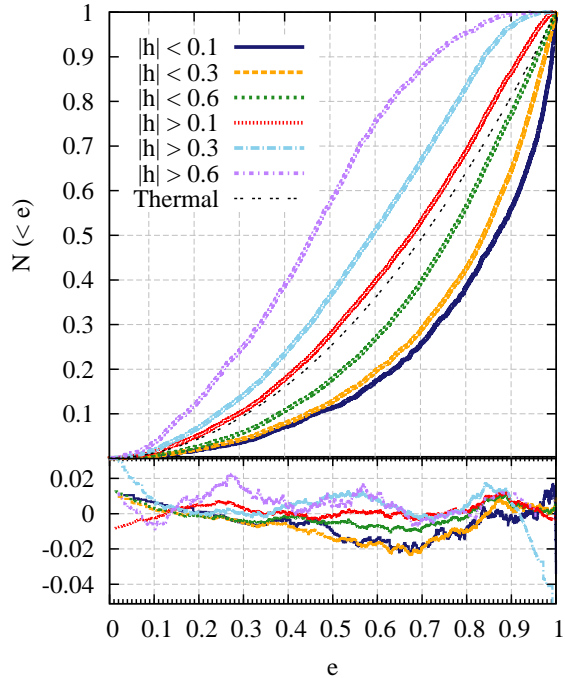} 
\caption{Cumulative plots of stellar orbital eccentricity for different ranges in $|h|$. The full stellar distribution has a thermal eccentricity profile, $f(e) = e^2$, also plotted for comparison. In Table \ref{tab:cdf_params} the residuals from the simulated cumulative distribution functions and the fits to the function in Equation \ref{eq:fit} are plotted.}
\label{fig:cum_e_iso}
\end{figure*}

\begin{table}[h]
\centering
\caption{\label{tab:cdf_params} Fits to Equation (\ref{eq:fit})}
\begin{tabular}{ccccccc}
\hline
\hline 
        \noalign{\smallskip}
$|h|^{1}$ & $\mu$ & $\sigma^2$  & $\beta$  & $\gamma$  & $\delta$  & $\epsilon$ \tabularnewline
\hline 
        \noalign{\smallskip}
$< 0.1$ & $0.96 $ & $0.17$ & $-1.06$ & $-1.22$ & $1.20$ & $1.10$ \tabularnewline
$< 0.3$ & $1.06$ & $0.01$ & $4.22$ & $0.15$ & $-0.16$ & $-4.09$\tabularnewline
$< 0.6$ & $1.66$ & $0.16$ & 5.43 & $0.09$ & $-0.10$ & $-5.26$\tabularnewline
$> 0.1$ & $0.34$ & $0.13$ & 5.75 & $0.29$ & $ -0.45$ & $-6.52$\tabularnewline
$> 0.3$ & $0.71$ & $0.07$ & -0.62 & $0.01$ & $ -0.05$ & $1.00$\tabularnewline
$> 0.6$ & $0.51$ & $0.05$ & -2.80 & $-0.54$ & $ 0.51$ & $3.15$ \tabularnewline
\hline 
\hline
\multicolumn{7}{l}{{\scriptsize {\bf Notes.}}}\tabularnewline
\multicolumn{7}{l}{{\scriptsize $^{1}$ Absolute value of $h$ corresponding to distributions in Figure \ref{fig:cum_e_iso}.}}\tabularnewline
        \noalign{\smallskip}
                \noalign{\smallskip}
\end{tabular}
\end{table}

Statistical constraints are placed on the mapping of orbital eccentricities to $|h|$ for an isotropically distributed thermal cluster of stars. We plot the cumulative distribution of orbital eccentricities in specific $|h|$ ranges in Figure \ref{fig:cum_e_iso}. $50 \%$ of stars with $|h| < 0. 1 (0.3)$ have an orbital eccentricity $e > 0.87 (0.84)$, while $90 \%$ of stars with $|h| > 0.6$ have an orbital eccentricity $e < 0.73$. We fit the cumulative distributions shown in Figure \ref{fig:cum_e_iso} with the following formula
\begin{equation}\label{eq:fit}
\begin{split}
f(x) &= \dfrac{1}{2} \left[ 1 + {\rm erf} \left( \dfrac{x - \mu}{\sqrt{2\sigma^2}} \right) \right]  \\
&+ \beta x + \gamma \sqrt{1 - x^2} + \delta \cos(x) + \epsilon \sin(x), 
\end{split}
\end{equation}
where $\mu$ and $\sigma^2$ are the mean and variance of a cumulative Gaussian distribution, $\beta$, $\gamma$, $\delta$ and $\epsilon$ are coefficients and $x = e$. The trigonometric terms are chosen ad-hoc; they are necessary to fit the extreme distributions $|h| >0.6$, $|h| < 0.1$. The fitted values for each line in Figure \ref{fig:cum_e_iso} are listed in Table \ref{tab:cdf_params}; the residuals from subtracting the functions from the distributions are plotted in the lower panel.

\begin{figure*}[t!]
\begin{minipage}[b]{0.47\linewidth}
\centering
    \includegraphics[trim=0cm 0cm 0.cm 0cm, clip=true, angle = -90, scale=0.4]{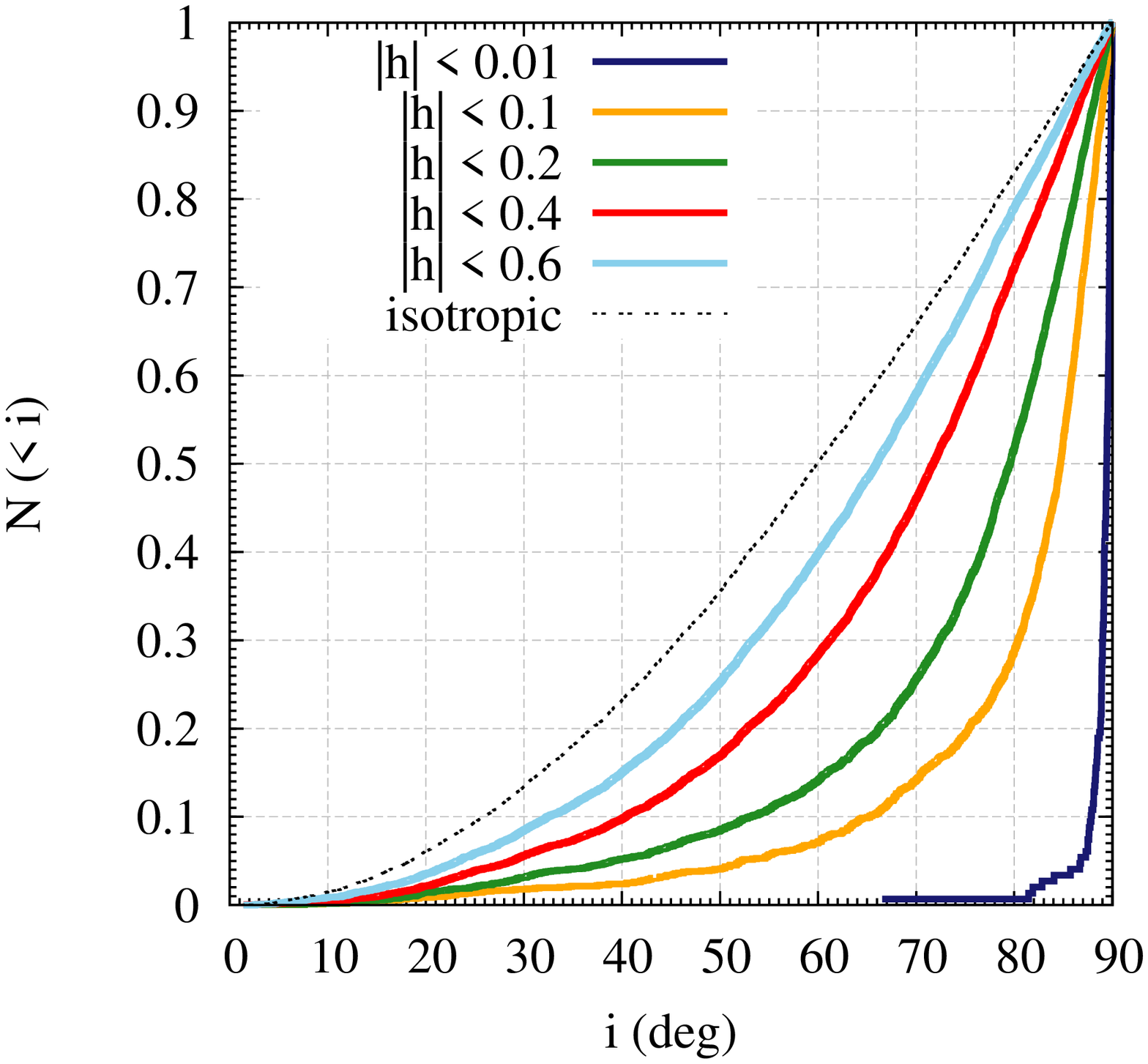}
\end{minipage}
\begin{minipage}[b]{0.47\linewidth}
\centering
    \includegraphics[trim=0cm 0cm 0.cm 0cm, clip=true, angle = -90, scale=0.4]{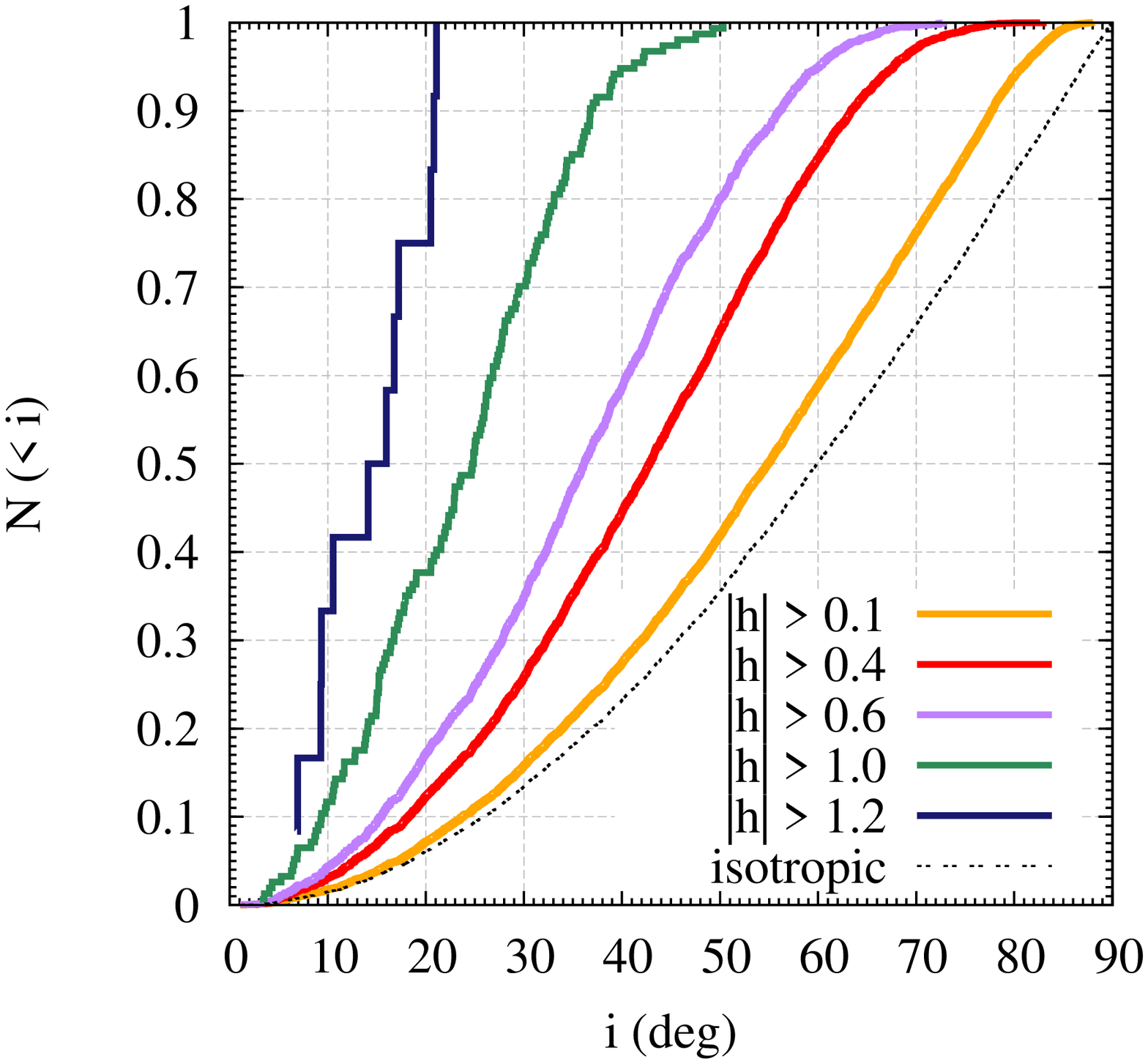}
\end{minipage}
\caption{Cumulative distribution function of stellar inclinations for different ranges of $|h|$ for an isotropic, thermal distribution of stellar orbits. $i = 0 \arcdeg$ ($i = 90 \arcdeg$) corresponds to a face-on (edge-on) orbit.} 
		\label{fig_inc_cum}
\end{figure*}

Statistical constraints are also placed on the mapping of orbital inclinations to $h$ for an isotropically distributed thermal cluster of stars. In Figure \ref{fig_inc_cum}, we plot the cumulative distribution function of the stellar inclinations for different $h$-values. Those with $|h| > 1$ (right) have orbital inclinations such that their angular momenta are aligned close to the $z$-axis (face-on orbits). Stars with $|h| > 1.2$ have almost face-on orbits. Stars with very low values of $|h|$ are likely to have angular momenta that are highly inclined to the $z$-axis. Almost all stars with $|h| < 0.01$ have inclinations of $\sim 90\arcdeg$ (edge-on orbit).

\section[]{Convergence Testing}

We run two {\it binary disruption} simulations with BD$_{1}$ parameters and individual stellar masses of $20 \Mo$ (BD$_{20}$) and $50 \Mo$ (BD$_{50}$) for convergence testing. For the $20 \Mo$ simulation, $\langle |h| \rangle = 0.105 (0.103)$, $\sigma_h = 0.057 (0.053)$,  $s_e = 0.003 (0.004)$ for stars with projected radii $p \geq 7(10) \arcsec$. For the $50 \Mo$ simulation, $\langle |h| \rangle = 0.112 (0.110)$, $\sigma_h = 0.068 (0.061)$,  $s_e = 0.007 (0.008)$ for $p \geq 7(10) \arcsec$. These values demonstrate that simulations with smaller stellar masses produce the same results (c.f. Table \ref{tab:sim_params}). In Figure \ref{fig_inc_cum} we compare the cumulative distribution functions of $|h|$-values drawn from a random viewing direction for each simulation. KS testing between the distributions yield $p_{\rm ks} = 0.84, 0.79, 0.98$ for BD$_{1}$ and BD$_{20}$, BD$_{1}$ and BD$_{50}$, BD$_{20}$ and BD$_{50}$ respectively. These results are insensitive to the chosen viewing direction.

\begin{figure}[t!]
\vspace{-0pt}  
  \centering
    \includegraphics[trim=0cm 0cm 0.cm 0cm, clip=true, angle = -90, width=0.58\textwidth]{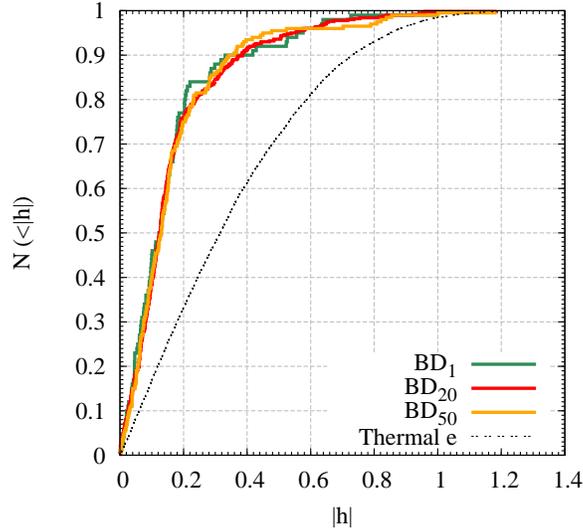}
    		\vspace{-0 pt}
		\caption{Cumulative $|h|$-distribution of {\it binary disruption} simulations BD$_{1}$, BD$_{20}$ and BD$_{50}$ (simulations with BD$_{1}$ parameters but with different individual stellar masses) with $|h|$-values drawn from a random viewing direction.} 
		\label{fig_converge}
\end{figure}


\begin{thebibliography}{77}

\bibitem[{{Agarwal} \& {Milosavljevi{\'c}}(2011)}]{Aga11}
{Agarwal}, M. \& {Milosavljevi{\'c}}, M. 2011, \apj, 729, 35

\bibitem[{{Alexander} {et~al.}(2007){Alexander}, {Begelman}, \&
  {Armitage}}]{Ale07}
{Alexander}, R.~D., {Begelman}, M.~C., \& {Armitage}, P.~J. 2007, \apj, 654,
  907

\bibitem[{{Antonini}(2013)}]{Ant13a}
{Antonini}, F. 2013, \apj, 763, 62

\bibitem[{{Antonini} {et~al.}(2012){Antonini}, {Capuzzo-Dolcetta},
  {Mastrobuono-Battisti}, \& {Merritt}}]{Ant12a}
{Antonini}, F., {Capuzzo-Dolcetta}, R., {Mastrobuono-Battisti}, A., \&
  {Merritt}, D. 2012, \apj, 750, 111

\bibitem[{{Antonini} \& {Merritt}(2013)}]{Ant13b}
{Antonini}, F. \& {Merritt}, D. 2013, \apjl, 763, L10

\bibitem[{{Bar-Or} {et~al.}(2013){Bar-Or}, {Kupi}, \& {Alexander}}]{Bar12}
{Bar-Or}, B., {Kupi}, G., \& {Alexander}, T. 2013, \apj, 764, 52

\bibitem[{{Bartko} {et~al.}(2009){Bartko}, {Martins}, {Fritz}, {Genzel},
  {Levin}, {Perets}, {Paumard}, {Nayakshin}, {Gerhard}, {Alexander},
  {Dodds-Eden}, {Eisenhauer}, {Gillessen}, {Mascetti}, {Ott}, {Perrin},
  {Pfuhl}, {Reid}, {Rouan}, {Sternberg}, \& {Trippe}}]{Bar09}
{Bartko}, H., {Martins}, F., {Fritz}, T.~K., et al. 2009, \apj, 697, 1741
  
\bibitem[{{Bartko} {et~al.}(2010){Bartko}, {Martins}, {Trippe}, {Fritz},
  {Genzel}, {Ott}, {Eisenhauer}, {Gillessen}, {Paumard}, {Alexander},
  {Dodds-Eden}, {Gerhard}, {Levin}, {Mascetti}, {Nayakshin}, {Perets},
  {Perrin}, {Pfuhl}, {Reid}, {Rouan}, {Zilka}, \& {Sternberg}}]{Bar10}
{Bartko}, H., {Martins}, F., {Trippe}, S., et al. 2010, \apj, 708, 834

\bibitem[{{Beloborodov} {et~al.}(2006){Beloborodov}, {Levin}, {Eisenhauer},
  {Genzel}, {Paumard}, {Gillessen}, \& {Ott}}]{Bel06}
{Beloborodov}, A.~M., {Levin}, Y., {Eisenhauer}, F., et al. 2006, \apj, 648, 405

\bibitem[{{Bertelli} {et~al.}(1994){Bertelli}, {Bressan}, {Chiosi}, {Fagotto},
  \& {Nasi}}]{Ber94}
{Bertelli}, G., {Bressan}, A., {Chiosi}, C., {Fagotto}, F., \& {Nasi}, E. 1994,
  \aaps, 106, 275

\bibitem[{{Blum} {et~al.}(2003){Blum}, {Ram{\'{\i}}rez}, {Sellgren}, \&
  {Olsen}}]{Blu03}
{Blum}, R.~D., {Ram{\'{\i}}rez}, S.~V., {Sellgren}, K., \& {Olsen}, K. 2003,
  \apj, 597, 323

\bibitem[{{Bonnell} \& {Rice}(2008)}]{BoR08}
{Bonnell}, I.~A. \& {Rice}, W.~K.~M. 2008, Science, 321, 1060

\bibitem[{{Bonnet} {et~al.}(2004){Bonnet}, {Abuter}, {Baker}, {Bornemann},
  {Brown}, {Castillo}, {Conzelmann}, {Damster}, {Davies}, {Delabre},
  {Donaldson}, {Dumas}, {Eisenhauer}, {Elswijk}, {Fedrigo}, {Finger},
  {Gemperlein}, {Genzel}, {Gilbert}, {Gillet}, {Goldbrunner}, {Horrobin}, {Ter
  Horst}, {Huber}, {Hubin}, {Iserlohe}, {Kaufer}, {Kissler-Patig}, {Kragt},
  {Kroes}, {Lehnert}, {Lieb}, {Liske}, {Lizon}, {Lutz}, {Modigliani}, {Monnet},
  {Nesvadba}, {Patig}, {Pragt}, {Reunanen}, {R{\"o}hrle}, {Rossi}, {Schmutzer},
  {Schoenmaker}, {Schreiber}, {Stroebele}, {Szeifert}, {Tacconi}, {Tecza},
  {Thatte}, {Tordo}, {van der Werf}, \& {Weisz}}]{Bon04}
{Bonnet}, H., {Abuter}, R., {Baker}, A., et al. 2004, The Messenger, 117, 17

\bibitem[{{Brown} {et~al.}(2012){Brown}, {Cohen}, {Geller}, \&
  {Kenyon}}]{Bro12b}
{Brown}, W.~R., {Cohen}, J.~G., {Geller}, M.~J., \& {Kenyon}, S.~J. 2012,
  \apjl, 754, L2

\bibitem[{{Capuzzo-Dolcetta}(1993)}]{Cap93}
{Capuzzo-Dolcetta}, R. 1993, \apj, 415, 616

\bibitem[{{Cox}(2000)}]{Cox00}
{Cox}, A.~N. 2000, {Allen's astrophysical quantities} (New York: AIP Press;
  Springer, 4th ed.)

\bibitem[{{Danby}(1992)}]{Dan92}
{Danby}, J.~M.~A. 1992, {Fundamentals of celestial mechanics} (Richmond:
  Willman-Bell, |c1992, 2nd ed.)

\bibitem[{{Do} {et~al.}(2013){Do}, {Lu}, {Ghez}, {Morris}, {Yelda}, {Martinez},
  {Wright}, \& {Matthews}}]{Do13}
{Do}, T., {Lu}, J.~R., {Ghez}, A.~M., et al. 2013, \apj, 764, 154

\bibitem[{{Do} {et~al.}(2013){Do}, {Martinez}, {Yelda}, {Ghez}, {Bullock}, {Kaplinghat},
  {Lu}, {Peter}, \&  {Phifer}}]{Do13b}
{Do}, T., {Martinez}, G.~D., {Yelda}, S., et al. 2013, \apjl, 779, L6

\bibitem[{{Eisenhauer} {et~al.}(2003){Eisenhauer}, {Sch{\"o}del}, {Genzel},
  {Ott}, {Tecza}, {Abuter}, {Eckart}, \& {Alexander}}]{Eis03}
{Eisenhauer}, F., {Sch{\"o}del}, R., {Genzel}, R., et al. 2003, \apjl, 597, L121

\bibitem[{{Eisenhauer} {et~al.}(2005)}]{Eis05}
{Eisenhauer}, F. {et~al.} 2005, \apj, 628, 246

\bibitem[{{Fritz} {et~al.}(2011){Fritz}, {Gillessen}, {Dodds-Eden}, {Lutz},
  {Genzel}, {Raab}, {Ott}, {Pfuhl}, {Eisenhauer}, \& {Yusef-Zadeh}}]{Fri11}
{Fritz}, T.~K., {Gillessen}, S., {Dodds-Eden}, K., et al.
  2011, \apj, 737, 73

\bibitem[{{Genzel} {et~al.}(2010){Genzel}, {Eisenhauer}, \&
  {Gillessen}}]{Gen11}
{Genzel}, R., {Eisenhauer}, F., \& {Gillessen}, S. 2010, Reviews of Modern
  Physics, 82, 3121

\bibitem[{{Genzel} {et~al.}(2003)}]{Gen03a}
{Genzel}, R., {Sch{\"o}del}, R., {Ott}, T.  {et~al.} 2003, \apj, 594, 812

\bibitem[{{Ghez} {et~al.}(2005){Ghez}, {Salim}, {Hornstein}, {Tanner}, {Lu},
  {Morris}, {Becklin}, \& {Duch{\^ e}ne}}]{Ghe05}
{Ghez}, A.~M., {Salim}, S., {Hornstein}, S.~D., et al. 2005, \apj, 620, 744

\bibitem[{{Ghez} {et~al.}(2008){Ghez}, {Salim}, {Weinberg}, {Lu}, {Do}, {Dunn},
  {Matthews}, {Morris}, {Yelda}, {Becklin}, {Kremenek}, {Milosavljevic}, \&
  {Naiman}}]{Ghe08a}
{Ghez}, A.~M., {Salim}, S., {Weinberg}, N.~N., et al. 2008, \apj, 689, 1044

\bibitem[{{Ghez} {et~al.}(2003)}]{Ghe03a}
{Ghez}, A.~M. {et~al.} 2003, \apjl, 586, L127

\bibitem[{{Gillessen} {et~al.}(2009{\natexlab{a}}){Gillessen}, {Eisenhauer},
  {Fritz}, {Bartko}, {Dodds-Eden}, {Pfuhl}, {Ott}, \& {Genzel}}]{Gil09a}
{Gillessen}, S., {Eisenhauer}, F., {Fritz}, T.~K.,et al. 2009{\natexlab{a}}, \apjl, 707,
  L114

\bibitem[{{Gillessen} {et~al.}(2009{\natexlab{b}}){Gillessen}, {Eisenhauer},
  {Trippe}, {Alexander}, {Genzel}, {Martins}, \& {Ott}}]{Gil09b}
{Gillessen}, S., {Eisenhauer}, F., {Trippe}, S., et al. 2009{\natexlab{b}}, \apj, 692, 1075

\bibitem[{{Gould} \& {Quillen}(2003)}]{Gou03}
{Gould}, A. \& {Quillen}, A.~C. 2003, \apj, 592, 935

\bibitem[{{Gualandris} {et~al.}(2012){Gualandris}, {Mapelli}, \&
  {Perets}}]{Gua12}
{Gualandris}, A., {Mapelli}, M., \& {Perets}, H.~B. 2012, \mnras, 427, 1793

\bibitem[{{Hartmann} {et~al.}(2011){Hartmann}, {Debattista}, {Seth},
  {Cappellari}, \& {Quinn}}]{Har11}
{Hartmann}, M., {Debattista}, V.~P., {Seth}, A., {Cappellari}, M., \& {Quinn},
  T.~R. 2011, \mnras, 418, 2697

\bibitem[{{Hartung} {et~al.}(2003){Hartung}, {Lenzen}, {Hofmann}, {B{\"o}hm},
  {Brandner}, {Finger}, {Fusco}, {Lacombe}, {Laun}, {Granier}, {Storz}, \&
  {Wagner}}]{Har03}
{Hartung}, M., {Lenzen}, R., {Hofmann}, R., et al. 2003, in Society of Photo-Optical Instrumentation
  Engineers (SPIE) Conference Series, Vol. 4841, Society of Photo-Optical
  Instrumentation Engineers (SPIE) Conference Series, ed. {M.~Iye \&
  A.~F.~M.~Moorwood}, 425--436

\bibitem[{{Hills}(1988)}]{Hil88}
{Hills}, J.~G. 1988, \nat, 331, 687

\bibitem[{{Hills}(1991)}]{Hil91}
---. 1991, \aj, 102, 704

\bibitem[{{Hopkins} \& {Quataert}(2010{\natexlab{a}})}]{Hop10a}
{Hopkins}, P.~F. \& {Quataert}, E. 2010{\natexlab{a}}, \mnras, 407, 1529

\bibitem[{{Hopkins} \& {Quataert}(2010{\natexlab{b}})}]{Hop10b}
---. 2010{\natexlab{b}}, \mnras, 405, L41

\bibitem[{{Ivanov} {et~al.}(2005){Ivanov}, {Polnarev}, \& {Saha}}]{Iva05}
{Ivanov}, P.~B., {Polnarev}, A.~G., \& {Saha}, P. 2005, \mnras, 358, 1361

\bibitem[{{Kinoshita} {et~al.}(1991){Kinoshita}, {Yoshida}, \& {Nakai}}]{Kin91}
{Kinoshita}, H., {Yoshida}, H., \& {Nakai}, H. 1991, Celestial Mechanics and
  Dynamical Astronomy, 50, 59

\bibitem[{{Kocsis} \& {Tremaine}(2011)}]{Koc11}
{Kocsis}, B. \& {Tremaine}, S. 2011, \mnras, 412, 187

\bibitem[{{Leigh} {et~al.}(2012){Leigh}, {B{\"o}ker}, \& {Knigge}}]{Lei12}
{Leigh}, N., {B{\"o}ker}, T., \& {Knigge}, C. 2012, \mnras, 424, 2130

\bibitem[{{Levin}(2007)}]{Lev07}
{Levin}, Y. 2007, \mnras, 374, 515

\bibitem[{{Levin} \& {Beloborodov}(2003)}]{Lev03}
{Levin}, Y. \& {Beloborodov}, A.~M. 2003, \apjl, 590, L33

\bibitem[{{Lu} {et~al.}(2013){Lu}, {Do}, {Ghez}, {Morris}, {Yelda}, \&
  {Matthews}}]{Lu13}
{Lu}, J.~R., {Do}, T., {Ghez}, A.~M., et al. 2013, \apj, 764, 155

\bibitem[{{Lu} {et~al.}(2009){Lu}, {Ghez}, {Hornstein}, {Morris}, {Becklin}, \&
  {Matthews}}]{Lu09}
{Lu}, J.~R., {Ghez}, A.~M., {Hornstein}, S.~D., et al. 2009, \apj, 690, 1463

\bibitem[{{Madigan} {et~al.}(2011){Madigan}, {Hopman}, \& {Levin}}]{Mad11b}
{Madigan}, A.-M., {Hopman}, C., \& {Levin}, Y. 2011, \apj, 738, 99

\bibitem[{{Madigan} \& {Levin}(2012)}]{Mad12}
{Madigan}, A.-M. \& {Levin}, Y. 2012, \apj, 754, 42

\bibitem[{{Madigan} {et~al.}(2009){Madigan}, {Levin}, \& {Hopman}}]{Mad09}
{Madigan}, A.-M., {Levin}, Y., \& {Hopman}, C. 2009, \apjl, 697, L44

\bibitem[{{Maeder} \& {Meynet}(2004)}]{Mae04}
{Maeder}, A. \& {Meynet}, G. 2004, in IAU Symposium, Vol. 215, Stellar
  Rotation, ed. A.~{Maeder} \& P.~{Eenens}, 500

\bibitem[{Martins {et~al.}(2007)Martins, Genzel, Hillier, Eisenhauer, Paumard,
  Gillessen, Ott, \& Trippe}]{Mar07}
Martins, F., Genzel, R., Hillier, D.~J., et al. 2007, A\&A, 468, 233

\bibitem[{{Martins} {et~al.}(2008){Martins}, {Gillessen}, {Eisenhauer},
  {Genzel}, {Ott}, \& {Trippe}}]{Mar08}
{Martins}, F., {Gillessen}, S., {Eisenhauer}, F., et al. 2008, \apjl, 672, L119

\bibitem[{{Merritt}(2010)}]{Mer10}
{Merritt}, D. 2010, \apj, 718, 739

\bibitem[{{Milosavljevi{\'c}}(2004)}]{Mil04b}
{Milosavljevi{\'c}}, M. 2004, \apjl, 605, L13

\bibitem[{{Morris}(1993)}]{Mor93}
{Morris}, M. 1993, \apj, 408, 496

\bibitem[{{Nayakshin} \& {Cuadra}(2005)}]{Nay05b}
{Nayakshin}, S. \& {Cuadra}, J. 2005, \aap, 437, 437

\bibitem[{{Nishiyama} \& {Sch{\"o}del}(2013)}]{Nis12}
{Nishiyama}, S. \& {Sch{\"o}del}, R. 2013, \aap, 549, A57

\bibitem[{{Paumard} {et~al.}(2006){Paumard}, {Genzel}, {Martins}, {Nayakshin},
  {Beloborodov}, {Levin}, {Trippe}, {Eisenhauer}, {Ott}, {Gillessen}, {Abuter},
  {Cuadra}, {Alexander}, \& {Sternberg}}]{Pau06}
{Paumard}, T., {Genzel}, R., {Martins}, F., et al. 2006,
  \apj, 643, 1011

\bibitem[{{Perets} \& {Gualandris}(2010)}]{Per10}
{Perets}, H.~B. \& {Gualandris}, A. 2010, \apj, 719, 220

\bibitem[{{Perets} {et~al.}(2009){Perets}, {Gualandris}, {Kupi}, {Merritt}, \&
  {Alexander}}]{Per09b}
{Perets}, H.~B., {Gualandris}, A., {Kupi}, G., {Merritt}, D., \& {Alexander},
  T. 2009, \apj, 702, 884

\bibitem[{{Perets} {et~al.}(2007){Perets}, {Hopman}, \& {Alexander}}]{Per07}
{Perets}, H.~B., {Hopman}, C., \& {Alexander}, T. 2007, \apj, 656, 709

\bibitem[{{Pfahl}(2005)}]{Pfa05}
{Pfahl}, E. 2005, \apj, 626, 849

\bibitem[{{Pfuhl} {et~al.}(2011){Pfuhl}, {Fritz}, {Zilka}, {Maness},
  {Eisenhauer}, {Genzel}, {Gillessen}, {Ott}, {Dodds-Eden}, \&
  {Sternberg}}]{Pfu11}
{Pfuhl}, O., {Fritz}, T.~K., {Zilka}, M., et al. 2011, \apj, 741, 108

\bibitem[{{Rauch} \& {Tremaine}(1996)}]{Rau96}
{Rauch}, K.~P. \& {Tremaine}, S. 1996, New Astronomy, 1, 149

\bibitem[{{Rousset} {et~al.}(2003){Rousset}, {Lacombe}, {Puget}, {Hubin},
  {Gendron}, {Fusco}, {Arsenault}, {Charton}, {Feautrier}, {Gigan}, {Kern},
  {Lagrange}, {Madec}, {Mouillet}, {Rabaud}, {Rabou}, {Stadler}, \&
  {Zins}}]{Rou03}
{Rousset}, G., {Lacombe}, F., {Puget}, P., et al. 2003, in Society of
  Photo-Optical Instrumentation Engineers (SPIE) Conference Series, Vol. 4839,
  Society of Photo-Optical Instrumentation Engineers (SPIE) Conference Series,
  ed. {P.~L.~Wizinowich \& D.~Bonaccini}, 140--149

\bibitem[{{Saha} \& {Tremaine}(1992)}]{Sah92}
{Saha}, P. \& {Tremaine}, S. 1992, \aj, 104, 1633

\bibitem[{{Salaris} \& {Cassisi}(2006)}]{Sal06}
{Salaris}, M. \& {Cassisi}, S. 2006, {Evolution of Stars and Stellar
  Populations}

\bibitem[{{Sanders}(1998)}]{San98}
{Sanders}, R.~H. 1998, \mnras, 294, 35

\bibitem[{{Sch{\"o}del} {et~al.}(2007){Sch{\"o}del}, {Eckart}, {Alexander},
  {Merritt}, {Genzel}, {Sternberg}, {Meyer}, {Kul}, {Moultaka}, {Ott}, \&
  {Straubmeier}}]{Sch07}
{Sch{\"o}del}, R., {Eckart}, A., {Alexander}, T., et al. 2007, \aap, 469, 125

\bibitem[{{Sch{\"o}del} {et~al.}(2009){Sch{\"o}del}, {Merritt}, \&
  {Eckart}}]{Sch09}
{Sch{\"o}del}, R., {Merritt}, D., \& {Eckart}, A. 2009, \aap, 502, 91

\bibitem[{{Seth} {et~al.}(2006){Seth}, {Dalcanton}, {Hodge}, \&
  {Debattista}}]{Set06}
{Seth}, A.~C., {Dalcanton}, J.~J., {Hodge}, P.~W., \& {Debattista}, V.~P. 2006,
  \aj, 132, 2539

\bibitem[{{Stetson}(1987)}]{Ste87}
{Stetson}, P.~B. 1987, \pasp, 99, 191

\bibitem[{{Touma} \& {Sridhar}(2011)}]{Tou12}
{Touma}, J.~R. \& {Sridhar}, S. 2012, \mnras, 423, 2083

\bibitem[{{Tremaine} {et~al.}(1975){Tremaine}, {Ostriker}, \&
  {Spitzer}}]{Tre75}
{Tremaine}, S.~D., {Ostriker}, J.~P., \& {Spitzer}, Jr., L. 1975, \apj, 196,
  407

\bibitem[{{Trippe} {et~al.}(2008){Trippe}, {Gillessen}, {Gerhard}, {Bartko},
  {Fritz}, {Maness}, {Eisenhauer}, {Martins}, {Ott}, {Dodds-Eden}, \&
  {Genzel}}]{Tri08}
{Trippe}, S., {Gillessen}, S., {Gerhard}, O.~E., et al. 2008, \aap, 492, 419

\bibitem[{{{\v S}ubr} {et~al.}(2009){{\v S}ubr}, {Schovancov{\'a}}, \&
  {Kroupa}}]{Sub09}
{{\v S}ubr}, L.~., {Schovancov{\'a}}, J., \& {Kroupa}, P. 2009, \aap, 496, 695

\bibitem[{{Wardle} \& {Yusef-Zadeh}(2008)}]{War08}
{Wardle}, M. \& {Yusef-Zadeh}, F. 2008, \apjl, 683, L37

\bibitem[{{Wardle} \& {Yusef-Zadeh}(2012)}]{War12}
---. 2012, \apjl, 750, L38

\bibitem[{{Wisdom} \& {Holman}(1991)}]{Wis91}
{Wisdom}, J. \& {Holman}, M. 1991, \aj, 102, 1528

\end{thebibliography}
\end{document}